\documentclass[english,11pt,a4paper]{article}
\pdfoutput=1
\usepackage{jheppub}

\usepackage[T1]{fontenc}
\usepackage[utf8]{inputenc}
\usepackage{lmodern, amsmath, xcolor, slashed, xfrac}
\usepackage[normalem]{ulem}
\def\figureautorefname~#1\null{fig.\,#1\null}

\newcommand{\subappref}[1]{\hyperref[#1]{appendix~\ref{#1}}}

\def\equationautorefname~#1\null{eq.\,(#1)\null}
\renewcommand{\eqref}[1]{\hyperref[#1]{(\ref{#1})}}
\newcommand{\eqs}[3][--]{\hyperref[#2]{eqs.~(\ref{#2})#1}\eqref{#3}}

\newcommand{\chir}[1]{\if#100\else\if#1+R\else\if#1-L\else\ifx#1\mp\sfrac{L\!}{\!R\,}\else\ifx#1\pm\sfrac{R\!}{\!L}\else#1\fi\fi\fi\fi\fi}

\allowdisplaybreaks[4]
\def\beq#1\eeq{\begin{align}#1\end{align}}
\newcommand{\beqa}{\begin{eqnarray}}
\newcommand{\eeqa}{\end{eqnarray}}
\newcommand{\nn}{\nonumber}

\newcommand{\eqndot}{{\,.}}
\newcommand{\eqncomma}{{\,,}}

\DeclareMathOperator{\CP}{CP}

\DeclareMathOperator{\diag}{diag}

\DeclareMathOperator{\Order}{\mathcal{O}}

\newcommand{\bpm}{\begin{pmatrix}}
\newcommand{\epm}{\end{pmatrix}}
\newcommand{\ket}[1]{{|#1\rangle}}
\newcommand{\bs}[1]{{\boldsymbol{#1}}}
\newcommand{\bra}[1]{{\langle #1|}}
\newcommand{\Ket}[1]{{|#1]}}
\newcommand{\Bra}[1]{{[#1|}}

\newcommand{\parn}[1]{\left(#1\right)}

\newcommand{\psic}{{\psi^c}}
\newcommand{\Lam}{{\bar\Lambda}}

\newcommand{\amp}[3]{($#1${}$#2${}$#3$)}
\newcommand{\sq}[1]{\sqrt{#1}}
\newcommand{\rzw}{r_{\!\scriptscriptstyle{ZW}}}

\newcommand{\sqb}[1]{[#1]}
\newcommand{\anb}[1]{\langle#1\rangle}

\newcommand{\ketBra}[3]{\ket{#1}\Bra{#2}}
\newcommand{\ketbra}[2]{\ket{#1}\bra{#2}}
\newcommand{\Ketbra}[3]{\Ket{#1}\bra{#2}}
\newcommand{\KetBra}[2]{\Ket{#1}\Bra{#2}}

\newcommand{\splitFunc}[1]{\mathrm{Split}}

\makeatletter
\def\EE{\@ifnextchar-{\@@EE}{\@EE}}
\def\@EE#1{\ifnum#1=1 \times10 \else \times10^{#1}\fi}
\def\@@EE#1#2{\times10^{-#2}}
\makeatother

\newcommand{\SL}[2]{{SL}\left(#1,#2\right)}

\newcommand{\Slash}[1]{{\ooalign{\hfil/\hfil\crcr$#1$}}}

\newcommand{\eg}{{\em e.g.}}
\newcommand{\ie}{{\em i.e.}}

\makeatletter\g@addto@macro\bfseries{\boldmath}\makeatother

\title{The electroweak effective field theory from on-shell amplitudes}

\author[a]{Gauthier Durieux,}
\author[a,b,c]{Teppei Kitahara,}
\author[a]{Yael Shadmi,}
\author[a]{and Yaniv Weiss}

\affiliation[a]{Physics Department, Technion---Israel Institute of Technology,\\
Technion city, Haifa 3200003, Israel}
\affiliation[b]{Institute for Advanced Research, Nagoya University,\\ Furo-cho Chikusa-ku, Nagoya 464-8601, Japan}
\affiliation[c]{Kobayashi-Maskawa Institute for the Origin of Particles and the Universe, Nagoya University, Furo-cho Chikusa-ku, Nagoya 464-8602, Japan}
\emailAdd{durieux@campus.technion.ac.il}
\emailAdd{teppeik@kmi.nagoya-u.ac.jp}
\emailAdd{yshadmi@physics.technion.ac.il}
\emailAdd{yanivwe@campus.technion.ac.il}

\abstract{%
We apply on-shell methods to the bottom-up construction of electroweak amplitudes, allowing for both renormalizable and non-renormalizable interactions.
We use the little-group covariant massive-spinor formalism, and flesh out some of its details along the way.
Thanks to the compact form of the resulting amplitudes, many of their properties, and in particular the constraints of perturbative unitarity, are easily seen in this formalism.
Our approach is purely bottom-up, assuming just the standard-model electroweak spectrum as well as  the conservation of electric charge and fermion number.
The most general massive three-point amplitudes consistent with these symmetries are derived and studied in detail, as the primary building blocks for the construction of scattering amplitudes.
We employ a simple argument, based on tree-level unitarity of four-point amplitudes, to identify the three-point amplitudes that are non-renormalizable at tree level.
This bottom-up analysis remarkably reproduces many low-energy relations implied by electroweak symmetry through the standard-model Higgs mechanism and beyond it.
We then discuss four-point amplitudes.
The gluing of three-point amplitudes into four-point amplitudes in the massive spinor  helicity formalism is clarified.
As an example, we work out the $\psic\psi Zh$ amplitude, including also the non-factorizable part.
The latter is an all-order expression in the effective-field-theory expansion.
Further constraints on the couplings are obtained by requiring perturbative unitarity.
In the $\psic\psi Zh$ example, one for instance obtains the renormalizable-level relations between vector and fermion masses and gauge and Yukawa couplings.
We supplement our bottom-up derivations with  a matching of three- and four-point amplitude coefficients onto the standard-model effective field theory (SMEFT) in the broken electroweak phase.
This establishes the correspondence with the usual Lagrangian approach and paves the way for SMEFT computations in the on-shell formalism.
}

\keywords{Scattering Amplitudes, Effective Field Theories, Spontaneous Symmetry Breaking}

\usepackage{ifpdf,feynmp}
\ifpdf\fi
\AtBeginDocument{\begin{fmffile}{fgraph}% close this env before the bib
\fmfcmd{%
arrow_ang := 11;
arrow_len := 3.5thick;
curly_len := 2.5thick;}
}

\begin{document}
\maketitle

%----------------------------------------------------------------------------------------
%	Section1
%----------------------------------------------------------------------------------------
\section{Introduction}
\label{sec:intro}
The bottom-up construction of effective-field-theory (EFT) Lagrangians proceeds from a field content and imposed---spacetime, local, and global---symmetries~\cite{Weinberg:1978kz}.
Similarly, the structure of on-shell amplitudes is stringently constrained by Lorentz and global internal symmetries, combined with locality and unitarity (for reviews see, \eg, refs.~\cite{Elvang:2013cua, Dixon:2013uaa, Schwartz:2013pla, Henn:2014yza, Elvang:2015rqa, Cheung:2017pzi}).
This parallel motivates the application of on-shell methods to the direct derivation of EFT amplitudes.
The advantages of this approach include the absence of gauge or operator redundancies, and the prospect of further leveraging the power of on-shell techniques in EFT contexts.
As a first step in this direction, tree-level amplitudes involving a massive color singlet of spin $0$ or $1$ and up to three gluons were derived using on-shell methods in ref.~\cite{Shadmi:2018xan}, including the contributions of non-renormalizable operators up to dimension $13$.
This procedure also gives a simple counting and classification of independent EFT operators at a given dimension, providing an alternative to EFT Lagrangian constructions~\cite{Leung:1984ni, Buchmuller:1985jz, Jenkins:2009dy, Grzadkowski:2010es, Lehman:2015coa, Henning:2015alf, Lehman:2015via, Henning:2017fpj, Gripaios:2018zrz, Henning:2019enq, Henning:2019mcv, Ruhdorfer:2019qmk}.
In this paper, we apply these methods to the electroweak sector of the standard model (SM).

A recent addition to the on-shell toolbox, which is essential for SM applications, is the systematic treatment of massive spinning particles introduced in ref.~\cite{Arkani-Hamed:2017jhn}.
Each massive momentum is decomposed as the sum of two lightlike vectors, which transform under the $SU(2)$ little group.
The corresponding spinors then carry an $SU(2)$ little-group index, and symmetric combinations of these spinors are employed to represent particles of any spin.
Little-group covariance implies powerful selection rules on the allowed structure of amplitudes.
For related work and previous applications see refs.~\cite{Conde:2016vxs,Christensen:2018zcq, Shadmi:2018xan, Ma:2019gtx, Herderschee:2019ofc,Herderschee:2019dmc,Aoude:2019tzn, Ochirov:2018uyq, Bonifacio:2018vzv, Chung:2018kqs, Chen:2019dkx, Alonso:2019ptb}.

Spontaneous electroweak symmetry breaking (EWSB) through the Brout-Englert-Higgs mechanism~\cite{Englert:1964et, Higgs:1964pj} is a crucial ingredient of the SM.
In a bottom-up construction of on-shell amplitudes, the natural objects to consider are the physical SM particles in the broken phase.
One of the main goals of this paper is to explicitly see how EWSB emerges from this construction.
Given the electroweak particle content, the $SU(2)_L\times U(1)_Y$ local symmetry follows from perturbative tree-level unitarity imposed on a sufficient number of four- and higher-point amplitudes~\cite{LlewellynSmith:1973yud,Joglekar:1973hh, Cornwall:1974km,Vayonakis:1976vz,Lee:1977eg,Chanowitz:1985hj}.
The relations between couplings and masses predicted by the SM Higgs mechanism should emerge in this way.
At the non-renormalizable level, perturbative unitarity would also yield relations between various couplings, and in particular, between  amplitudes differing only by the numbers of external Higgs legs (see sec.~3.3 of ref.~\cite{Maltoni:2019aot} for an example featuring a fermion dipole operator).
The standard-model effective field theory (SMEFT) should emerge below cutoffs parametrically higher than the electroweak scale.
Alternatively, by relaxing the requirement of perturbative unitarity at the non-renormalizable level, our bottom-up construction can be used to explore general non-renormalizable interactions, beyond the SMEFT, such as non-linear realizations of $SU(2)_L\times U(1)_Y$~\cite{Alonso:2012px,Buchalla:2013rka, Brivio:2013pma, deFlorian:2016spz}.

A related study recently appeared in ref.~\cite{Aoude:2019tzn}, where broken-phase amplitudes involving certain bosonic SMEFT operators were computed.
We instead adopt a bottom-up approach, without imposing the $SU(2)_L\times U(1)_Y$ symmetry at the outset, and also include fermion amplitudes.
Massive three-point SM amplitudes were studied in ref.~\cite{Christensen:2018zcq}, focusing on renormalizable interactions.\footnote{While submitting this paper, we also noted the new ref.~\cite{Christensen:2019mch} where decay processes are computed from the renormalizable three-point amplitudes of ref.~\cite{Christensen:2018zcq}.}
There are some discrepancies between our results that of ref.~\cite{Christensen:2018zcq} where non-renormalizable contributions were argued to vanish in the massless limit.
Three-point amplitudes of supersymmetric theories and their high-energy limits were studied in ref.~\cite{Herderschee:2019ofc}, and used as building blocks for the construction of higher-point amplitudes in theories with ${\cal N}=4$ supersymmetry~\cite{Herderschee:2019dmc}.  
It was demonstrated in ref.~\cite{Ma:2019gtx} that the on-shell construction of amplitudes in the unbroken phase efficiently reproduces the counting of dimension-six SMEFT operators.

We adopt a simplified electroweak-like spectrum with a pair of massive Dirac fermions $\psi,\psi'$ of electric charges differing by one unit (plus the corresponding anti-fermions), the massive $Z$ and $W^\pm$ electroweak bosons, the Higgs $h$, and the photon.
The fermion pair can represent the up- and down-type quarks of any generation, or a charged lepton and neutrino.
Gluons are not considered and trivial color structures are understood where necessary.
The most general amplitudes consistent with electric charge and fermion number conservation are constructed, including both renormalizable and non-renormalizable interactions.
These assumptions let us focus on the most phenomenologically relevant amplitudes but could easily be relaxed.
Considering Majorana particles would  be more involved.
Besides its particle spectrum, the resulting theory is characterized by the coefficients of a finite number of three-point amplitudes and of an infinite set of higher-point contact terms.
These correspond to the coefficients of independent non-renormalizable operators.

We first construct and analyze all the three-point amplitudes.
These depend on several different scales, including the particle masses and the cutoff $\Lam$ of the broken-phase theory.
As usual, three-point amplitudes can be formulated \emph{on-shell} for any external masses by analytically continuing the momenta to the complex plane.
 General prescriptions for constructing the allowed spinorial structures in three-point amplitudes of different masses and spins were given in ref.~\cite{Arkani-Hamed:2017jhn}.
In some cases, most notably the $WWZ$ amplitude, it is actually easier to directly build the list of independent spinor structures.
As a cross-check,  we use a simple group-theory argument to determine the number of independent structures in each three-point amplitude.
With three external particles, this number is given by the number of ways to form overall angular-momentum singlets from the combinations of three spins.

Having obtained the three-point amplitudes, our next task is to identify the non-renormalizable tree-level terms.
For this purpose, one needs to analyze the high-energy growth of physical scattering amplitudes, starting with the four-point amplitudes.
Terms growing as $E$ or faster must be suppressed by an appropriate power of the cutoff $\Lam$.
It suffices however to examine the three-point amplitudes.
Since these are defined in the complex plane, one can consider their \emph{high-energy} or massless limit, which encodes information about the high-energy behavior of physical scattering amplitudes.

This high-energy limit was used in ref.~\cite{Arkani-Hamed:2017jhn} to uncover many properties of the Higgs mechanism in two renormalizable examples: a fully Higgsed $U(1)$, and a fully-Higgsed non-abelian gauge theory.
In our electroweak toy model, the high-energy limit of the amplitudes easily reveals many known features of EWSB.
Thus for example, we see that fermions with a vector-like coupling to the $Z$ do not interact with its longitudinal component in this limit;
conversely, for fermions with a chiral coupling to the $Z$, the coupling to the longitudinal component is proportional to the fermion mass.
While this high-energy limit is equivalent to taking all the masses to zero at the same rate---preserving the particle spectrum---different scenarios can be studied by taking the external masses to zero at different rates.
The fermion-fermion-$Z$ amplitude for instance tells us that the vector mass cannot be taken to zero faster than the fermion mass, unless the coupling is vector-like (see \autoref{sec:ppZ}).
The $WWZ$ amplitude is of special significance, and the underlying $SU(2)$ gauge-group structure follows from the consideration of this three-point amplitude on its own.
For three degenerate bosons (\ie, for $m_Z=m_W$), we show that the three vector coupling is proportional to $\epsilon^{abc}$.
Unlike in the massless case, the three vectors can be taken to have identical polarizations in this case, so that Bose symmetry under permutations of all three bosons is readily seen.
For $m_Z\neq m_W$, the renormalizable amplitude is fully determined by a single coupling and the $W$, $Z$ mass ratio, and automatically preserves $U(1)_{\rm EM}$.
Furthermore, only two types of non-renormalizable couplings are allowed.
The first is associated with the three-field-strength dimension-six operator.
The other vanishes in  the limit $m_Z\to m_W$.
At tree level it is therefore proportional to $m_W^2-m_Z^2$.
Indeed, in the SM, this term is proportional to the electromagnetic coupling.

We then turn to the calculation of four-point tree-level amplitudes, and derive the $\psic\psi Zh$ amplitude as an example.
Generically, these amplitudes have factorizable parts, featuring single-particle poles, and non-factorizable parts, with no poles.
To derive the amplitudes one can write down the full list of kinematic structures, consistent with Lorentz symmetry, and the little group in particular, allowing for single-particle poles.
For the massive four-point amplitudes we consider, each term contains a single pole, corresponding to a particular factorization channel.
The residue at each pole is given by the product of the two on-shell, three-point  amplitudes associated with this factorization channel.
This uniquely determines the factorizable part of the tree-level amplitude.
This procedure can also be described as the \emph{gluing} of two three-point amplitudes.
Note that, since we only work at tree-level, we can only consistently glue together two tree-level three-point amplitudes.
In our bottom-up derivation, the remaining, non-factorizable part features arbitrary coefficients, corresponding to the unknown Wilson coefficients of non-renormalizable operators.
Combinations of operators giving rise to three-point amplitudes that vanish on-shell do not appear in this gluing procedure.
Their contributions are accounted for by the \emph{non-factorizable} terms in the on-shell approach.
The constructibility of EFT amplitudes was studied in ref.~\cite{Cohen:2010mi}.
A judicious choice of momentum shifts, with all-line holomorphic or antiholomorphic shifts, was shown to give vanishing boundary terms in these examples.
The main result is very intuitive: contact-term contributions can only be inferred from lower-point amplitudes if they are related to them by a symmetry~\cite{Cohen:2010mi}.
Earlier applications of factorization to the derivation of tree-level EFT amplitudes were studied in refs.~\cite{Dixon:1993xd,Dixon:2004za}.

The $\psic\psi Zh$ amplitude is relevant for Higgs decay as well as $Zh$ production at hadron and lepton colliders.
We derive the factorizable and non-factorizable components of this amplitude at the tree level.
As explained above, the non-factorizable part is determined by Lorentz symmetry, and in particular by little-group considerations.
The general, all-order result is given by $12$ independent spinorial structures.
As in ref.~\cite{Shadmi:2018xan}, the expansion of non-factorizable amplitude coefficients in $(p_i\cdot p_j)/\Lam^2$ can be used to infer the number of independent operators contributing at each $\Lam$ order.
At the leading order in the EFT expansion, there are only four contact-terms.
Imposing perturbative unitarity on four-point (and higher-point) amplitudes would restrict the allowed contact terms, and reproduce the constraints of $SU(2)_L\times U(1)_Y$.
Here we restrict our attention to the renormalizable part of the $\psic\psi Zh$ amplitude.
Perturbative unitarity determines the fermion mass in terms of the $Z$-boson mass, the Yukawa coupling and the gauge coupling.

Finally, we provide the matching of all the unknown coefficients appearing in the three-point and $\psic\psi Zh$ amplitudes to the broken-phase SMEFT (in the conventions of ref.~\cite{Dedes:2017zog}).
This completes the results obtained in ref.~\cite{Aoude:2019tzn} with fermionic three-point amplitudes and the four-point $\psic\psi Zh$ one.
With this, all elements needed for computing the factorizable contributions to other electroweak four-point amplitudes are available.

This paper is organized as follows.
We describe the essentials of the  massive-spinor formalism~\cite{Arkani-Hamed:2017jhn} in \autoref{sec:bolded}.
The construction and discussion of three-point amplitudes are addressed in \autoref{sec:3pt}.
At the end of this section, we comment on the matching to the amplitudes of the high-energy unbroken theory, and note how the unbroken $SU(2)_L\times U(1)_Y$ emerges.
Tree-level factorizable and non-factorizable contributions to the $\psic\psi Zh$ four-point amplitude are then obtained in \autoref{sec:4-pt}, including the leading non-renormalizable terms.
We derive in particular the advertised relation between the gauge and Yukawa couplings and the $Z$-boson and fermion masses.
Our conventions and notations are defined in \autoref{sec:formalism}, where we also clarify some elements of the massive spinor formalism.
Subsequent appendices contain a matching of amplitude coefficients to the SMEFT in the broken phase (\autoref{sec:matching}), a compendium of massless amplitudes (\autoref{sec:massless}), and a derivation of the non-factorizable $\psic\psi Zh$ amplitudes for massless fermions (\autoref{sec:ffZh_EFT_derivation}).

%----------------------------------------------------------------------------------------
%	Section2
%----------------------------------------------------------------------------------------
\section{Massive spinor formalism}
\label{sec:bolded}
Spinor helicity variables have been widely used for the computation of massless helicity amplitudes (for reviews, see \eg\ refs.~\cite{Mangano:1990by, Elvang:2015rqa, Schwartz:2013pla}).
With  the four-momenta and external polarizations written in terms of the basic building blocks of the Lorentz group, namely, massless spinors, the amplitudes take compact forms, and their little-group structure is manifest.
Here we briefly summarize the essential elements of the spinor variables we will use.
For details, we refer the reader to \autoref{sec:formalism}.

A lightlike momentum is written in terms of  a pair of chiral and anti-chiral spinors,
\begin{align}\label{eqn:massless_momenta_spinors}
	p_{\alpha\dot{\alpha}} \equiv p_\mu \sigma^\mu_{\alpha\dot{\alpha}} = \lambda_\alpha \tilde\lambda_{\dot\alpha} \equiv \ketBra{p}{p} \,,~~~~ 
	~~~~
	\bar{p}^{\dot\alpha \alpha} \equiv p_\mu \bar\sigma^{\mu \dot\alpha\alpha}
	= \tilde\lambda^{\dot\alpha} \lambda^{\alpha} \equiv \Ketbra{p}{p}\,,
\end{align}
where the half-brackets are defined in \autoref{sec:formalism}. 
These spinors carry undotted or dotted indices and transform as the $(1/2,0)$ and $(0,1/2)$ representations of the $\SL{2}{\mathbb{C}}$ Lorentz group, respectively.
Here and in the following, we interchangeably refer to these spinors as dotted and undotted, or square- and angle-spinors.
Little-group transformations, which form a $U(1)$ in the massless case, correspond to opposite  phase rotations of square- and angle-spinors,%
\footnote{Note that $\xi$ is real for real Lorentzian momenta, such that \autoref{eqn:massless_momenta_spinors} is Hermitian.
When the momenta are complex the parameter $\xi$ can be complex.}
\begin{align}
	\ket{p}\to e^{-i\xi}\ket{p} \,,~~~~ \text{and} ~~~~ \Bra{p} \to e^{
	+i\xi}\Bra{p}\,.
\end{align}
These leave the momentum invariant, but phase-rotate the polarization associated with the external fermion or vector of momentum $p$.
Amplitudes of particles with non-zero spin thus carry little-group charges, or weights, and these lead to selection rules on the allowed spinor structures.

Different generalizations of this formalism for the treatment of massive particles have been explored (see, \eg, refs.~\cite{Kleiss:1985yh,Dittmaier:1998nn,Arkani-Hamed:2017jhn}).
Here we adopt the method of ref.~\cite{Arkani-Hamed:2017jhn}.
A massive momentum ($p^2 = m^2 > 0$) can be decomposed in terms of chiral and anti-chiral spinors similarly to \autoref{eqn:massless_momenta_spinors}, but two pairs of massless spinors are now required, 
\begin{align}\label{eqn:massive_momenta_spinors}
	\bs{p}_{\alpha\dot{\alpha}} = \bs{\lambda}^I_\alpha \bs{\tilde\lambda}_{I\dot\alpha} = \ketBra{\bs{p}^I}{\bs{p}_I} \,,~~~~ \text{and} ~~~~
 \bar{\bs{p}}^{\dot\alpha \alpha} = -\bs{\tilde\lambda}^{I \dot\alpha} \bs{\lambda}_{I}^{\alpha} = -\Ketbra{\bs{p}^I}{\bs{p}_I}\,.
\end{align}
Here $I=1,2$, and boldface is used to denote massive momenta and their corresponding spinors.
The equation of motion then reads,
\beq\label{eqn:eom}
\bs{p} \Ket{\bs{p}^{I }}
=   m \ket{\bs{p}^I}\,,  ~~
\bar{\bs{p}} \ket{ \bs{p}^I }= 
m \Ket{ \bs{p}^{I }}\,,~~
\Bra{\bs{p}^{I}} \bar{\bs{p}}=  -  m \bra{\bs{p}^{I}}\,,  ~~
 \bra{\bs{p}^{I }} \bs{ p} =    -  m \Bra{\bs{p}^{I}} \,.
\eeq

Little-group transformations, which form an $SU(2)$ in the massive case, rotate the vectors $\bs{p}^I$, and leave $\bs{p}$ invariant,
\begin{align}
	\ket{\bs{p}^I} \to W^I_{\, J}\ket{\bs{p}^J} \,,~~~~ \text{and} ~~~~ \Bra{\bs{p}_I} \to \parn{W^{-1}}^{\, J}_{I}\Bra{\bs{p}_J} \,.
\end{align}
For real Lorentzian momenta, $W^I_J$ is a matrix in the $SU(2)$ subgroup of $\SL{2}{\mathbb{C}}$.

Since any spin $s$ can be obtained by combining spin-1/2 representations, the polarizations of massive particles can be expressed as completely symmetric $SU(2)$ tensors of rank $2s$.
An amplitude with a massive spin-$s$ particle of momentum $\bs{p}$ can then be written as (see ref.~\cite{Arkani-Hamed:2017jhn}),
\begin{align}
	\mathcal{M}^{I_1 \dots I_{2s}}
= \ket{\bs{p}}^{I_1}_{\alpha_{1}} \dots \ket{\bs{p}}^{I_{2s}}_{\alpha_{2s}} M^{\left\lbrace \alpha_{1} \dots \alpha_{2s} \right\rbrace } 
\;
= \Ket{\bs{p}}^{I_1}_{\dot\alpha_{1}} \dots \Ket{\bs{p}}^{I_{2s}}_{\dot\alpha_{2s}} \tilde{M}^{\left\lbrace \dot\alpha_{1} \dots \dot\alpha_{2s} \right\rbrace } \eqncomma
\label{eqn:spin_amp}
\end{align}
where $I_1\dots I_{2s}$ are massive little-group indices.
The reduced amplitudes $M$ and $\tilde{M}$ are  equivalent, as one can switch between the dotted and undotted spinors with~\autoref{eqn:eom}. 
They are also completely symmetric in the $\SL{2}{\mathbb{C}}$ indices $\alpha$ and $\dot{\alpha}$.

Thus, the amplitudes we write below are given in terms of angle- and square-spinors, and as usual, we abbreviate the spinor associated with external leg $i$ by $|p_i]\to |i]$ or $|p_i\rangle\to |i\rangle$.
The multiple occurrences of boldfaced $|\bs{i}]$, $|\bs{i}\rangle$ spinors in an amplitude
 are understood to carry $SU(2)$ little-group indices which are completely symmetrized.
For a fermion external leg, there is a single such spinor, while a  vector external leg carries two spinors whose indices are symmetrized.
Thus for example, if particle $\bs{1}$ is a massive vector, the amplitude could feature $|\bs{1}]|\bs{1}]\equiv|\bs{1}^{\{I_1}]|\bs{1}^{I_2\}}]$.
Note that we define the symmetric states using the appropriate Clebsch-Gordan coefficients (see \autoref{eqn:bold_clebsch}).
This convention is convenient for handling physical states.

Obviously, the choice of the pair of vectors $\bs{p}^I$ is arbitrary.
Generically, this can be used to choose arbitrary spin-quantization axes, and our general expressions do not assume any particular direction.
Different values for the $I_1\dots I_{2s}$ indices in \autoref{eqn:spin_amp} yield the amplitudes for the different spin polarizations along the chosen direction.
Massive-vector amplitudes for transverse polarizations are for instance obtained with $I_1=I_2=1$ and $2$, while $I_1\neq I_2$ yield the amplitude for zero polarization.
For considering the high-energy limit where helicity amplitudes are the natural choice, it will be convenient to choose the spatial direction of $\bs{p}^{I}$ along the direction of the particle momentum.
The leading contribution in the high-energy limit can then simply be obtained by unbolding the spinors~\cite{Arkani-Hamed:2017jhn}.

Further details of this formalism and the high-energy limit are discussed in  \autoref{sec:formalism}.

%----------------------------------------------------------------------------------------
%	Section3
%----------------------------------------------------------------------------------------
%%%%%%%%%%%%%%%%%%%%%%%%%%%%%%%%%%%%%
\section{Three-point amplitudes}
\label{sec:3pt}
%%%%%%%%%%%%%%%%%%%%%%%%%%%%%%%%%%%%%

In this section, we study the massive three-point amplitudes of our simplified electroweak theory, including non-renormalizable contributions.
In our bottom-up approach, the only dimensionful parameters are the masses and the EFT cutoff scale $\Lam$.
The Higgs vacuum expectation value never appears explicitly.
We distinguish $\Lam$ in the broken-phase theory from $\Lambda$ in the unbroken SMEFT.
The latter only appears explicitly in the matching of \autoref{sec:matching}.
Terms suppressed by $\Lam^{\bar n}$ correspond to SMEFT contributions of order $m^{n}/\Lambda^{\bar n+n}$ where $m$ generically denotes external particle masses.

For each amplitude, we derive the most general set of independent kinematic structures allowed.
Structures generated by non-renormalizable couplings at the tree level can be identified based on simple dimensional analysis in the high-energy limit.
This limit should of course be interpreted in the complex plane, where the amplitudes can be defined on-shell independently of external particle masses.
Angle- and square-spinors are then no longer related by complex conjugation, and the massless three-point kinematics can be smoothly approached, with square-spinor bilinears scaling as $E$, and angle-spinor bilinears scaling as $m^2/E$, or vice versa.
Here $m$ stands for any of the external-state masses.

Assuming that the cutoff $\Lam$ is parametrically higher than $m$, terms growing as $E^2$ in three-point amplitudes correspond to non-renormalizable tree-level interactions and are therefore assigned a prefactor $1/\Lam$.
This simple rule of thumb follows from the high-energy behavior of the four-point amplitudes at tree-level.
Unitarity restricts the energy growth of these amplitudes.
Terms growing as $E$ or faster should be suppressed by the cutoff $\Lam$, where perturbative unitarity is lost.
Since the tree-level four-point amplitudes are obtained by gluing two three-point amplitudes with a propagator scaling as $1/E^2$, three-point amplitudes growing as $E^2$ should be suppressed by $\Lam$ as well.
Indeed, three-point amplitudes have mass dimension $1$, so $\mathcal{O}(E)$ terms are consistent with renormalizable couplings, while $\mathcal{O}(E^2)$ growths must be accompanied by an inverse mass.
Such terms can of course be generated by the renormalizable SM couplings at the loop level, but these contributions cannot be consistently included in the tree-level gluing we rely on.
In full one-loop computations, the high-energy growth of loop three-point functions 
 shuts-off in the massless limit.

The implicit assumption in the discussion above is that the new physics at $\Lam$ is decoupling and can be integrated out.
If it is not, $\Lam$ cannot be taken to be much higher than $m$, and its precise value depends on the measured values of the different couplings.
In all cases considered, our results are consistent with the SMEFT.
Eventually, the systematic evaluation of four- and higher-point amplitudes should be performed to determine the exact constraints applying to three-point amplitude coefficients in a bottom-up way.
For our toy electroweak theory, this task is however left for future work.

%%%%%%%%%%%%%%%%%%%%%%%%%%%%%%%%%%
\subsection{\texorpdfstring{$\psic\psi Z$}{ffZ}}
\label{sec:ppZ}
%%%%%%%%%%%%%%%%%%%%%%%%%%%%%%%%%%

\newcommand{\mffz}[3]{\mathcal{M}(1_\psic^#1,2_\psi^#2,3_Z^#3)}
\newcommand{\cffz}[3]{c_{\psic\psi Z}^{\chir{#1}\chir{#2}\chir{#3}}}

We first consider the massive fermion-fermion-vector amplitude, with equal-mass fermions.
Since this is our first example, we will cover it in detail.
It is not difficult to show that the $\mathcal{M}(\bs{1}_\psic, \bs{2}_{\psi},\bs{3}_{Z})$ amplitude is fully characterized by the following spinorial structures:
\begin{equation}
\mathcal{M}(\bs{1}_\psic, \bs{2}_{\psi},\bs{3}_{Z})=
  \frac{\cffz+++}{\Lam}	\sqb{\bs{13}}\sqb{\bs{23}}
+ \frac{\cffz-+0}{m_Z}	\anb{\bs{13}}\sqb{\bs{23}}
+ \frac{\cffz+-0}{m_Z}	\sqb{\bs{13}}\anb{\bs{23}}
+ \frac{\cffz---}{\Lam}	\anb{\bs{13}}\anb{\bs{23}}\,.
\label{eqn:mffz}
\end{equation}
Our specific choice of dimensional normalization will
prove appropriate for our tree-level description.
The $\cffz{}{}{}$ coefficients are dimensionless and their superscripts
refer to chiralities.%
\footnote{Recall that all particles are incoming.
For our choice of massive momenta decomposition, the fermion chirality label corresponds to the fermion helicity in the massless limit.
Thus, for $\bs1$ and $\bs2$ being charge conjugates of each other, the second and third terms with $LR$ and $RL$ give rise to the chirality preserving gauge couplings, and the $RR$ and $LL$ terms are chirality violating.
The vector labels are not as intuitive, with the gauge coupling originating from the terms with $0$ vector label.}
Different helicity amplitudes generically get contributions from all four terms, with appropriate mass suppressions.
Having assumed that $\bs1_{\psic}$ and $\bs2_{\psi}$ are conjugate of each others (they have the same mass and opposite charges), $\CP$ would interchange their labels in addition to flipping all helicities (\ie, interchanging square and angle brackets).\footnote{%
To be precise, parity sends 
$\ket{\lambda^I} \leftrightarrow \Ket{\tilde \lambda_I} $ and $\Ket{\tilde \lambda^I} \leftrightarrow - \ket{\lambda_I} $, or equivalently $\ket{k} \leftrightarrow \Ket{k}$ and $\ket{q} \leftrightarrow \Ket{q}$ in the two-lightlike spinors notation.
}
The renormalizable spinor structures are thus self-conjugate, while the non-renormalizable ones (of dipole type) are exchanged under $\CP$.
Since the overall phase of an amplitude is not physical, statements about $\CP$ conservation have to be made at the squared amplitude level.
Considering for instance the interference of a renormalizable contribution with dipole ones, we learn that the coefficients of the dipole terms are equal if $\CP$ is conserved,
since these terms are exchanged by $\CP$.

In three-point amplitudes, Mandelstam variables are combinations of external masses squared.
All kinematics is therefore encoded in the spinor structures, and their coefficients are pure constants.
These coefficients, together with the external particle masses, fully specify the amplitude.
The number of independent kinematic structures can actually be inferred from angular momentum considerations.
In the $\psic\psi Z$ case, there are four distinct ways to form a singlet out of two spin-$1/2$ and one massive spin-$1$ particle: adding two spin $1/2$, we can form states with total spin $0$ and orbital angular momentum $1$, or total spin $1$ and orbital angular momentum $0$, $1$ or $2$, each of which admits a total angular momentum $J=1$ matching the $Z$-boson spin.
In contrast, higher-point amplitudes involve additional relative orbital angular momenta, which allow for infinite ways to form invariants.
Indeed, these amplitudes generically depend on an infinite set of parameters, multiplying arbitrary high powers of the Mandelstam variables.
For three-point amplitudes, this counting is equivalent to that of the number of irreducible representations appearing in the sum of the three spins, since the orbital angular momentum configuration can always be adjusted to form an overall singlet.
In the present $\psic\psi Z$ case, labeling irreducible representations by their dimensions one indeed finds four of them: $2\otimes2\otimes3=1\oplus3\oplus3\oplus5$.

Following the reasoning outlined in the introduction to this section, we would like to determine which spinorial structures can be associated with renormalizable interactions at the tree level.
For this purpose, we consider the massless limit of the theory, for constant $\Lam$, with all mass ratios held fixed.
As mentioned above, for complex momenta, this can be loosely thought of as the high-energy limit, with all masses held constant, and either square or angle spinor products scaling as $E\to\infty$, with $E/\Lam$ fixed.

\newcommand{\affz}[3]{a^{\chir{#1}\chir{#2}\chir{#3}}}

The spinorial structures leading to the fastest energy growth in the $\psic\psi Z$ amplitude of \autoref{eqn:mffz} are the first and fourth ones.
They generate $\mffz+++$ and $\mffz---$ amplitudes scaling as $E^2$:%
\footnote{Here, either $\sqb{ij}\sim E$ and $\anb{ij}\sim m^2/E$, or $\anb{ij}\sim E$ and $\sqb{ij}\sim m^2/E$.
Reference~\cite{Christensen:2018zcq} argued that these amplitudes, and more generally, other non-renormalizable ones, vanish in the massless limit, see \eg, eqs.~(50) and (51).
We disagree with these claims.}
\begin{gather}\label{highe}
\mffz+++ \to \sqb{13}\sqb{23}	\;\cffz+++ /\Lam\,,\qquad
\mffz--- \to \anb{13}\anb{23}	\;\cffz--- /\Lam\,.
\end{gather}
As discussed above, this behavior should be accompanied at the tree level by a factor of $1/\Lam$. 
This justifies our choice of normalization of these terms.
Indeed, these amplitudes are generated by SM loops and, at tree-level, by dipole operators in the SMEFT (see matching in \subappref{sec:ffZ}).%
\footnote{In the SMEFT, these terms arise at dimension six, suppressed by $1/\Lambda^2$.
From a bottom-up perspective, a well-behaved high-energy limit for the $\psic\psi ZZ$ amplitude requires this $1/\Lambda^2$ suppression in a chiral theory where $\cffz-+0\ne\cffz+-0$.
Dipole operators of dimension five, suppressed only by $1/\Lambda$, would however be allowed for fermions having vector-like $\cffz-+0=\cffz+-0$ coupling.
The presentation of an explicit derivation of this result is left for future work.}

No such constraint applies to the $\anb{\bs{13}}\sqb{\bs{23}}$ and $\sqb{\bs{13}}\anb{\bs{23}}$ spinor structures.
These do not give rise to $E^2$ terms in the high-energy limit since, for instance, $\anb{13}\sqb{23} = -\langle 132] = \langle 1(1+2)2] = 0$ for massless kinematics.
They only generate amplitudes of acceptable linear energy growth:
\begin{equation}
\begin{aligned}
  \mffz--0 &\to + \anb{12}\; (\cffz-+0-\cffz+-0)\; m_\psi/\sqrt{2}m_Z
\,,
\\\mffz-+- &\to - \frac{\anb{13}^2}{\anb{12}} \cffz-+0
\,,
\\\mffz-++ &\to + \frac{\sqb{23}^2}{\sqb{12}} \cffz-+0
\,,
\\\mffz+-- &\to + \frac{\anb{23}^2}{\anb{12}} \cffz+-0
\,,
\\\mffz+-+ &\to - \frac{\sqb{13}^2}{\sqb{12}} \cffz+-0
\,,
\\\mffz++0 &\to - \sqb{12}\; (\cffz-+0-\cffz+-0)\; m_\psi/\sqrt{2}m_Z
\,,
\end{aligned}
\label{eqn:mffz_HE}
\end{equation}
up to corrections of order $m_Z^2/E^2$ and $m_\psi^2/E^2$.
Since parity flips chiralities by exchanging angle and square brackets, we learn, by comparing the \amp++0 and \amp--0 amplitudes, that the longitudinal component of the $Z$ boson has a pseudo-scalar, or parity-odd, coupling to fermions in the high-energy limit.
The high-energy relations between coefficients of longitudinal and transverse amplitudes are also remarkable.
They imply that vector-like fermions, having parity conserving couplings to the transverse components of the $Z$ boson ($\cffz-+0=\cffz+-0$),
do not couple to its longitudinal component in the high-energy limit.
On the other hand, chiral fermions for which $\cffz-+0\ne \cffz+-0$ do couple to the longitudinal component of the $Z$, with a strength proportional to their mass, in the high-energy limit.

The \amp++-, \amp+-0, \amp-+0, \amp--+ amplitudes only arise at order $\cffz\pm\mp0\:m$ and $\cffz\pm\pm\pm\:m^2/\Lam$ and are therefore subleading in the high-energy limit.

\paragraph{Limits to other amplitudes and theories}
Going to the high-energy limit by keeping all mass ratios constant allowed us to constrain the form of the $\psic\psi Z$ amplitudes.
Taking instead $m_Z/E$ and $m_\psi/E$ to zero at different rates (such that the mass ratios are modified), allows us to make contact with the $\psic\psi\gamma$ amplitudes of the next section, and with a theory of massless fermions.

Let us first consider the limit in which $m_Z/E$ goes to zero faster than $m_\psi/E$.
We can then expand the amplitude coefficients in powers of $m_Z/m_\psi$:
\begin{equation}
\cffz\pm\mp0 = \sum_n\left(\frac{m_Z}{m_\psi}\right)^n\:  \affz\pm\mp0_n(m/\Lam).
\end{equation}
A smooth $m_Z/m_\psi\to0$ limit for the transverse amplitudes of \autoref{eqn:mffz_HE} require $\affz\pm\mp0_{n<0}=0$.
In this limit, all $\affz\pm\mp0_{n>0}$ terms moreover give vanishing contributions to these transverse amplitudes, which thus only receive finite contributions from the $n=0$ term.
Additionally, a smooth limit for the longitudinal amplitudes of \autoref{eqn:mffz_HE} requires $\affz+-0_0=\affz-+0_0$.
We therefore learn that massive fermions only have vector-like renormalizable couplings to massless vectors in the high-energy limit.
Note that the longitudinal amplitudes were crucial in reaching this conclusion, although they are absent if one starts off with a massless vector.
In the next section, we will directly consider the massless vector amplitude and obtain the same result.
Conversely, the mass of a vector having chiral $\affz+-0_0\ne\affz-+0_0$ couplings to a fermion cannot be consistently taken to zero faster than the fermion mass.

On the other hand, if $m_\psi/E$ goes to zero faster than $m_Z/E$, we must impose $\affz\pm\mp0_{n>0}=0$ for the transverse amplitudes of \autoref{eqn:mffz_HE} to be well behaved.
All $\affz\pm\mp0_{n<1}$ terms moreover give vanishing contributions to the longitudinal amplitudes in this limit.
We therefore learn that the coupling of a massless fermion to the longitudinal component of a vector vanishes in the high-energy limit.
The corresponding amplitude is suppressed by $m_Z/E$ compared to transverse ones.
This is in line with what we expect from the Higgs mechanism.

\paragraph{Summary}
To summarize, by considering the high-energy limit of the $\psic\psi Z$ amplitude, we learn that:
\begin{itemize}
\item The $\sqb{\bs{13}}\sqb{\bs{23}}$ and $\anb{\bs{13}}\anb{\bs{23}}$ spinorial structures are generated by non-renormalizable interactions at tree level.
\item In the high-energy limit, the fermion coupling to the longitudinal $Z$ boson has a pseudo-scalar spinor structure.
\item In the high-energy limit, fermions with a vector-like $LR0/RL0$ coupling to the $Z$ boson do not interact with its longitudinal component.
\item Fermions with chiral $LR0/RL0$ couplings to the $Z$ boson, do interact with its longitudinal component in the high-energy limit, with a strength proportional to their mass.
\item A massive fermion only has vector-like $LR0/RL0$ coupling to a massless vector.
\item The coupling of a massless fermion to the longitudinal component of a massive vector vanishes in the high-energy limit.
\item The mass of a fermion with chiral $LR0/RL0$ couplings to a massive vector, can only be taken to vanish at least as fast as the vector mass.
\end{itemize}

This is a nice bottom-up derivation of results expected from the Higgs mechanism, and specifically from the Goldstone boson equivalence theorem.
Indeed, vector-like fermions do not obtain their mass from the Higgs mechanism, and therefore do not couple to the Goldstone component of the massive vector to leading order in the high-energy expansion.
Their masses are furthermore independent of the vector boson masses.
In contrast, chiral fermions must get their mass solely from some Higgs field, and thus couple to the longitudinal $Z$ boson with a coupling proportional to this mass.
At this stage, however, no relation between the $\psic\psi h$ and longitudinal $\psic\psi Z$ couplings is forced upon us.
We will get back to this point and derive this relation when considering the $\psic\psi Zh$ amplitude in \autoref{sec:4-pt}.

\subsection{\texorpdfstring{$\psic\psi\gamma$}{ffA}}
\label{sec:ffA}
\newcommand{\cffa}[3]{c_{\psic\psi\gamma}^{\chir{#1}\chir{#2}\chir{#3}}}

As discussed in ref.~\cite{Arkani-Hamed:2017jhn}, the construction of three-point amplitudes with one massless particle and two particles of the same mass requires the introduction of an ancillary spinor, which  we denote by $\zeta$.
In such cases, external states can indeed only be used to form one independent spinor instead of two.
Labeling the massive particles by $1,2$, and the massless particle by $3$, the spinors $\ket{3}$ and $\bs1\Ket{3}$ are in particular not independent since $\langle3\bs13]=2\,(p_\bs1\cdot p_3) = m_2^2-m_1^2 = 0$.
The proportionality constant between these two spinors can be derived by contracting them with $\bra{\zeta}$.
It is given by $\langle\zeta\bs13]/\anb{\zeta3}$.
Up to normalization, this is the $x$-factor of ref.~\cite{Arkani-Hamed:2017jhn}.
Note that it carries helicity weight $+1$ for particle $3$.
(Similarly, the spinors $\Ket{3}$ and $\bar{\bs1}\ket{3}$ are the same, up to the proportionality factor $[\zeta\bs13\rangle/\sqb{\zeta3}$, which carries helicity weight $-1$ for particle $3$.)

The construction of the negative-helicity $\psic\psi\gamma^-$ amplitudes can then proceed by contracting the $\bra{\bs1},\bra{\bs2}$ massive spinors with $\epsilon_{\alpha \beta} [\zeta\bs13\rangle/\sqb{\zeta3}$ or with $\ket{3},\ket{3}$ structures.
One thus obtains that the $\psic\psi\gamma^-$ amplitude can be expressed as a linear combination of $\sqb{\bs1\bs2}\langle\zeta\bs13]/\anb{\zeta3}$ and $\sqb{\bs13}\sqb{\bs23}$ (and similarly for the positive helicity amplitude, with angle and square brackets exchanged).
The Schouten identity and momentum conservation ($\bs1+\bs2+3=0$) however lead to:
\begin{equation}
\begin{aligned}
\sqb{\bs1\bs2}\langle\zeta\bs13]/\anb{\zeta3}
&=m
(\anb{\zeta\bs1}\sqb{3\bs2}
+\anb{\zeta\bs2}\sqb{3\bs1})/\anb{\zeta3}
+\sqb{\bs13}\sqb{\bs23}\,,
\\
\anb{\bs1\bs2}[\zeta\bs13\rangle/\sqb{\zeta3}
&=m
(\sqb{\zeta\bs1}\anb{3\bs2}
+\sqb{\zeta\bs2}\anb{3\bs1})/\sqb{\zeta3}
+\anb{\bs13}\anb{\bs23}\,,
\\
\anb{\bs1\bs2}\langle\zeta\bs13]/\anb{\zeta3}
&=m
(\anb{\zeta\bs1}\sqb{3\bs2}
+\anb{\zeta\bs2}\sqb{3\bs1})/\anb{\zeta3}
\,,
\\
\sqb{\bs1\bs2}[\zeta\bs13\rangle/\sqb{\zeta3}
&=m
(\sqb{\zeta\bs1}\anb{3\bs2}
+\sqb{\zeta\bs2}\anb{3\bs1})/\sqb{\zeta3}
\,,
\end{aligned}
\label{eqn:x_factor_equalities}
\end{equation}
where $m=m_1=m_2$ (see \autoref{sec:identities} for a detailed derivation).
Using the massless polarization vectors defined in \autoref{eqn:polarimassless1} and \eqref{eqn:polarimassless2}, the first two equalities allow us to write the positive-helicity amplitude as a linear combination of $(\langle\bs1\varepsilon_3^+\bs2]+\langle\bs2\varepsilon_3^+\bs1])$ and $\sqb{\bs13}\sqb{\bs23}$ and similarly for the negative-helicity amplitude.
Thus we obtain:
\begin{equation}
\begin{aligned}
\mathcal{M}(\bs1_\psic,\bs2_\psi,3_\gamma^+)
	&= \cffa{}{}{} (\langle\bs1\varepsilon_3^+\bs2]+\langle\bs2\varepsilon_3^+\bs1])
	+ \frac{\cffa+++}{\Lam}
	\sqb{\bs13}\sqb{\bs23}\,,
\\
\mathcal{M}(\bs1_\psic,\bs2_\psi,3_\gamma^-)
	&= \cffa{}{}{} (\langle\bs1\varepsilon_3^-\bs2]+\langle\bs2\varepsilon_3^-\bs1])
	+ \frac{\cffa---}{\Lam}
	\anb{\bs13}\anb{\bs23}\,,
\end{aligned}
\label{eqn:mffa}
\end{equation}
where $\cffa{}{}{}$, $\cffa+++$, $\cffa---$ are dimensionless functions of $m/\Lam$ and mass ratios.
In particular, the first terms on the right-hand side of \autoref{eqn:mffa} are the electromagnetic gauge interactions and correctly reproduce the $g=2$ factor (see~\subappref{sec:g2}).
An alternative notation involves the $\anb{\bs1\bs2}\langle\zeta\bs13]/\anb{\zeta3},\,\sqb{\bs13}\sqb{\bs23}$ and $\sqb{\bs1\bs2}[\zeta\bs13\rangle/\sqb{\zeta3},\,\anb{\bs13}\anb{\bs23}$ spinor structures.
It is obtained from the above one by applying (backwards) the last two equalities of \autoref{eqn:x_factor_equalities}.
The high-energy limits of the $\psic\psi\gamma$ amplitudes have the same form as the transverse $\psic\psi Z$ ones provided in \autoref{eqn:mffz_HE}.
The exact expressions are obtained with the following replacements:
\begin{equation}
\cffz-+0\to \cffz+-0\to -\sq2\: \cffa{}{}{}\,.
\label{eqn:ffz2ffa_replacement}
\end{equation}
As suggested by our choice of normalization, the $\sqb{\bs13}\sqb{\bs23}$ and $\anb{\bs13}\anb{\bs23}$ spinorial structures are thus expected to only arise at the non-renormalizable (or loop) level.

It is remarkable that the amplitude formalism forces the $c_{\psic\psi\gamma}$ terms to be purely vectorial, without an axial-vector component.
This confirms the conclusion derived in the previous section from massive vector amplitudes, and their longitudinal components in particular.
The coefficients of the $\langle\bs1\varepsilon_3^+\bs2]$ and $\langle\bs1\varepsilon_3^-\bs2]$ terms (as well as $\langle\bs2\varepsilon_3^+\bs1]$ and $\langle\bs2\varepsilon_3^-\bs1]$ ones) are moreover the same since the two photon helicities are parts of the same irreducible representation of the Lorentz group.
Altogether, one thus obtains that $c_{\psic\psi\gamma}$ terms conserve parity which for instance sends the $\langle\bs1\varepsilon_3^+\bs2]$ spinorial structure onto the $\langle\bs2\varepsilon_3^-\bs1]$ one.

The presence of three independent coefficients is also confirmed by angular-momentum considerations similar to those exposed in the previous section.
Placing the two components of the massless photon in a fundamental representation, the decomposition of the $2\otimes2\otimes2$ product indeed involves three irreducible representations: $2\oplus2\oplus4$.

%%%%%%%%%%%%%%%%%%%%%%%%%%%%%%%%%%%%%%%%%%%%%%%%%%%%%%%%%%%
\subsection{\texorpdfstring{$\psic\psi' W$}{ff'W}}
\label{sec:ffW}
%%%%%%%%%%%%%%%%%%%%%%%%%%%%%%%%%%%%%%%%%%%%%%%%%%%%%%%%%%%

\newcommand{\mffw}[3]{\mathcal{M}(1_\psic^#1,2_{\psi'}^#2,3_W^#3)}
\newcommand{\cffw}[3]{c_{\psic\psi' W}^{\chir{#1}\chir{#2}\chir{#3}}}

The $\psic\psi'W$ amplitude takes the same form as the $\psic\psi Z$ one:
\begin{equation}
\mathcal{M}(\bs{1}_\psic, \bs{2}_{\psi'},\bs{3}_{W})=
  \frac{\cffw+++}{\Lam}	\sqb{\bs{13}}\sqb{\bs{23}}
+ \frac{\cffw-+0}{m_W}	\anb{\bs{13}}\sqb{\bs{23}}
+ \frac{\cffw+-0}{m_W}	\sqb{\bs{13}}\anb{\bs{23}}
+ \frac{\cffw---}{\Lam}	\anb{\bs{13}}\anb{\bs{23}}
\,,
\label{eqn:mffw}
\end{equation}
where $c_{\psic\psi'W}$ are again dimensionless.
Because $m_\psi\ne m_{\psi'}$ in general, the high-energy-limit expressions for \amp\pm\pm0\ amplitudes become:
\begin{align}
  \mffw--0 &\to + \anb{12}\; (m_{\psi'}\, \cffw-+0 - m_{\psi}\,\cffw+-0)/\sqrt{2}m_W
\,,	\label{eqn:mffw_mm0}
\\\mffw++0 &\to - \sqb{12}\; (m_\psi\, \cffw-+0 - m_{\psi'}\,\cffw+-0)/\sqrt{2}m_W
\,.	\label{eqn:mffw_pp0}
\end{align}

\paragraph{Limits to other amplitudes and theories}

\newcommand{\affw}[3]{a^{\chir{#1}\chir{#2}\chir{#3}}}

Taking $m_W/E$ to zero faster than both $m_{\psi}/E$ and $m_{\psi'}/E$, the expansion of $\cffw\pm\mp0$ coefficients in powers of $m_W/m_\psi$ is
\begin{equation}
\cffw\pm\mp0
	= \sum_n \left(\frac{m_W}{m_\psi}\right)^n \affw\pm\mp0_n(m/\Lam)\,.
\end{equation}
A smooth limit for transverse amplitudes (analogous to that of \autoref{eqn:mffz_HE} in the $\psic\psi Z$ case) requires $\affw\pm\mp0_{n<0}=0$.
The transverse amplitudes therefore receive finite contributions from $\affw\pm\mp0_{n=0}$ only.
A smooth limit for the longitudinal amplitudes however requires $m_\psi\, \affw-+0_{n=0} = m_{\psi'}\,\affw+-0_{n=0}$.
A fairly peculiar relation between the $\chir+${}$\chir-0$ and $\chir-${}$\chir+0$ couplings is therefore obtained in theories where a massless vector couples to two fermions of different non-vanishing masses.

%%%%%%%%%%%%%%%%%%%%%%%%%%%%%%%%%%
\subsection{\texorpdfstring{$\psic\psi h$}{ffh}}
%%%%%%%%%%%%%%%%%%%%%%%%%%%%%%%%%%

\newcommand{\cffh}[2]{c_{\psic\psi h}^{\chir{#1}\chir{#2}}}
\newcommand{\mffh}[2]{\mathcal{M}({1}_\psic^{#1},{2}_\psi^{#1}, \bs{3}_{h})}

It is straightforward to obtain the general form of the massive fermion-fermion-scalar amplitude:
\beq
\mathcal{M}(\bs{1}_{\psic},\bs{2}_{\psi}, \bs{3}_{h})=
	\cffh++ \sqb{\bs{12}} + \cffh--\anb{\bs{12}}\,.
\label{eqn:mffh}
\eeq
This form does not depend on whether the scalar is massive or massless.
The high-energy limit does not lead to any further insights.
For a massless fermion,
$
\mffh++
	\to \cffh++  \sqb{12}
$, $\;
\mffh--
	\to \cffh-- \anb{12}
$
where $\cffh\pm\pm$ are now evaluated towards $m_\psi\to0$.
We will however show in \autoref{sec:4-pt} that perturbative unitarity requires that these couplings actually vanish for a massless fermion, up to non-renormalizable contributions.

%%%%%%%%%%%%%%%%%%%%%%%%%%%%%%%%%%
\subsection{\texorpdfstring{$ZZh$ and $WWh$}{ZZh and WWh}}
\label{sec:vvh}
%%%%%%%%%%%%%%%%%%%%%%%%%%%%%%%%%%

\newcommand{\cvvh}[2]{c_{VVh}^{\chir#1\chir#2}}
\newcommand{\czzh}[2]{c_{ZZh}^{\chir#1\chir#2}}
\newcommand{\cwwh}[2]{c_{WWh}^{\chir#1\chir#2}}

The massive $VVh$ amplitude can be parametrized as
\begin{equation}
\mathcal{M}(\bs1_V, \bs2_V, \bs3_h)
	= \frac{\cvvh++}{\Lam}	\sqb{\bs1\bs2}^2
	+ \frac{\cvvh00}{m_V}	\sqb{\bs1\bs2}\anb{\bs1\bs2}
	+ \frac{\cvvh--}{\Lam}	\anb{\bs1\bs2}^2\,,
\label{eqn:mvvh}
\end{equation}
where $VV$ stands for either  $ZZ$ or  $W^+W^-$, and $\cvvh{}{}$ are dimensionless.
The presence of three independent terms is compatible with the angular-momentum considerations introduced in \autoref{sec:ppZ} since three irreducible representations are found in the product of two spin-$1$ and one spin-$0$ representations: $3\otimes3\otimes1 = 1\oplus3\oplus5$.
Similarly to other amplitudes involving scalars, this form does not depend on whether $h$ is massive or not.
Given that all spinor structures in \autoref{eqn:mvvh} are symmetric under the exchange of $1\leftrightarrow 2$ spinor labels, the $W^+W^-h$ amplitudes are automatically invariant under charge conjugation.

\newcommand{\mzzh}[2]{\mathcal{M}(1_V^#1,2_V^#2,3_h)}

In the high-energy limit (with $m_V/E\to 0$ for constant $E/\Lam$ and mass ratios) the fastest growing amplitudes are generated by the $\sqb{\bs1\bs2}^2$ and $\anb{\bs1\bs2}^2$ terms:
\begin{equation}
\begin{aligned}
\mzzh++ &\to \cvvh++ \sqb{12}^2	/\Lam,\\
\mzzh-- &\to \cvvh-- \anb{12}^2	/\Lam.
\end{aligned}
\label{eqn:mvvh_HE}
\end{equation}
which scale as $E^2$.
This justifies the choice of $1/\Lam$ for their normalizations.
The $\sqb{\bs1\bs2}^2$ and $\anb{\bs1\bs2}^2$ terms should thus only arise at the non-renormalizable or loop level.

Subleading amplitudes, which grow like $E$, are generated by the $\sqb{\bs1\bs2}\anb{\bs1\bs2}$ term for \amp\pm00 and \amp0\pm0 helicities:
\begin{equation}
\begin{aligned}
\mzzh-0	&\to 
	+\anb{12}\anb{13}/\anb{23}	\;(\cvvh00 - 2\cvvh-- m_Z/\Lam)	/\sq2
	,\\
\mzzh0- &\to
	-\anb{12}\anb{23}/\anb{13}	\;(\cvvh00 - 2\cvvh-- m_Z/\Lam)	/\sq2
	,\\
\mzzh0+	&\to
	-\sqb{12}\sqb{23}/\sqb{13}	\;(\cvvh00 - 2\cvvh++ m_Z/\Lam)	/\sq2
	,\\
\mzzh+0	&\to
	+\sqb{12}\sqb{13}/\sqb{23}	\;(\cvvh00 - 2\cvvh++ m_Z/\Lam)	/\sq2
	.
\end{aligned}
\label{eqn:mvvh_E0}
\end{equation}
Thus, the high-energy helicity amplitudes with a longitudinal $Z$ boson have the same form as amplitudes with the longitudinal $Z$ replaced by a scalar (see \autoref{eqn:ssv}), consistent with the Goldstone boson equivalence theorem.

The \amp\pm\mp0 amplitudes only receive contributions scaling as $m_V\,\cvvh00$ or $m_V^2\,\cvvh\pm\pm/\Lam$ and are therefore subleading in the high-energy limit:
\begin{equation}
\begin{aligned}
\mzzh+- &\to -\sqb{13}^2/\sqb{23}^2\; (\cvvh00 - (\cvvh+++\cvvh--)\: m_Z/\Lam)\:m_Z\,,\\
\mzzh-+ &\to -\sqb{23}^2/\sqb{13}^2\; (\cvvh00 - (\cvvh+++\cvvh--)\: m_Z/\Lam)\:m_Z\,.\\
\end{aligned}
\end{equation}
Finally, for purely longitudinal \amp000\ amplitude, we get
\begin{equation}
\mzzh00 \to - \cvvh00\: (m_h^2-2m_V^2)/2m_V - (\cvvh++ + \cvvh--) m_V^2/\Lam\,.
\label{eqn:mvvh_00}
\end{equation}
This is the only amplitude in which the spinorial structure gives rise, in the high-energy limit, to a factor of $m_h$, leaving a factor of $m_V$ in the denominator for dimensional reasons.

\paragraph{Limits to other amplitudes and theories}
If $m_V/E$ goes to zero faster than $m_h/E$, the expansion of $\cvvh00$ in powers of $m_V/m_h$,
\begin{equation}
\cvvh00 = \sum_n \left(\frac{m_V}{m_h}\right)^n a_n(m_h/\Lam)\,,
\end{equation}
is forced by the \amp000 amplitude of \autoref{eqn:mvvh_00} to start at least at $n=1$ (\ie, $a_{n<1}=0$).
The \amp0\pm0, \amp\pm00 amplitudes of \autoref{eqn:mvvh_E0} then vanish in this limit.
Conversely, requiring them to be non-vanishing constrains $m_h/E$ to go to zero at least as fast as $m_V/E$.

In the opposite limit, if $m_h/E$ goes to zero faster than $m_V/E$, a smooth limit for the \amp0\pm0, \amp\pm00 amplitudes of \autoref{eqn:mvvh_E0} requires $a_{n>0}=0$.

\paragraph{Summary}
To summarize, we learn that,
\begin{itemize}
\item The coupling of a scalar $h$ to two transverse vectors of equal polarizations, only arises at the non-renormalizable or loop level.
\item
At the renormalizable level, the tree $VVh$ amplitude is controlled by a single coupling. Vector bosons of opposite polarizations are involved.
\item The only renormalizable amplitudes that remain non-zero in the high-energy limit are the \amp0\pm0 and \amp\pm00 ones, which involve one transverse and one longitudinal vector.
\item The \amp\pm\mp0 and \amp000 amplitudes are proportional to vector and scalar masses in the high-energy limit.
\item Allowing mass ratio to vary, the renormalizable \amp0\pm0 and \amp\pm00 amplitudes only remain non-vanishing provided $m_h/E\to0$ at least as fast as $m_V/E$.
\end{itemize}
These conclusions are again compatible with our knowledge about the Goldstone boson equivalence theorem and of the Higgs mechanism.

\subsection{\texorpdfstring{$\gamma\gamma h$}{AAh} and \texorpdfstring{$\gamma Zh$}{AZh}}

\newcommand{\caah}[2]{c_{\gamma\gamma h}^{\chir#1\chir#2}}
\newcommand{\cazh}[2]{c_{\gamma Z h}^{\chir#1\chir#2}}

The $\gamma\gamma h$ amplitude can be obtained from the $VVh$ one of the previous subsection, in the $m_V/m_h\to0$ limit.
The $\cvvh00$ coupling then vanishes identically.
So one obtains in general:
\begin{equation}
\mathcal{M}(1^+_\gamma,2^+_\gamma,\bs3_h) = \frac{\caah++}{\Lam} \sqb{12}^2,
\qquad\text{and}\qquad
\mathcal{M}(1^-_\gamma,2^-_\gamma,\bs3_h) = \frac{\caah--}{\Lam} \anb{12}^2.
\end{equation}
Thus there is no renormalizable  $\gamma\gamma h$ amplitude at tree-level.

Similarly for $\gamma Zh$ amplitudes, one can only write the following spinorial structures:
\begin{equation}
\mathcal{M}(1^+_\gamma,\bs2_Z,\bs3_h) = \frac{\cazh++}{\Lam} \sqb{1\bs2}^2,
\qquad\text{and}\qquad
\mathcal{M}(1^-_\gamma,\bs2_Z,\bs3_h) = \frac{\cazh--}{\Lam} \anb{1\bs2}^2.
\label{eqn:mazh}
\end{equation}

\subsection{\texorpdfstring{$hhZ$}{hhZ} and \texorpdfstring{$hh\gamma$}{hhA}}
\label{sec:hhZhhA}
The vector coupling to two identical scalars vanishes because of Bose symmetry.
To see this, note first that the $\sqb{\bs{33}}$ and $\anb{\bs{33}}$ bilinears vanish.
The $SU(2)$ little-group indices must appear in a symmetric combination, but $\sqb{3^I3^J}=\epsilon^{IJ}m_3=-\anb{3^I3^J}$ is antisymmetric.
Thus the $\mathcal{M}(\bs1_h,\bs2_h,\bs3_Z)$ amplitude can only contain the $\langle \bs3(\bs1-\bs2)\bs3]$ spinorial structure.
For identical scalars, the amplitude is however required to be symmetric under the $\bs1\leftrightarrow\bs2$ exchange.
Since no symmetric structure can be written down, one concludes that the $hhZ$ amplitude vanishes.%
\footnote{This gives a short proof that $\rho^0 \to 2 \pi^0$ is forbidden.}

Likewise, starting directly with a massless vector, the only possible structure is $[3(\bs{1} - \bs{2})\zeta\rangle/\anb{3 \zeta}$, with some auxiliary spinor $\zeta$  (or similarly with angle and square brackets exchanged).
This spinorial structure is again antisymmetric under the $\bs1\leftrightarrow\bs2$ exchange.
There is thus no $\mathcal{M}(\bs1_h,\bs2_h,3_\gamma)$ amplitude either.

\subsection{\texorpdfstring{$hhh$}{hhh}}
The triple-Higgs amplitude is trivial:
\begin{equation}
\mathcal{M}(\bs1_h,\bs2_h,\bs3_h) = m_h\;c_{hhh}\,.
\end{equation}

\subsection{\texorpdfstring{$WWZ$}{WWZ}}
\label{sec:wwz}

\newcommand{\cwwz}[4][]{c_{WWZ}^{\chir{#2}\chir{#3}\chir{#4}}}
The $WWZ$ amplitude has a central role in the theory, as it determines the underlying non-abelian gauge structure.
The derivation below only relies on the Lorentz structure, namely, the fact that the three particles have spin $1$, and that two of them are mass-degenerate. Indeed, with no information about the remaining amplitudes, we have no notion of charge, and the amplitude we write down below can be interpreted either as the $W^+ W^- Z$ amplitude, with distinguishable $W$s, or as the amplitude of two indistinguishable $W$s carrying an index $a=1,2$. 
As we will see below, the only renormalizable coupling allowed preserves $U(1)_{{\rm EM}}$.
This is related to the fact that only three massive vector bosons participate  in the interaction, corresponding to an $SU(2)$ gauge theory.
It is also instructive to consider the amplitude for $m_Z=m_W$. The $SU(2)$  group theory structure follows in this case solely from Bose statistics of the three-point amplitude.

Constructing the massive $WWZ$ amplitude following the prescription of eq.\,(4.27) in ref.~\cite{Arkani-Hamed:2017jhn} is somewhat tedious.
From the eleven spinorial structures originally generated, eight independent ones survive after extensive use of the Schouten identity, momentum conservation, and on-shell conditions.
Another way to obtain the same result is to list independent spinor contractions, and construct products of appropriate helicity weights.
All possible spinor bilinears can be employed:
\begin{equation}
\begin{array}{ccc}
\anb{\bs{12}},	&\anb{\bs{13}},	&\anb{\bs{23}},\\
\sqb{\bs{12}},	&\sqb{\bs{13}},	&\sqb{\bs{23}}.
\end{array}
\label{eqn:wwz_bilinears}
\end{equation}
The $[\bs{ijk}\rangle$ trilinears involving all three spinors can be reduced to bilinears by imposing momentum conservation ($\bs j=-\bs i-\bs k$) and using on-shell conditions like $\bs{k}\ket{\bs{k}} = + m_k \Ket{\bs{k}}$.
Quadrilinears and higher multilinears can be reduced in the same way.
The $[\bs i (\bs j-\bs k) \bs i\rangle = \langle\bs i(\bs j-\bs k) \bs i]$ trilinears involving identical bra and ket spinors could have been expressed as dot products of momenta if $\bs i$ was massless.
They are equivalent to $[\bs i \bs j\bs i\rangle$ or $[\bs i \bs k\bs i\rangle$ structures, up to terms involving bilinear products.
In a triple-vector amplitude, since each spinor should appear twice, a $[\bs i \cdot\bs i\rangle$ trilinear would necessarily be multiplied by bilinears involving both $\bs j$ and $\bs k$, or by $[\bs j \cdot\bs j\rangle$ and $[\bs k \cdot\bs k\rangle$ trilinears.
Both kinds of products are however reducible to products of bilinears using the Schouten identity, momentum conservation and on-shell conditions.
One for instance obtains:
\begin{equation}
\begin{array}{rcl}
	  [\bs{iji}\rangle \anb{\bs{kj}} 
	=&[\bs{ijk}\rangle \anb{\bs{ij}}
	- [\bs{ijj}\rangle \anb{\bs{ik}}
	&= m_i \anb{\bs{ik}}\anb{\bs{ij}}
	- m_k \sqb{\bs{ik}}\anb{\bs{ij}}
	- m_j \sqb{\bs{ij}}\anb{\bs{ik}}
,\\[.5ex]
	  [\bs{iji}\rangle [\bs{jij}\rangle
	=&-\sqb{\bs{ijij}}\anb{\bs{ij}}
	+[\bs{ijj}\rangle \langle\bs{iij}]
	&= -2(p_i\cdot p_j) \sqb{\bs{ij}} \anb{\bs{ij}}
	-m_i m_j (\anb{\bs{ij}}^2+\sqb{\bs{ij}}^2)
.
\end{array}
\end{equation}
So one concludes that the amplitudes for three massive vectors can all be constructed from products of the bilinears in \autoref{eqn:wwz_bilinears}.
The restriction that each spinor appears exactly twice leads to eight independent terms:
\begin{equation}
\begin{aligned}
\mathcal{M}(\bs 1_W,\bs 2_W,\bs 3_Z)
	&= \anb{\bs{12}} \anb{\bs{13}} \anb{\bs{23}} \;\cwwz---/\Lam^2
\\	&+ \anb{\bs{12}} \anb{\bs{13}} \sqb{\bs{23}} \;\cwwz-00/m_W m_Z
\\	&+ \anb{\bs{12}} \sqb{\bs{13}} \anb{\bs{23}} \;\cwwz0-0/m_W m_Z
\\	&+ \sqb{\bs{12}} \anb{\bs{13}} \anb{\bs{23}} \;\cwwz00-/m_W^2
\\	&+ \anb{\bs{12}} \sqb{\bs{13}} \sqb{\bs{23}} \;\cwwz00+/m_W^2
\\	&+ \sqb{\bs{12}} \anb{\bs{13}} \sqb{\bs{23}} \;\cwwz0+0/m_W m_Z
\\	&+ \sqb{\bs{12}} \sqb{\bs{13}} \anb{\bs{23}} \;\cwwz+00/m_W m_Z
\\	&+ \sqb{\bs{12}} \sqb{\bs{13}} \sqb{\bs{23}} \;\cwwz+++/\Lam^2\,,
\end{aligned}
\hspace{1cm}\text{\raisebox{2.5cm}{\rotatebox{-90}{(superseded by \autoref{eqn:mwwz})}}}
\label{eqn:mwwz_preliminary}
\end{equation}
where the $\cwwz{}{}{}$'s are dimensionless coefficients, and their specific normalizations are chosen for later convenience.
Not all of these structures are actually independent.
One relation will be found between them.
There are indeed seven distinct ways to combine three spin-one representations.
Labeling representations by their dimensions, one indeed finds $3\otimes3\otimes3=1\oplus3\oplus3\oplus3\oplus5\oplus5\oplus7$.
Additional constraints, not apparent in \autoref{eqn:mwwz_preliminary}, appear in the high-energy limit ($m_Z/E\to 0$, for $m_W/m_Z$ and $E/\Lam$ fixed).
Eventually, \autoref{eqn:mwwz_preliminary} will be replaced by \autoref{eqn:mwwz}.

\newcommand{\mwwz}[3]{\mathcal{M}(1_W^#1,2_W^#2,3_Z^#3)}

The amplitudes with the fastest energy growth ($E^3$) are,
\begin{equation}
\begin{aligned}
\mwwz--- &\to \anb{12}\anb{13}\anb{23} \;\cwwz---/\Lam^2,\\
\mwwz+++ &\to \sqb{12}\sqb{13}\sqb{23} \;\cwwz+++/\Lam^2.
\end{aligned}
\end{equation}
The $\anb{\bs{12}} \anb{\bs{13}} \anb{\bs{23}}$ and $\anb{\bs{12}} \sqb{\bs{13}} \sqb{\bs{23}}$ spinorial structure should thus only arise at the non-renormalizable (or loop) level.
This motivates our initial choice of normalization for those terms.
Amplitudes growing like $E^2$ are produced by other terms which are thus also expected to only arise at the non-renormalizable (or loop) level.
They are:
\begin{equation}
\begin{aligned}
\mwwz--0 &\to +\anb{12}^2\; (\cwwz-00 - \cwwz0-0)	/\sq2 m_Z
,\\
\mwwz-0- &\to -\anb{13}^2\; (\cwwz-00 - \cwwz00- )	/\sq2 m_W
,\\
\mwwz0-- &\to +\anb{23}^2\; (\cwwz0-0 - \cwwz00- )	/\sq2 m_W
,\\
\mwwz0++ &\to +\sqb{23}^2\; (\cwwz0+0 - \cwwz00+ )	/\sq2 m_W
,\\
\mwwz+0+ &\to -\sqb{13}^2\; (\cwwz+00 - \cwwz00+ )	/\sq2 m_W
,\\
\mwwz++0 &\to +\sqb{12}^2\; (\cwwz+00 - \cwwz0+0)	/\sq2 m_Z
.
\end{aligned}
\label{eqn:mwwz_E2_preliminary}
\hspace{1cm}\text{\raisebox{2.3cm}{\rotatebox{-90}{(superseded by \autoref{eqn:mwwz_E2})}}}
\end{equation}
A well-behaved high-energy limit can be obtained with the following ansatz:
\newcommand{\LA}{[\!L0]0}%
\newcommand{\LS}{\{\!L0\}0}%
\newcommand{\RA}{[\!R0]0}%
\newcommand{\RS}{\{\!R0\}0}%
\newcommand{\cwwzLA}{c_{WWZ}^{\LA}}%
\newcommand{\cwwzLS}{c_{WWZ}^{\LS}}%
\newcommand{\cwwzRA}{c_{WWZ}^{\RA}}%
\newcommand{\cwwzRS}{c_{WWZ}^{\RS}}%
\begin{equation}
\begin{aligned}
(\cwwz-00 - \cwwz0-0) /m_Z	&\equiv 2\cwwzLA m_W/m_Z\Lam\,,\\
(\cwwz-00 - \cwwz00-) /m_W	&\equiv (\cwwzLS + \cwwzLA)/\Lam\,,\\
(\cwwz0-0 - \cwwz00-) /m_W	&\equiv (\cwwzLS - \cwwzLA)/\Lam\,,\\
(\cwwz0+0 - \cwwz00+) /m_W	&\equiv (\cwwzRS + \cwwzRA)/\Lam\,,\\
(\cwwz+00 - \cwwz00+) /m_W	&\equiv (\cwwzRS - \cwwzRA)/\Lam\,,\\
(\cwwz0+0 - \cwwz+00) /m_Z	&\equiv 2\cwwzRA m_W/m_Z\Lam\,.
\end{aligned}
\label{eqn:mwwz_he_constraints}
\end{equation}
The $\cwwz-00$, $\cwwz0-0$, $\cwwz0+0$ and $\cwwz+00$ coefficients can then for instance be eliminated in favor of the four new ones appearing on the right-hand side of \autoref{eqn:mwwz_he_constraints}.
The $\cwwz00\pm$ terms are then
\begin{equation}
\begin{aligned}
	&+ (\sqb{\bs{12}} \anb{\bs{13}} \anb{\bs{23}} \;m_Z/m_W
	   +\anb{\bs{12}} \sqb{\bs{13}} \anb{\bs{23}}
	   +\anb{\bs{12}} \anb{\bs{13}} \sqb{\bs{23}})
	   \;\cwwz00-/m_Z m_W
\\	&+ (\anb{\bs{12}} \sqb{\bs{13}} \sqb{\bs{23}} \;m_Z/m_W \;
	   +\sqb{\bs{12}} \anb{\bs{13}} \sqb{\bs{23}} \;
	   +\sqb{\bs{12}} \sqb{\bs{13}} \anb{\bs{23}})\,
	   \;\cwwz00+/m_Z m_W\,.
\end{aligned}
\label{eqn:cwwz_pm}
\end{equation}
These two spinorial structures are however equivalent.
Using the Schouten identity in a form similar to that of \autoref{eqn:x_factor_equalities}, one obtains:
\begin{equation}
\begin{aligned}
\langle\bs{313}]\sqb{\bs{12}}
&=m_1\anb{\bs{13}}\sqb{\bs{23}}
 +m_2\sqb{\bs{13}}\anb{\bs{23}}
 -m_3\sqb{\bs{13}}\sqb{\bs{23}}\,,
\\
[\bs{313}\rangle\anb{\bs{12}}
&=m_1\sqb{\bs{13}}\anb{\bs{23}}
 +m_2\anb{\bs{13}}\sqb{\bs{23}}
 -m_3\anb{\bs{13}}\anb{\bs{23}}.
\end{aligned}
\label{eqn:equalities_wwz}
\end{equation}
The left-hand side of the first equality times $\anb{\bs{12}}$ is equal to that of the second one times $\sqb{\bs{12}}$.
One therefore obtains an identity involving only products of bilinears:
\begin{equation}
\begin{aligned}
&m_1\anb{\bs{12}}\anb{\bs{13}}\sqb{\bs{23}}
+m_2\anb{\bs{12}}\sqb{\bs{13}}\anb{\bs{23}}
+m_3\sqb{\bs{12}}\anb{\bs{13}}\anb{\bs{23}}\\
=\:
&m_1\sqb{\bs{12}}\sqb{\bs{13}}\anb{\bs{23}}
\:+m_2\sqb{\bs{12}}\anb{\bs{13}}\sqb{\bs{23}}
\:+m_3\anb{\bs{12}}\sqb{\bs{13}}\sqb{\bs{23}}\,,
\end{aligned}
\label{eqn:wwz_bilinear_id}
\end{equation}
which indeed demonstrates that the two spinorial structures in \autoref{eqn:cwwz_pm} are equivalent.
We choose to keep them both and give them the same coefficient $\cwwz{}{}{}=\cwwz00+=\cwwz00-$.
Eventually, one therefore obtains that, at the renormalizable level, the $WWZ$ amplitude features a single coupling, namely $\cwwz{}{}{}$,
\begin{equation}
\begin{aligned}
\mathcal{M}(\bs 1_W,\bs 2_W,\bs 3_Z) =
	&+ \anb{\bs{12}} \anb{\bs{13}} \anb{\bs{23}}	\;\cwwz---/\Lam^2
\\	&+ \anb{\bs{12}} (\anb{\bs{13}} \sqb{\bs{23}} 
		- \sqb{\bs{13}} \anb{\bs{23}}	)	\;\cwwzLA/m_Z\Lam
\\	&+ \anb{\bs{12}} (\anb{\bs{13}} \sqb{\bs{23}} 
		+ \sqb{\bs{13}} \anb{\bs{23}}	)	\;\cwwzLS/m_Z\Lam
\\[-1mm]& + \Big(\sqb{\bs{12}} \anb{\bs{13}} \anb{\bs{23}} \;\rzw
	   +\anb{\bs{12}} \sqb{\bs{13}} \anb{\bs{23}}
	   +\anb{\bs{12}} \anb{\bs{13}} \sqb{\bs{23}}
\\[-2mm]	&+ \anb{\bs{12}} \sqb{\bs{13}} \sqb{\bs{23}} \;\rzw \;
	   +\sqb{\bs{12}} \anb{\bs{13}} \sqb{\bs{23}} \;
	   +\sqb{\bs{12}} \sqb{\bs{13}} \anb{\bs{23}}\Big)\,
                    \;\cwwz{}{}{}/m_Z m_W
\\[-1mm]	&+ \sqb{\bs{12}} (\anb{\bs{13}} \sqb{\bs{23}}
		- \sqb{\bs{13}} \anb{\bs{23}}	)	\;\cwwzRA/m_Z\Lam
\\	&+ \sqb{\bs{12}} (\anb{\bs{13}} \sqb{\bs{23}}
		+ \sqb{\bs{13}} \anb{\bs{23}}	)	\;\cwwzRS/m_Z\Lam
\\	&+ \sqb{\bs{12}} \sqb{\bs{13}} \sqb{\bs{23}} \;\cwwz+++/\Lam^2\,,
\end{aligned}
\label{eqn:mwwz}
\end{equation}
where  we defined $\rzw\equiv m_Z/m_W$, and as we will now see,  $\cwwzLA=\cwwzRA=0$.
It is remarkable that the non-trivial structure of the single renormalizable term is entirely dictated by purely bottom-up considerations, without imposing electroweak symmetry.
It is parity invariant, and just flips sign under either $\text{C}$ or $\CP$.
All spinorial structures except the $\cwwzLA$ and $\cwwzRA$ ones are actually odd under the $1\leftrightarrow2$ exchange associated with charge conjugation.
A $\CP$ conserving interference between, for instance, the renormalizable and $LLL/RRR$ terms is thus obtained if $\cwwz+++=\cwwz---$.
Since the renormalizable term is antisymmetric  under $1\leftrightarrow2$ exchange, it also preserves $U(1)_{{\rm EM}}$: amplitudes with same-sign $W$s vanish.
In contrast, the even character of $\cwwzLA$ and $\cwwzRA$ terms implies that their interferences with the renormalizable term violate $\text{C}$.
Since the $W^+W^-Z$ amplitude is self-conjugate, we conclude that $\cwwzLA$ and $\cwwzRA$ must vanish.
An explicit matching to the SMEFT confirms this conclusion (see~\autoref{eqn:WWZA}).

\newcommand{\cwww}[4][]{c_{WWW}^{\chir{#2}\chir{#3}\chir{#4}}}

\paragraph{Massive degenerate vectors and Lie group structure constants.}
It is interesting to consider our starting point, \autoref{eqn:mwwz_preliminary} in the $m_W=m_Z$ limit.
Labeling the three vectors by $W^{a=1,2,3}$ and choosing them all to have the same helicity (with say, $I=J=1$ for all), Bose symmetry implies,%
\footnote{To see this, one can assign a three-index tensor prefactor to each of the spinor structures, and impose overall Bose symmetry under, each particle exchange and a cyclic permutation of all three.}
\begin{equation}
\begin{aligned}
\mathcal{M}(\bs 1_W^a,\bs 2_W^b,\bs 3_W^c)
	= 
	\epsilon^{abc}\,\bigg\{
	&+\anb{\bs{12}} \anb{\bs{13}} \anb{\bs{23}} \; \cwww---/\Lam^2
\\[-2mm]	&+ \Big(\anb{\bs{12}} \anb{\bs{13}} \sqb{\bs{23}} 
+ \anb{\bs{12}} \sqb{\bs{13}} \anb{\bs{23}} 
+ \sqb{\bs{12}} \anb{\bs{13}} \anb{\bs{23}} 
\\[-2mm]	&+ 
\anb{\bs{12}} \sqb{\bs{13}} \sqb{\bs{23}}
	+ \sqb{\bs{12}} \anb{\bs{13}} \sqb{\bs{23}}
+ \sqb{\bs{12}} \sqb{\bs{13}} \anb{\bs{23}} \Big)\;  \cwww{}{}{}/m_W^2
\\[-2mm]	&+ \sqb{\bs{12}} \sqb{\bs{13}} \sqb{\bs{23}} \;\cwww+++/\Lam^2
\bigg\}
\,,
\end{aligned}
\label{eqn:mwww}
\end{equation}
Thus, with only three vectors, the $SU(2)$ structure constants emerge already at the level of the three-point function. 
More generally, with $n>3$ degenerate vectors, the $\epsilon^{abc}$ is replaced by completely antisymmetric tensors $C^{abc}$.
At this point, we can appeal to the four-point function in the massless limit to obtain the Jacobi identity, which establishes the identification of $C^{abc}$ with the Lie algebra structure constants $f^{abc}$.
Restricting attention to any subset of three out of the $n$ vectors, we would again recover $\epsilon^{abc}$; indeed, any three generators form an $SU(2)$ sub-algebra.
Note that deriving the $f^{abc}$ structure in the massive case is in some sense more straightforward that in the massless case.
Here all three vectors can have the same spin polarization, while in the massless case, the gauge coupling corresponds to \amp\pm\pm\mp\ helicity amplitudes.

Since the four $\cwwzLS$, $\cwwzRS$, $\cwwzLA$, $\cwwzRA$ terms are forbidden for vectors of identical masses, they must be proportional to the $m_Z-m_W$ mass difference in the $m_Z \to m_W$ limit.
Indeed, in the SM, the source of $\cwwzLS$, $\cwwzRS$ terms (which preserve $\text{C}$) is the additional $U(1)_Y$ (see~\autoref{eqn:WWZS}).
Additionally, in a gauge theory featuring a single $SU(2)$, spontaneous symmetry breaking leads to three degenerate massive bosons.

\paragraph{High-energy limit}
In the high-energy limit, the amplitude of \autoref{eqn:mwwz} has non-renorma\-lizable terms growing as  $E^3$ and $E^2$, as well as renormalizable terms scaling as $E$,
\begin{align}
&\begin{aligned}
\mwwz--- &\to \anb{12}\anb{13}\anb{23} \;\cwwz---/\Lam^2,\\
\mwwz+++ &\to \sqb{12}\sqb{13}\sqb{23} \;\cwwz+++/\Lam^2,
\end{aligned}
\label{eqn:mwwz_E3}\\
\intertext{and}
&\begin{aligned}
\mwwz--0 &\to +\anb{12}^2\; \cwwzLA \sq2/\rzw\Lam
,\\
\mwwz-0- &\to -\anb{13}^2\; (\cwwzLS + \cwwzLA)/\sq2\Lam
,\\
\mwwz0-- &\to +\anb{23}^2\; (\cwwzLS - \cwwzLA)/\sq2\Lam
,\\
\mwwz0++ &\to +\sqb{23}^2\; (\cwwzRS + \cwwzRA)/\sq2\Lam
,\\
\mwwz+0+ &\to -\sqb{13}^2\; (\cwwzRS - \cwwzRA)/\sq2\Lam
,\\
\mwwz++0 &\to -\sqb{12}^2\; \cwwzRA \sq2/\rzw\Lam
,
\end{aligned}
\label{eqn:mwwz_E2}\\
\intertext{and}
&\begin{aligned}
\mwwz--+
	&\to - {\anb{12}^3}/{\anb{13}\anb{23}}
	\;\:\cwwz{}{}{}
	,\\
\mwwz-00
	&\to + {\anb{12}\anb{13}}/{\anb{23}}
	\;\:(\rzw\,\cwwz{}{}{} + (\cwwzRA - \cwwzLA)\: m_W/\rzw\Lam)
	,\\
\mwwz-+-
	&\to - {\anb{13}^3}/{\anb{12}\anb{23}}
	\;\: (\cwwz{}{}{} + (\cwwzRS +\cwwzRA)\,m_W/\Lam)
	,\\
\mwwz-++
	&\to - {\sqb{23}^3}/{\sqb{12}\sqb{13}}
	\;\: (\cwwz{}{}{} + (\cwwzLS+\cwwzLA)\,m_W/\Lam)
	,\\
\end{aligned}\nn\\
&\begin{aligned}
\mwwz0-0
	&\to + {\anb{12}\anb{23}}/{\anb{13}}
	\;\:(\rzw\,\cwwz{}{}{} - (\cwwzRA - \cwwzLA) m_W/\rzw\Lam)
	,\\
\mwwz00-
	&\to + {\anb{13}\anb{23}}/{\anb{12}}
	\;\:((2 - \rzw^2)\cwwz{}{}{} + (\cwwzRS+\cwwzLS)\: m_W/\Lam)
	,\hspace*{-4cm}\\
\mwwz00+
	&\to + {\sqb{13}\sqb{23}}/{\sqb{12}}
	\;\:((2 - \rzw^2)\cwwz{}{}{} + (\cwwzRS+\cwwzLS)\: m_W/\Lam)
	,\hspace*{-4cm}\\
\mwwz0+0
	&\to + {\sqb{12}\sqb{23}}/{\sqb{13}}
	\;\:(\rzw\,\cwwz{}{}{} - (\cwwzRA - \cwwzLA) m_W/\rzw\Lam)
	,\\
\end{aligned}
\label{eqn:mwwz_E1}\\
&\begin{aligned}
\mwwz+--
	&\to - {\anb{23}^3}/{\anb{12}\anb{13}}
	\;\: (\cwwz{}{}{} + (\cwwzRS-\cwwzRA)\,m_W/\Lam)
	,\\
\mwwz+-+
	&\to - {\sqb{13}^3}/{\sqb{12}\sqb{23}}
	\;\: (\cwwz{}{}{} + (\cwwzLS-\cwwzLA)\,m_W/\Lam)
	,\\
\mwwz+00
	&\to + {\sqb{12}\sqb{13}}/{\sqb{23}}
	\;\:(\rzw\,\cwwz{}{}{} + (\cwwzRA - \cwwzLA)\: m_W/\rzw\Lam
	,\\
\mwwz++-
	&\to - {\sqb{12}^3}/{\sqb{13}\sqb{23}}
	\;\:\cwwz{}{}{}
	.\\
\end{aligned}
\nn
\end{align}

\paragraph{Limits to other amplitudes and theories}
Taking $m_W$ to zero faster than $m_Z$ ($1/\rzw=m_W/m_Z\to0$) one can expand $\cwwz{}{}{}$, like other coefficients, in powers of $m_W/m_Z$,
\begin{equation}
\cwwz{}{}{} = \sum_n \left(\frac{m_W}{m_Z}\right)^n \:a_n(m/\Lam)\,,
\end{equation}
A smooth limit for \amp00\pm\ amplitudes requires $a_{n<2}=0$.
They only receive finite contributions from $a_{n=2}$.
The $a_{n\ge2}$ terms are however irrelevant for all other amplitudes in this limit.
The purely non-renormalizable coefficients have $m_W/m_Z$ expansions which are forced by $E^3$ and $E^2$ amplitudes to only contain non-negative powers.
The \amp\pm\pm0 amplitudes therefore vanish in this limit.

Taking $m_Z/E$ to zero faster than $m_W/E$, the fully transverse amplitudes of \autoref{eqn:mwwz_E1} require $a_{n>0}=0$ and only receive finite contributions from $a_{n=0}$.
Lower powers are irrelevant for all other amplitudes.
The purely non-renormalizable terms have a $m_Z/m_W=\rzw$ expansion which is forced by $E^3$ and $E^2$ amplitudes to contain only non-negative powers.
The $\cwwzLA$ and $\cwwzRA$ coefficients are moreover required by \amp\pm\pm0 amplitudes to start with a strictly positive power of $m_Z/m_W=\rzw$.

\subsection{\texorpdfstring{$WW\gamma$}{WWA}}
\label{sec:wwa}
\newcommand{\cwwa}[3]{c_{WW\gamma}^{\chir{#1}\chir{#2}\chir{#3}}}
Instead of constructing the $WW\gamma$ amplitude from scratch, it is instructive to obtain it from the massless $Z$-boson limit of the $WWZ$ amplitude.
For the positive helicity photon amplitude, $\mathcal{M}(\bs 1_W,\bs 2_W,3_\gamma^+)$, one obtains from \autoref{eqn:mwwz}:
\begin{equation}
\begin{aligned}
&+\anb{\bs{12}} \anb{\bs1 3_q} \anb{\bs2 3_q}	\;\cwwz---/\Lam^2	\\
&+\sqb{\bs{12}} \anb{\bs1 3_q} \anb{\bs2 3_q}	\;\cwwz{}{}{}/m_W^2	\\
&+\anb{\bs{12}} (\anb{\bs1 3_q} \sqb{\bs2 3} - \anb{\bs2 3_q} \sqb{\bs1 3}) \;\cwwzLA/m_Z\Lam	\\
&+\sqb{\bs{12}} (\anb{\bs1 3_q} \sqb{\bs2 3} - \anb{\bs2 3_q} \sqb{\bs1 3}) \;\cwwzRA/m_Z\Lam	\\
&+\anb{\bs{12}} (\anb{\bs1 3_q} \sqb{\bs2 3} + \anb{\bs2 3_q} \sqb{\bs1 3}) \;(\cwwz{}{}{} + \cwwzLS\:m_W/\Lam)/m_Zm_W	\\
&+\sqb{\bs{12}} (\anb{\bs1 3_q} \sqb{\bs2 3} + \anb{\bs2 3_q} \sqb{\bs1 3}) \;(\cwwz{}{}{} + \cwwzRS\:m_W/\Lam)/m_Zm_W	\\
&+\anb{\bs{12}} \sqb{\bs1 3} \sqb{\bs2 3}	\;\cwwz{}{}{}/m_W^2	\\
&+\sqb{\bs{12}} \sqb{\bs1 3} \sqb{\bs2 3}	\;\cwwz+++/\Lam^2\,.	
\end{aligned}
\label{eqn:mwwz_massless}
\end{equation}
According to our discussion of the $m_Z/m_W\to0$ limit in the previous subsection, the expansion of all $\cwwz{}{}{}$ involves only non-negative powers of $m_Z/m_W$.
The expansion of $\cwwzLA$ and $\cwwzRA$ coefficients also starts with a strictly positive power.
Given that $\ket{3_q}$ scales as $m_Z/\sqrt{E}$, one obtains that the first four terms of \autoref{eqn:mwwz_massless} vanish in the $m_Z/m_W\to 0$ limit.

Introducing an auxiliary spinor $\zeta$, one can identify $\ket{3_q}/m_Z\to \ket{\zeta}/\anb{3\zeta}$ (and similarly $\Ket{3_q}/m_Z\to \Ket{\zeta}/\sqb{\zeta3}$).
Using the equalities of \autoref{eqn:x_factor_equalities}, one can also rewrite:
\begin{equation}
\begin{aligned}
\anb{\bs{12}} \sqb{\bs1 3} \sqb{\bs2 3}
&= m_W(\anb{\bs{12}}-\sqb{\bs{12}})
(\anb{\bs1\zeta}\sqb{\bs23}
+\anb{\bs2\zeta}\sqb{\bs13})/\anb{3\zeta}
,
\\
\sqb{\bs{12}} \anb{\bs1 3} \anb{\bs2 3}
&= m_W(\sqb{\bs{12}}-\anb{\bs{12}})
(\sqb{\bs1\zeta}\anb{\bs23}
+\sqb{\bs2\zeta}\anb{\bs13})/\sqb{3\zeta}
.
\end{aligned}
\label{eqn:x_factor_consequence}
\end{equation}
The spinorial structures in the fifth to seventh lines in \autoref{eqn:mwwz_massless} thus reduce to two independent structures.
One can also introduce the massless polarization vector of \autoref{eqn:polarimassless1} and \eqref{eqn:polarimassless2} to rewrite:
\begin{equation}
\begin{aligned}
(\anb{\bs1\zeta}\sqb{\bs23}
+\anb{\bs2\zeta}\sqb{\bs13})/\anb{3\zeta} &=
-(\langle\bs1\varepsilon^+_3 \bs2]
+ \langle\bs2\varepsilon^+_3 \bs1])/\sq2 \,,
\\
(\sqb{\bs1\zeta}\anb{\bs23}
+\sqb{\bs2\zeta}\anb{\bs13})/\sqb{3\zeta}\; &=
-(\langle\bs1\varepsilon^-_3 \bs2]
+ \langle\bs2\varepsilon^-_3 \bs1])/\sq2 
\,.
\end{aligned}
\end{equation}
Thus, the massless $Z$-boson limit of the $WWZ$ amplitude yields only three terms with a positive-helicity $Z$:
\begin{equation}
\begin{aligned}
&-\anb{\bs{12}} (\langle\bs1\varepsilon_3^+\bs2] + \langle\bs2\varepsilon_3^+\bs1]) \;(2\cwwz{}{}{} + \cwwzLS\:m_W/\Lam)/\sq2 m_W	\\
&-\sqb{\bs{12}} (\langle\bs1\varepsilon_3^+\bs2] + \langle\bs2\varepsilon_3^+\bs1]) \;\cwwzRS/\sq2\Lam	\\[1mm]
&+\sqb{\bs{12}} \sqb{\bs1 3} \sqb{\bs2 3}	\;\cwwz+++/\Lam^2
.
\end{aligned}
\label{eqn:cwwz_to_cwwA}
\end{equation}
One can finally use \autoref{eqn:x_factor_consequence} to trade $\sqb{\bs{12}} (\langle\bs1\varepsilon_3^+\bs2] + \langle\bs2\varepsilon_3^+\bs1])$ for a combination of $\anb{\bs{12}} (\langle\bs1\varepsilon_3^+\bs2] + \langle\bs2\varepsilon_3^+\bs1])$ and $\anb{\bs{12}} \sqb{\bs1 3} \sqb{\bs2 3}/m_W$.
Following the same reasoning for the $\mathcal{M}(\bs 1_W,\bs 2_W, 3_\gamma^-)$ amplitude, and introducing new $\cwwa{}{}{}$ coefficients, one thus finally obtains:
\begin{equation}
\begin{aligned}
\mathcal{M}(\bs 1_W,\bs 2_W, 3_\gamma^+) =
	&+\sqb{\bs{12}} \sqb{\bs1 3} \sqb{\bs2 3} \;\cwwa+++/\Lam^2	\\
	&+\anb{\bs{12}} \sqb{\bs1 3} \sqb{\bs2 3} \;\cwwa00+/m_W\Lam	\\
	&+\anb{\bs{12}} (\langle\bs1\varepsilon^+_3\bs2]+\langle\bs2\varepsilon^+_3\bs1]) \; \cwwa{}{}{}/m_W\,,
	\\
\mathcal{M}(\bs 1_W,\bs 2_W, 3_\gamma^-) =
	&+\anb{\bs{12}} \anb{\bs1 3} \anb{\bs2 3} \;\cwwa---/\Lam^2	\\
	&+\sqb{\bs{12}} \anb{\bs1 3} \anb{\bs2 3} \;\cwwa00-/m_W\Lam	\\
	&+\sqb{\bs{12}}\: (\langle\bs1\varepsilon^-_3\bs2]+\langle\bs2\varepsilon^-_3\bs1]) \; \cwwa{}{}{}/m_W
\,,
\end{aligned}
\label{eqn:mwwa}
\end{equation}
with, in particular,
\begin{equation}
-(2\cwwz{}{}{} + (\cwwzLS + \cwwzRS) m_W/\Lam)/\sq2 \to \cwwa{}{}{}\,.
\label{eqn:cwwz2cwwa}
\end{equation}
The same result can also be obtained from eq.~(4.12) of ref.~\cite{Arkani-Hamed:2017jhn}.

The presence of five independent coefficients is compatible with the angular-momentum counting introduced in \autoref{sec:ppZ}.
Placing the two photon components in a fundamental representation and labeling representations by their dimensions, one indeed obtains five terms in the $3\otimes3\otimes2= 2\oplus2\oplus4\oplus4\oplus6$ decomposition.

\subsection{\texorpdfstring{$ZZZ$, $ZZ\gamma$, $Z\gamma\gamma$, and $\gamma\gamma\gamma$}{ZZZ, ZZA, ZAA, and AAA}}

\newcommand{\mzzz}[3]{\mathcal{M}(1_Z^#1,2_Z^#2,3_Z^#3)}
\newcommand{\czzz}[3]{c_{ZZZ}^{#1#2#3}}

In our discussion of the $WWZ$ amplitude, we found that for three mass-degenerate vectors, the $V^a V^b V^c$ amplitude is proportional to $\epsilon^{abc}$.
The $ZZZ$ amplitude therefore vanishes identically.

In amplitudes involving three identical massless vectors, permutation symmetry must be enforced among the same helicity ones.
Amplitudes involving three massless vectors of identical helicity are formed as spinor bilinear products like $\anb{12}\anb{13}\anb{23}$.
No fully symmetric combination can be formed out of such structures.
Amplitudes involving two massless vectors of identical helicity take forms like the ${\anb{12}^3}/{\anb{13}\anb{23}}$ one for the \amp--+ case.
They are thus antisymmetric under the exchange of the same-helicity vectors and therefore also vanish if those are identical.
We therefore conclude that no $\gamma\gamma\gamma$ amplitude can be written down.

Similarly, the $ZZ\gamma$ amplitude can be obtained from the $WW \gamma$ one with a symmetrization under $1\leftrightarrow 2$ exchange.
All the spinorial structures of the $WW \gamma$ amplitude are however antisymmetric under that exchange.
So we conclude that no $ZZ\gamma$ amplitude can be written down.

The $\mathcal{M}(\bs 1_Z, 2^{h_2}_\gamma,3^{h_3}_\gamma)$ amplitude is restricted by little-group considerations to the following structures (see \eg\ refs.~\cite{Arkani-Hamed:2017jhn}): $[\bs{1} 2]^{1 + h_2  - h_3}[\bs{1} 3]^{1 + h_3  - h_2} [23]^{h_2 + h_3 -1}= [\bs{1}2][\bs{1}3][23]$, $\anb{\bs{1}2}\anb{\bs{1}3} \anb{23}$ which are also antisymmetric under the $2\leftrightarrow 3$ exchange and thus incompatible with Bose statistics.
So we conclude that no $Z\gamma\gamma$ amplitude can be written down in agreement with the Landau--Yang theorem~\cite{Landau:1948kw, Yang:1950rg}.

%----------------------------------------------------------------------------------------
%	Section3
%----------------------------------------------------------------------------------------

%%%%%%%%%%%%%%%%%%%%%%%%%%%%%%%%%%%%%
\subsection{High-energy theory and \texorpdfstring{$SU(2)_L\times U(1)_Y$}{SU(2)L x U(1)Y}}
\label{sec:hesu2}
%%%%%%%%%%%%%%%%%%%%%%%%%%%
Our approach throughout this paper is purely bottom-up.
We only assume the unbroken symmetries of the SM, and we expect the $SU(2)_L\times U(1)_Y$ structure to emerge once unitarity is imposed on the four- and higher-point amplitudes of the SM particles.
It is instructive however to briefly comment on how this gauge structure emerges if we instead assume that the theory has a smooth massless limit, and make contact with the high-energy theory.
The high-energy theory comprises four massless spin-one degrees of freedom arising as linear combinations of the $W^\pm$, $Z$ and the photon, as well as four scalars arising from the Higgs $h$ and the longitudinal components of the $W^\pm$ and $Z$.

The $WWZ$ amplitude then includes interactions between three massless transverse vectors, and we have already seen that, for degenerate vectors $V^{a=1,2,3}$, it is proportional to $\epsilon^{abc}$.
We can furthermore isolate a $U(1)_Y$ gauge boson, by defining the combination $B \propto  \sqrt{2} c_{WWZ} \,  \gamma  +  c_{WW\gamma}\, Z$, such that the renormalizable $WWB$ amplitude  vanishes in the high-energy limit for transverse polarizations.%
\footnote{The appearance of this specific combination can be traced back to the renormalizable part of the \autoref{eqn:cwwz2cwwa} redefinition performed when taking the limit of the $WWZ$ amplitude to the $WW\gamma$ one.}

The high-energy limit of the $\psic\psi'W$ amplitudes then implies that the fermions come in specific $SU(2)_L$ representations.
Similarly, the coupling of the scalar to the longitudinal components of the vector implies that they should be embedded in a $SU(2)_L$ multiplet in the high-energy theory.
With one scalar and three longitudinal vector components, one can form a complex doublet.
A consistent factorization of the massless fermion-fermion-vector-vector or scalar-scalar-vector-vector amplitudes moreover relate the fermion-vector, scalar-vector, and triple-vector couplings~\cite{Arkani-Hamed:2017jhn} for both $SU(2)_L$ and $U(1)_Y$ vectors.

The $SU(2)_L\times U(1)_Y$ symmetry of the massless high-energy theory can thus be inferred from known results mostly concerning massless amplitudes.
We expect that the same conclusion can be obtained by considering the high-energy limit of massive four- and higher-point amplitudes.
Their study is postponed to future work but we detail the specific example of the $\psic\psi Zh$ amplitude in the next section.

%----------------------------------------------------------------------------------------
%	Section4
%----------------------------------------------------------------------------------------

%%%%%%%%%%%%%%%%%%%%%%%%%%%%%%%%%%%%%
\section{Four-point \texorpdfstring{$\psic \psi Z h$}{ffZh} amplitude}
\label{sec:4-pt}
%%%%%%%%%%%%%%%%%%%%%%%%%%%%%%%%%%%%%

We now turn to the construction of four-point, tree-level amplitudes.
These generically have factorizable parts, featuring single-particle poles, 
and non-factorizable parts, or contact-terms, with no negative powers of the kinematic invariants.
For the massive external states considered here, constructing the factorizable part is straightforward: each pole is associated with a particular factorization channel, and its residue is given by the product of two on-shell three-point functions.
This part of the amplitude can be simply obtained by gluing two three-point amplitudes, with the appropriate propagator.
Thus for example, gluing two three-point amplitudes along a massive vector line gives, 
\begin{equation}
-
\frac{1}{p^2-m^2}
\left[
\sum_{I,J,\tilde{I},\tilde{J}=1,2}
	\frac{\mathcal{M}^{IJ}_1(p) + \mathcal{M}^{JI}_1(p)}{2}\;
	\epsilon_{I\tilde I}\;\epsilon_{J\tilde J}
	\frac{\mathcal{M}^{\tilde I\tilde J}_2(\tilde p) + \mathcal{M}^{\tilde J\tilde I}_2(\tilde p)}{2}
\right]_{p^2=m^2}\,,
\label{eqn:gluing_prescription}
\end{equation}
where $m$ is the vector mass, and $\tilde p$ stands for an {\sl outgoing} momentum.
Pictorially,
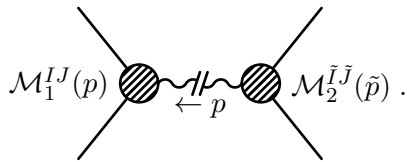
\begin{center}
\begin{fmfgraph*}(40,20)
\fmfleft{l1,l2}
\fmfright{r1,r2}
\fmf{vanilla}{l1,v1,l2}
\fmf{photon,tension=1, lab=$\leftarrow p$}{v1,v2}
\fmf{vanilla}{r1,v2,r2}
\fmfv{d.shape=circle,d.fill=shaded, lab=$\mathcal{M}_1^{IJ}(p)$,lab.angle=180, lab.dist=4mm}{v1}
\fmfv{d.shape=circle,d.fill=shaded, lab=$\mathcal{M}_2^{\tilde{I}\tilde{J}}(\tilde{p})\;.$,lab.angle=0, lab.dist=4mm}{v2}
\fmffreeze
\fmfcmd{
vdraw;
pair mid;
path p;
mid := .5[vloc(__v1),vloc(__v2)];
p := (-.5mm,-1.5mm)--(.5mm,1.5mm);
draw (p shifted mid) withpen (pencircle xscaled 1mm) withcolor white;
draw ((p shifted mid) shifted (.5mm,0));
draw ((p shifted mid) shifted (-.5mm,0));
vlist.last := 0;
}
\end{fmfgraph*}
\end{center}
Recall that so far we have taken all momenta to be incoming.
When gluing together two three-point amplitudes, we need to consistently define the contractions of bolded spinors for incoming and outgoing momenta.
These are discussed in \subappref{sec:massive_gluing}.

To obtain the non-factorizable component, we write the most general kinematic structures involving positive powers of the spinor products, that are consistent with little-group constraints, similarly to the derivation of the three-point amplitudes.
These structures appear with a-priori arbitrary coefficients.
Combining the factorizable and non-factorizable pieces we thus get the most general amplitude which consistently factorizes on single particle poles.
Imposing $SU(2)_L\times U(1)_Y$, or alternatively, requiring perturbative unitarity, would imply specific relations between the contact-term coefficients and the couplings of three-point amplitudes.

For definiteness, we will focus on the $\psic \psi Z h$ amplitudes relevant for Higgs decay and associated production at both hadron and lepton colliders.
The high-energy limit of non-factorizable contributions does not lead to energy growths that are not matched by appropriate power of the cutoff $\Lam$.
This does however occur for factorizable contributions, which feature negative powers of the masses, and thus allow us to derive relations between amplitude coefficients, satisfied up to $m/\Lam$ corrections.

\subsection{Non-factorizable contributions}
\newcommand{\cffzh}[4][]{c_{\psic\psi Z h}^{\chir{#2}\chir{#3}\chir{#4}#1}}
\newcommand{\nf}{^\text{nf}}

Using momentum conservation, Schouten identities, on-shell conditions and Dirac matrix algebra, it is possible to show that all non-factorizable contributions to the fully massive $\psic\psi Zh$ four-point amplitude can be reduced to the following twelve terms:
\begin{equation}
\begin{aligned}
\mathcal{M}\nf(\bs1_\psic,\bs2_\psi,\bs3_Z,\bs4_h)
&=	\frac{\cffzh+++}{\Lam^2}	\sqb{\bs{13}}\sqb{\bs{23}}
+	\frac{\sqb{\bs{12}}}{\Lam^3} \langle\bs3\;\{
		 \cffzh[_A]++0 (\bs1+\bs2)
		+\cffzh[_S]++0 (\bs1-\bs2)
		\}\;\bs3]
\\&+	\frac{\cffzh+-0}{\Lam^2}\sqb{\bs{13}}\anb{\bs{23}}
+	\frac{\cffzh+-+}{\Lam^3}[\bs{312}\rangle \sqb{\bs{13}}
+	\frac{\cffzh+--}{\Lam^3}\langle\bs{321}] \anb{\bs{23}}
\\&+	\frac{\cffzh-+0}{\Lam^2}\anb{\bs{13}}\sqb{\bs{23}}
+	\frac{\cffzh-++}{\Lam^3}[\bs{321}\rangle \sqb{\bs{23}}
+	\frac{\cffzh-+-}{\Lam^3}\langle\bs{312}] \anb{\bs{13}}
\\&+	\frac{\cffzh---}{\Lam^2}	\anb{\bs{13}}\anb{\bs{23}}
+	\frac{\anb{\bs{12}}}{\Lam^3} \langle\bs3\;\{
		 \cffzh[_A]--0 (\bs1+\bs2)
		+\cffzh[_S]--0 (\bs1-\bs2)
		\}\;\bs3]
.
\end{aligned}
\label{eqn:mffzh}
\end{equation}
In the case of three-point amplitudes, all kinematics dependence was included in the spinor structures, leaving only constant coefficients.
For four-point amplitudes such as that of \autoref{eqn:mffzh}, the twelve $\cffzh{}{}{}$ coefficients are functions of $s_{ij}$.
At tree-level, each of them is given as an expansion in positive powers of the $s_{ij}$.

In the construction of \autoref{eqn:mffzh}, $p_4$ is eliminated using momentum conservation.
Each term must contain a single angle- or square-bracket spinor for particles $1$ and $2$, as well as two such spinors for particle $3$, and therefore involves two spinor products.
In order to find the independent structures, one need only start with terms involving two spinor products and up to one insertion of each of the momenta, since multiple insertions can be eliminated using anti-commutation relations and/or on-shell conditions.
We have verified explicitly that these structures can be systematically reduced to \autoref{eqn:mffzh}.
The following equality is for instance used to eliminate the $\chir+\chir+\chir-$ term (and the $\chir-\chir-\chir+$ one by exchanging square and angle brackets):
\begin{multline}
\sqb{\bs{12}}\anb{\bs{3123}} =\:
	2\; \sqb{\bs{12}} \;
		\langle\bs3\;\{
		 \bs1\: (p_2\cdot p_3)
		-\bs2\: (p_1\cdot p_3)\}\;
		 \bs3] /m_3
	\\- 2(p_1\cdot p_2) \sqb{\bs{13}}\sqb{\bs{23}}
	- m_1 [\bs{321}\rangle \sqb{\bs{23}}
	- m_2 [\bs{312}\rangle \sqb{\bs{13}}.
\label{eqn:ffzh++-}
\end{multline}
Remarkably, the first term on the right-hand side has a well-defined $m_3\to 0$ limit since the leading $(p_2\cdot p_3) \langle3 1 3 ] -  (p_1\cdot p_3) \langle 3 2 3]$ term vanishes identically.
We also note that the second line is equal to $\sqb{\bs{12}}\sqb{\bs{3213}}$, a structure that is thus also reducible to the ones of \autoref{eqn:mffzh}.
The massless fermion amplitude can be obtained by simply unbolding the fermion spinors.
A direct derivation of the massless fermion amplitude is provided in \autoref{sec:ffZh_EFT_derivation}.
The agreement between the results of \autoref{eqn:mffzh}, \autoref{eqn:ffZh_EFT_helicity_amp_appendix} in the massless fermion limit establishes the independence of the twelve terms of \autoref{eqn:mffzh}.
This independence can also be verified numerically, by considering vectors formed by the numerical values of the twelve spinor structures for arbitrary choices of momenta, and verifying that they are linearly independent.

\newcommand{\mffzh}[4][]{\mathcal{M}#1(1_\psic^#2,2_\psi^#3,3_Z^#4,4_h)}

The leading high-energy contributions are obtained by simple \emph{unbolding} of the spinorial structures in \autoref{eqn:mffzh} (see also a direct derivation from $U(1)$ little group in \subappref{sec:massless_ffzh}):
\begin{equation}
\begin{aligned}
\mffzh[\nf]+++	&\to \sqb{13}\sqb{23}		\;\cffzh+++	/\Lam^2	,\\
\mffzh[\nf]++0	&\to 2\;\sqb{12} 
		\;\{ \cffzh[_A]++0\; p_3\cdot(p_1+p_2) + \cffzh[_S]++0\; p_3\cdot(p_1-p_2)\}
								/\Lam^3	,\\
\mffzh[\nf]+-0	&\to -[132\rangle		\;\cffzh+-0	/\Lam^2	,\\
\mffzh[\nf]+-+	&\to - \langle 1 2\rangle \sqb{13}^2	\;\cffzh+-+	/\Lam^3	,\\
\mffzh[\nf]+--	&\to  [12] \anb{23}^2	\;\cffzh+--	/\Lam^3 ,\\
\mffzh[\nf]-+0	&\to -[231\rangle		\;\cffzh-+0	/\Lam^2	,\\
\mffzh[\nf]-++	&\to  \langle12 \rangle\sqb{23}^2	\;\cffzh-++ 	/\Lam^3	,\\
\mffzh[\nf]-+-	&\to - [12] \anb{13}^2	\;\cffzh-+- 	/\Lam^3	,\\
\mffzh[\nf]---	&\to \anb{13}\anb{23}		\;\cffzh---	/\Lam^2	,\\
\mffzh[\nf]--0	&\to  2\;\anb{12} 
		\;\{ \cffzh[_A]--0\; p_3\cdot(p_1+p_2) + \cffzh[_S]--0\; p_3\cdot(p_1-p_2)\}
								/\Lam^3	.
\end{aligned}
\end{equation}
This justifies the normalization adopted with inverse powers of $\Lam$ (at tree level).
In the SMEFT, dimension-six operators (dipoles and vectors) lead to the four $1/\Lam^2$ structures (see \subappref{sec:matching_ffzh}).
The missing $\mffzh[\nf]++-\;\text{ and }\;\mffzh[\nf]--+$ amplitudes are generated at orders $1/\Lam^4$ (see \autoref{eqn:massless_FFVS}) by the specific linear combination in the right-hand side of \autoref{eqn:ffzh++-} and its analogue with square and angle brackets interchanged.

In the soft-Higgs limit, with the four-momentum $p_\bs4\to 0$, only the four $1/\Lam^2$ terms of \autoref{eqn:mffzh} survive, reproducing the list of structures found for the $\psic\psi Z$ amplitude in \autoref{eqn:mffz}.
Indeed, at the level of the non-factorizable terms, this limit corresponds to setting the Higgs field to its vacuum expectation value.

%%%%%%%%%%%%%%%%%%%%%%%%%%%%%%%%%%%%%
\subsection{Factorizable contributions}
%%%%%%%%%%%%%%%%%%%%%%%%%%%%%%%%%%%%%

Given the field content of our electroweak theory, the possible factorizable contributions to the $\psic\psi Zh$ amplitudes are:
\begin{equation}\raisebox{-10mm}{%
\fmfframe(5,5)(5,5){\begin{fmfgraph*}(20,10)
\fmfleft{l1,l2}
\fmfright{r1,r2}
\fmf{fermion}{l1,v1,l2}	\fmflabel{$\bs2_\psi$}{l1}	\fmflabel{$\bs1_\psic$}{l2}
\fmf{photon,tens=1,lab=$\bs p_{Z,,\gamma}$}{v1,v2}
\fmf{photon}{v2,r2}	\fmflabel{$\bs3_Z$}{r2}
\fmf{dashes}{v2,r1}	\fmflabel{$\bs4_h$}{r1}
\end{fmfgraph*}}
\quad
\fmfframe(5,5)(5,5){\begin{fmfgraph*}(20,10)
\fmfleft{l1,l2}
\fmfright{r1,r2}
\fmf{fermion}{l1,v1}
\fmf{fermion}{v2,l2}	\fmflabel{$\bs2_\psi$}{l1}	\fmflabel{$\bs1_\psic$}{l2}
\fmf{fermion,tens=.5,lab=$\bs p_\psi$}{v1,v2}
\fmf{photon}{v2,r2}	\fmflabel{$\bs3_Z$}{r2}
\fmf{dashes}{v1,r1}	\fmflabel{$\bs4_h$}{r1}
\end{fmfgraph*}}
\quad
\fmfframe(5,5)(5,5){\begin{fmfgraph*}(20,10)
\fmfleft{l1,l2}
\fmfright{r1,r2}
\fmf{fermion}{l1,v1}
\fmf{fermion}{v2,l2}	\fmflabel{$\bs2_\psi$}{l1}	\fmflabel{$\bs1_\psic$}{l2}
\fmf{fermion,tens=.5,lab=$\bs p_\psi$,lab.side=left}{v1,v2}
\fmf{phantom}{v2,r2}
\fmf{phantom}{v1,r1}
\fmffreeze
\fmf{photon}{v1,r2}	\fmflabel{$\bs3_Z$}{r2}
\fmf{dashes}{v2,r1}	\fmflabel{$\bs4_h$}{r1}
\end{fmfgraph*}}
\quad
\fmfframe(5,5)(5,5){\begin{fmfgraph*}(20,10)
\fmfleft{l1,l2}
\fmfright{r1,r2}
\fmf{fermion}{l1,v1,l2}	\fmflabel{$\bs2_\psi$}{l1}	\fmflabel{$\bs1_\psic$}{l2}
\fmf{dashes,tens=1,lab=$\bs p_h$}{v1,v2}
\fmf{photon}{v2,r2}	\fmflabel{$\bs3_Z$}{r2}
\fmf{dashes}{v2,r1}	\fmflabel{$\bs4_h$}{r1}
\end{fmfgraph*}}
}
\label{eqn:ffzh_diagrams}
.
\end{equation}
We know however that, for identical scalars, the $hhZ$ amplitude vanishes on-shell.
Thus, only $s$-channel $Z$ and photon exchanges, as well as $t$- and $u$-channel fermion ones have to be considered.
We treat each of those contributions in turn, employing the \emph{gluing} prescription detailed in \autoref{sec:massive_gluing}.

\paragraph{\texorpdfstring{$s$}{s}-channel \texorpdfstring{$Z$}{Z} exchange}

Let us start with the $s$-channel $Z$-boson exchange.
The three-point amplitudes to be joined are those of \autoref{eqn:mffz} and \eqref{eqn:mvvh}.
Applying the prescription of \autoref{eqn:gluing_prescription}, we obtain the following $s$-channel amplitude:
\begin{equation}
\begin{aligned}
\mathcal{M}^Z(\bs1_\psic,\bs2_\psi,\bs3_Z,\bs4_h) &\times (s-m_Z^2)=\\
	&+(\anb{\bs{12}} - \sqb{\bs{12}})	\;
	\langle\bs3(\bs1+\bs2)\bs3]		\;
	(\cffz+-0 - \cffz-+0)	\czzh00		\;
	m_\psi/2m_Z^2
\\
	&- \anb{\bs{13}} \sqb{\bs{23}}\; \cffz-+0 \czzh00
\\
	&- \anb{\bs{23}} \sqb{\bs{13}}\; \cffz+-0 \czzh00
\\[1.5mm]
	&+(\anb{\bs{12}}-\sqb{\bs{12}}) \langle\bs3 (\bs1-\bs2) \bs3] \;
	(\cffz--- - \cffz+++) \czzh00	\;
	/4\Lam
\\
	&+(\anb{\bs{12}}+\sqb{\bs{12}}) \langle\bs3 (\bs1-\bs2) \bs3] \;
	(\cffz--- + \cffz+++) \czzh00	\;
	/4\Lam
\\
	&-(\anb{\bs{13}} \sqb{\bs{23}} + \anb{\bs{23}} \sqb{\bs{13}}) 	\;
	(\cffz--- + \cffz+++) \czzh00	\;
	m_\psi/\Lam
\\[1.5mm]
	&+ \anb{\bs{13}} \langle\bs3\bs1\bs2]\; \cffz-+0 \czzh-- / \Lam
\\
	&+ \anb{\bs{23}} \langle\bs3\bs2\bs1]\; \cffz-+0 \czzh-- / \Lam
\\
	&- \anb{\bs{13}} \anb{\bs{23}}\;	(\cffz+-0 + \cffz-+0) \czzh-- m_\psi/\Lam
\\[1.5mm]
	&+ \sqb{\bs{13}} \langle\bs2\bs1\bs3]\; \cffz-+0 \czzh++ / \Lam
\\
	&+ \anb{\bs{23}} \langle\bs1\bs2\bs3]\; \cffz-+0 \czzh++ / \Lam
\\
	&- \sqb{\bs{13}} \sqb{\bs{23}}\;	(\cffz+-0 + \cffz-+0) \czzh++ m_\psi/\Lam\,,
\end{aligned}
\end{equation}
where only one non-renormalizable vertex has been allowed.

In the high-energy limit, the longitudinal component of the $\langle\bs3(\bs1+\bs2)\bs3]$ spinorial structure tends to $-s/\sq2$ and cancels the propagator up to mass corrections.
(The square root of two arises from the Clebsch-Gordan coefficient in the prescription of \autoref{eqn:bold_clebsch}).
The fastest growing $s$-channel amplitudes are thus:
\begin{equation}
\begin{aligned}
\mffzh[^Z]--0
	&\to -\anb{12} \;
	((\cffz+-0 - \cffz-+0)	\;
	m_\psi/m_Z^2
	-\cffz---\: (t-u)/s\Lam)\;	\czzh00	/2\sq2
,\\
\mffzh[^Z]-+-
	&\to +\anb{13}^2/\anb{12}\;
	\cffz-+0 \czzh-- /\Lam
,\\
\mffzh[^Z]-++
	&\to -\sqb{23}^2/\sqb{12}\;
	\cffz-+0 \czzh++ /\Lam
,\\
\mffzh[^Z]+--
	&\to -\anb{23}^2/\anb{12}\;
	\cffz+-0 \czzh-- /\Lam
,\\
\mffzh[^Z]+-+
	&\to +\sqb{13}^2/\sqb{12}\;
	\cffz+-0\czzh++ /\Lam
,\\
\mffzh[^Z]++0
	&\to + \sqb{12}	\;
	((\cffz+-0 - \cffz-+0)	\;
	m_\psi/m_Z^2
	+\cffz+++\: (t-u)/s\Lam)\;	\czzh00	/2\sq2
,\\
\end{aligned}
\end{equation}
where we have used $\sqb{12}/s\to -1/\anb{12}$ and  $\anb{12}/s\to -1/\sqb{12}$.
Compared to the expectation deriving from our simplified argument based three-point amplitudes, we note that the \amp--{00} and \amp++{00} receive contributions of unacceptable $E/m$ growths.
Amplitudes involving longitudinal vector bosons are well-known to give rise to such behaviors.

\paragraph{\texorpdfstring{$s$}{s}-channel photon exchange}
The $s$-channel photon exchange only arises at the non-renormalizable level, since the $\gamma Zh$ amplitude of \autoref{eqn:mazh} receives no renormalizable tree-level contribution.
The result can be obtained from the $m_Z\to 0$ limit of the $Z$-boson exchange with $\cffz-+0\to \cffz+-0\to -\sq2\: \cffa{}{}{}$ (following \autoref{eqn:ffz2ffa_replacement}):
\begin{equation}
\begin{aligned}
\mathcal{M}^A(\bs1_\psic,\bs2_\psi,\bs3_Z,\bs4_h) \times s=
	&- \sqrt{2}\anb{\bs{13}} \langle\bs3\bs1\bs2]\;  \cffa{}{}{} \cazh-- / \Lam
\\
	&- \sqrt{2}\anb{\bs{23}} \langle\bs3\bs2\bs1]\;  \cffa{}{}{} \cazh-- / \Lam
\\
	&+ 2 \sqrt{2} \anb{\bs{13}} \anb{\bs{23}}\;\cffa{}{}{} \cazh--  m_\psi/\Lam
\\[1.5mm]
	&- \sqrt{2}\sqb{\bs{13}} \langle\bs2\bs1\bs3]\; \cffa{}{}{} \cazh++ / \Lam
\\
	&- \sqrt{2}\sqb{\bs{23}} \langle\bs1\bs2\bs3]\; \cffa{}{}{} \cazh++ / \Lam
\\
	&+2 \sqrt{2} \sqb{\bs{13}} \sqb{\bs{23}}\;\cffa{}{}{}\cazh++ m_\psi/\Lam
.
\end{aligned}
\end{equation}
This form is also obtained from a direct gluing of the \autoref{eqn:mffa} and \eqref{eqn:mazh} photon amplitudes.

\paragraph{\texorpdfstring{$t$- and $u$-}{t- and u-}channel fermion exchanges}

The $t$- an $u$-channel diagrams of \autoref{eqn:ffzh_diagrams} involve the three-point amplitudes of \autoref{eqn:mffz} and \eqref{eqn:mffh}.
The prescription of \autoref{eqn:gluing_prescription} leads to the following $t$-channel amplitude:
\begin{equation}
\begin{aligned}
\mathcal{M}^t(\bs1_\psic,\bs2_\psi,\bs3_Z,\bs4_h) &\times (t-m_\psi^2)=\\
	&- \anb{\bs{12}} \langle\bs{313}]\;\cffh-- \cffz-+0 /m_Z
\\	&- \sqb{\bs{12}} \langle\bs{313}]\;\cffh++ \cffz+-0 /m_Z
\\	&+ \anb{\bs{13}} \sqb{\bs{23}}\; ((\cffz+-0-\cffz-+0) m_\psi/m_Z - \cffz--- m_Z/\Lam) \cffh++
\\	&- \anb{\bs{23}} \sqb{\bs{13}}\; ((\cffz+-0-\cffz-+0) m_\psi/m_Z + \cffz+++ m_Z/\Lam) \cffh--
\\[1.5mm]
	&-\anb{\bs{13}}\anb{\bs{23}}\; (\cffz-+0 + \cffz--- m_\psi/\Lam) \cffh--
\\	&-\anb{\bs{13}}\langle\bs{312}]\; \cffh++\cffz---/\Lam
\\[1.5mm]
	&-\sqb{\bs{13}}\sqb{\bs{23}}\; (\cffz+-0 + \cffz+++ m_\psi/\Lam) \cffh++
\\	&-\sqb{\bs{13}}\langle\bs{213}]\; \cffh--\cffz+++/\Lam\,.
\end{aligned}
\end{equation}
In the high-energy limit, the longitudinal component of the $\langle\bs{313}]$ spinor structure tends to $t/\sq2$.
The fastest growing amplitudes are then:
\begin{equation}
\begin{aligned}
\mffzh[^t]--0
	&\to -\anb{12} \;
	\cffz-+0\cffh-- \;
	/\sq2m_Z
\,,\\
\mffzh[^t]-+-
	&\to -\anb{13}^2/\anb{12}\; \cffz---\cffh++\; s/t\Lam
\,,\\
\mffzh[^t]+-+
	&\to -\sqb{13}^2/\sqb{12}\; \cffz+++\cffh--\; s/t\Lam
\,,\\
\mffzh[^t]++0
	&\to -\sqb{12} \;
	\cffz+-0\cffh++ \;
	/\sq2m_Z
\,,
\end{aligned}
\end{equation}
where again the \amp--{00} and \amp++{00} ones receive energy-growing contributions not naively expected from our simplified analysis of three-point amplitudes.
Similarly, the $u$-channel contribution is:
\begin{equation}
\begin{aligned}
\mathcal{M}^u(\bs1_\psic,\bs2_\psi,\bs3_Z,\bs4_h) &\times (u-m_\psi^2)=\\
	&+ \anb{\bs{12}} \langle\bs{323}]\;\cffh-- \cffz+-0 /m_Z
\\	&+ \sqb{\bs{12}} \langle\bs{323}]\;\cffh++ \cffz-+0 /m_Z
\\	&- \anb{\bs{23}} \sqb{\bs{13}}\; ((\cffz+-0-\cffz-+0) m_\psi/m_Z + \cffz--- m_Z/\Lam) \cffh++
\\	&+ \anb{\bs{13}} \sqb{\bs{23}}\; ((\cffz+-0-\cffz-+0) m_\psi/m_Z - \cffz+++ m_Z/\Lam) \cffh--
\\[1.5mm]
	&-\anb{\bs{13}}\anb{\bs{23}}\; (\cffz+-0 + \cffz--- m_\psi/\Lam) \cffh--
\\	&-\anb{\bs{23}}\langle\bs{321}]\; \cffh++\cffz---/\Lam
\\[1.5mm]
	&-\sqb{\bs{13}}\sqb{\bs{23}}\; (\cffz-+0 + \cffz+++ m_\psi/\Lam) \cffh++
\\	&-\sqb{\bs{23}}\langle\bs{123}]\; \cffh--\cffz+++/\Lam\,,
\end{aligned}
\end{equation}
where $\langle\bs{323}]$ tends to $u/\sq2$ in the high-energy limit.
The leading amplitudes are then:
\begin{equation}
\begin{aligned}
\mffzh[^u]--0
	&\to +\anb{12} \;
	\cffz+-0\cffh-- \;
	/\sq2m_Z
\,,\\
\mffzh[^u]-++
	&\to +\sqb{23}^2/\sqb{12} \; \cffz+++\cffh-- s/u\Lam
\,,\\
\mffzh[^u]+--
	&\to +\anb{23}^2/\anb{12} \; \cffz---\cffh++ s/u\Lam
\,,\\
\mffzh[^u]++0
	&\to +\sqb{12} \;
	\cffz-+0\cffh++ \;
	/\sq2m_Z
\,,
\end{aligned}
\end{equation}
and the \amp--{00} and \amp++{00} again display a growth not expected from the three-point amplitude analysis.
Note that the $u$-channel amplitude is obtained from the $t$-channel one by exchange of the fermion labels (including chirality labels, where needed).

\paragraph{Perturbative unitarity and the fermion mass.}

Perturbative unitarity constrains the parameters of the $\psi^c\psi Zh$ amplitude.
For the non-renormalizable terms, this gives non-trivial relations among the coefficients of the factorizable and non-factorizable terms.
The $s$-wave coefficients of each helicity amplitude growing like a power of $E/\Lam$, must be smaller than a some numerical value such that unitarity is only lost at $E=\Lam$.
Since the result of \autoref{eqn:mffzh} is an all-orders result, which can be expanded in powers of $p_i\cdot p_j/\Lam^2$, this implies quantitative relations between the coefficients in this expansion.
We leave a full exploration of these relations to future work.

Furthermore, at the renormalizable level, there is $E/m$ energy growth in the sum of contributions to the \amp\pm\pm{00} amplitudes.
To cancel this, one is forced to impose:
\begin{equation}
\begin{aligned}
(\cffz+-0 - \cffz-+0)\; ( \czzh00\:{m_\psi}/{2m_Z} - \cffh--)  &= 0 + \Order({m}/{\Lam})\,,\\
(\cffz+-0 - \cffz-+0)\; ( \czzh00\:{m_\psi}/{2m_Z} - \cffh++)  &= 0 + \Order({m}/{\Lam})\,.
\end{aligned}
\label{eqn:four-point-relation}
\end{equation}
One can distinguish two possibilities for satisfying this condition.
One is that
\beq
\cffz+-0 = \cffz-+0 + \Order(m/\Lam)\,.
\eeq
In this case, the fermion has a \emph{vector-like} coupling to the $Z$ boson, up to $m/\Lam$ corrections.
No relation is then forced between its mass and its Yukawa coupling.
The second possibility is that the fermion has \emph{chiral} couplings to the $Z$ boson, with $\cffz+-0 \ne \cffz-+0$ at zeroth order in $m/\Lam$.
Then, its mass is directly related to the Yukawa coupling,
\beq
\cffh++= \czzh00 m_\psi/2m_Z = \cffh--\,,
\eeq
up to $m/\Lam$ corrections (see also eq.\,(5.5) of ref.~\cite{Maltoni:2019aot}).

In the SM, it is of course this second possibility that is realized, and the requirements of \autoref{eqn:four-point-relation} are explicitly satisfied in the SMEFT (see \subappref{sec:four-point-relation-matching}).
In the presence of additional Higgses, which couple to $\psi$, this relation would be modified.
Then, the low energy theory would feature an additional scalar, $h^\prime$, and the $\psi^c\psi Zh$ amplitude would get a contribution from factorization on $\psi^c\psi h^\prime$ and $Z h h^\prime$. Indeed, with two different scalars, the latter can be nonzero.

%----------------------------------------------------------------------------------------
% 	Section5
%----------------------------------------------------------------------------------------

\section{Conclusions}
\label{sec:conclusion}

We adopted massive on-shell amplitude methods to examine a theory of the electroweak spectrum from the bottom up, including both renormalizable and non-renormalizable interactions.
Our main objectives were first to explicitly demonstrate the emergence of the patterns due to electroweak symmetry breaking and, second, to provide the necessary basis for on-shell amplitude computations equivalent to those obtained from the Lagrangian of the standard-model effective field theory (SMEFT).

We started by performing the bottom-up construction of all massive on-shell three-point amplitudes with the electroweak field content.
The only assumptions made concerned the particle spectrum, as well as electric-charge and fermion-number conservation.
A simplified argument based on the tree-level unitarity of factorizable four-point amplitudes was employed to distinguish renormalizable and non-renormalizable contributions to tree-level, three-point amplitudes, without reference to a Lagrangian.
The non-trivial structure of the renormalizable $\psic\psi\gamma$, $WWZ$ or $WW\gamma$ amplitudes could for instance be obtained in a purely bottom-up way.
The emergence of properties expected from electroweak symmetry breaking, the Higgs mechanism and the Goldstone equivalence theorem were highlighted.

Prescriptions for the treatment of massive spinors and the gluing of massive three-point amplitudes into tree-level higher-point ones were clarified.
The construction of both factorizable and non-factorizable contributions to the $\psic\psi Zh$ four-point amplitude was detailed as illustration.
Imposing tree-level unitarity up to an arbitrarily high cutoff scale led to the expected relation between renormalizable $ZZh$,\,$\psic\psi h$ couplings and $m_Z$,\,$m_\psi$ masses.

All renormalizable and non-renormalizable parameters appearing in the three-point and non-factorizable four-point amplitudes were defined and matched at tree-level to the broken-phase of the SMEFT.
This laid the basis for future SMEFT computations using on-shell amplitude methods.
Operator redundancies due to field redefinitions are absent in this approach.

Sticking to a bottom-up approach, all the four-fermion, two-fermion-two-boson and four-boson amplitudes would next need to be considered.
As in the $\psic\psi Zh$ case we discussed, the unitarity of these four-point amplitudes would impose further constraints on both the renormalizable and non-renormalizable parameters introduced here.
All relations deriving from the underlying $SU(2)_L\times U(1)_Y$ symmetry of the theory are expected to be obtained from four- and higher-point amplitudes.
For an electroweak-like particle spectrum, this is indeed the only consistent ultraviolet completion.
Before perturbative unitarity is imposed, our bottom-up construction also applies to more general theories featuring for instance a low-cutoff scale or non-linearly realized $SU(2)_L\times U(1)_Y$ symmetry.
The breakdown of unitarity in such theories may be best probed by amplitudes with $n>4$ vector boson and Higgs legs~\cite{Henning:2018kys,Chang:2019vez}.
The general, relatively compact expressions for $n$-point amplitudes obtained in our approach will be useful starting points for this study.
It will also be interesting to use these results in order to explore the relation between the high-energy behavior and analytic structure of the amplitudes, and the non-analytic Higgs potentials featured in EFTs beyond the SMEFT~\cite{Falkowski:2019tft}.

The soft limits of on-shell EFT amplitudes have yielded powerful restrictions on general EFTs~\cite{Cheung:2014dqa, Cheung:2015ota, Cheung:2015cba, Cheung:2016drk, Low:2017mlh, Cheung:2018oki, Low:2019ynd}, and it would be interesting to explore them in the case at hand.
Amplitudes with Goldstone bosons should vanish in this limit, while the soft-Higgs limit amounts to setting the Higgs to its vacuum expectation value~\cite{Dixon:2004za}.

%----------------------------------------------------------------------------------------
%	Acknowledgements
%----------------------------------------------------------------------------------------

\section*{Acknowledgments}
We thank Junmou Chen, Go Mishima, and Ofri Telem for valuable discussions.
This research is supported by the Israel Science Foundation (Grant No.~720/15), by the United-States-Israel Binational Science Foundation (BSF) (Grant No.~2014397), and by the ICORE Program of the Israel Planning and Budgeting Committee (Grant No.~1937/12).
The work of G.D.\ is supported in part at the Technion by a fellowship from the Lady Davis Foundation.
The work of T.K.\ is supported by the Japan Society for the Promotion of Science (JSPS) KAKENHI Grant Number 19K14706.
G.D.\ thanks the Aspen Center for Physics, which is supported by National Science Foundation grant PHY-1607611, where some of this work was performed.

%----------------------------------------------------------------------------------------
%	Appendix
%----------------------------------------------------------------------------------------

\appendix

%%%%%%%%%%%%%%%%%%%%%%%%%%%%%%%%%%%%%
\section{Massless and massive spinors}
\label{sec:notation}
\label{sec:formalism}
%%%%%%%%%%%%%%%%%%%%%%%%%%%%%%%%%%%%%

We use the massive spinor formalism introduced in ref.~\cite{Arkani-Hamed:2017jhn}.
We first introduce our notations and conventions.
Our conventions for two-component spinors mostly follow ref.~\cite{Dreiner:2008tw}.%
\footnote{Thus, our conventions for raising and lowering $SU(2)$ indices are consistent with refs.~\cite{Arkani-Hamed:2017jhn, Chung:2018kqs, Aoude:2019tzn}, but different from  ref.~\cite{Shadmi:2018xan}.}
The metric is 
\beq
\eta_{\mu\nu}= \eta^{\mu\nu}  = \diag \parn{ +, -, -, -}\,,
\eeq
and hence  $p^2 = m^2 > 0$ for massive particles.
The gamma matrices are defined as
\beq
\gamma^{\mu} = 
\begin{pmatrix}
0 & \left(\sigma^{\mu}\right)_{\alpha \dot\beta} \\
\left( \bar\sigma^{\mu} \right)^{\dot\alpha \beta}& 0 
\end{pmatrix}\,,
\eeq
where
 $\sigma^{\mu}_{\alpha \dot\beta} = ( \delta_{\alpha \dot\beta},\, \vec{\sigma}_{\alpha \dot\beta})$ and 
$\bar{\sigma}^{\mu  \dot\alpha \beta} = ( \delta^{\dot\alpha \beta},\, - \vec{\sigma}^{\dot\alpha \beta})$ and,
\beq
\sigma^{\mu}_{\alpha \dot \alpha} \bar{\sigma}_{\mu}^{\dot \beta \beta} = 2 \delta_{\alpha}^{\beta} \delta_{\dot \alpha}^{\dot \beta}\,.
\eeq
The commutator for the dipole operators is defined as
\beq
\sigma^{\mu \nu} = \frac{i}{2} [\gamma^{\mu},\gamma^{\nu}]\,.
\eeq
To simplify the notation we introduce the ``half-brackets'' 
\begin{equation}\label{eqn:spinor_helicity_variables_half_bracket_notation}
  \lambda^{\alpha}=\bra \lambda\,,
  \quad
  \lambda_{\alpha}=\ket \lambda\,,\quad 
\quad
  \tilde{\lambda}_{\dot{\alpha}}=\Bra{\tilde{\lambda}}\,,
  \quad
  \tilde{\lambda}^{\dot{\alpha}}=\Ket{\tilde{\lambda}}\,,
\end{equation}
where the $\lambda_{\alpha}$ and $\tilde{\lambda}_{\dot \alpha}$ 
are commuting and independent two-dimensional complex vectors 
(undotted and dotted index spinors, respectively).
Indices of the spinor brackets are contracted by
\beq
\anb{\lambda \chi}&= \epsilon_{\alpha \beta}\lambda^{\alpha} \chi^{\beta}= \lambda^{\alpha} \chi_{\alpha} = - \lambda_{\alpha} \chi^{\alpha} = - \chi^{\alpha}\lambda_{\alpha}=- \anb{\chi \lambda } \,,\nn\\
\sqb{\tilde{\lambda} \tilde{\chi}} 
&=  \epsilon^{ \dot \alpha \dot \beta} \tilde{\lambda}_{\dot\alpha} \tilde{\chi}_{\dot\beta}=
  \tilde{\lambda}_{\dot\alpha} \tilde{\chi}^{\dot\alpha} = - \tilde{\lambda}^{\dot\alpha} \tilde{\chi}_{\dot\alpha}
   = -  \tilde{\chi}_{\dot\alpha}\tilde{\lambda}^{\dot\alpha} = - \sqb{ \tilde{\chi}\tilde{\lambda}} \,,
\eeq
which satisfy
\beq
\langle \lambda \sigma^{\mu} \tilde{\chi}] = [ \tilde{\chi} \bar{\sigma}^{\mu} \lambda \rangle\,.
\eeq
The two-component Levi-Civita tensors are defined as in ref.~\cite{Dreiner:2008tw},%
\footnote{Note however that our convention for the totally antisymmetric tensor $\epsilon_{\mu \nu \rho \sigma}$ is opposite to that of ref.~\cite{Dreiner:2008tw}: we adopt $\epsilon_{0123}=+1$
as in ref.~\cite{Grzadkowski:2010es},  where
the Warsaw basis of SMEFT operators is defined.}
\beq
\epsilon^{IJ} = -\epsilon_{IJ} = \epsilon^{\alpha \beta} = - \epsilon_{\alpha \beta} = 
\epsilon^{\dot\alpha \dot\beta} = - \epsilon_{\dot\alpha \dot\beta}=
\begin{pmatrix}
0 & 1\\ 
-1 & 0
\end{pmatrix}\,.
\eeq

For massive particles, the momentum is given by the massive spinor brackets,
\begin{equation}
\begin{aligned}
\bs{p}_{\alpha \dot\alpha}& =\epsilon_{I J}  \lambda^I_{\alpha} \tilde\lambda^{J}_{\dot\alpha} =  \lambda^I_{\alpha} \tilde\lambda_{I \dot\alpha}= \ketBra{p^I}{p_I}\,,\\
\bar{\bs{p}}^{\dot\alpha \alpha} & =  \epsilon^{IJ} 
\tilde\lambda_I ^{ \dot\alpha}  \lambda_J^{ \alpha} = - \tilde\lambda^{I \dot\alpha}  \lambda_I^{\alpha} = - \Ketbra{p^I}{p_I} \,,
\end{aligned}
\label{eqn:spinor_momenta}
\end{equation}
which satisfy 
\beq
\det \bs{p}_{\alpha \dot\alpha} =\det \bar{\bs{p}}^{ \dot \alpha \alpha} = \frac{1}{2}  \bs{p}_{\alpha \dot\alpha}  \bar{\bs{p}}^{ \dot \alpha \alpha} = m^2 \,.
\eeq
In~\autoref{eqn:spinor_momenta} we used boldface for the  massive momentum.
At this point, this is just a matter of convenience, and we will use this notation throughout to help distinguish massive from massless momenta.
The significance of the bold spinor notation for treating massive particles of nonzero spin will be explained below.

The equations of motion are,
\beq
\bs{p} \Ket{\tilde{\lambda}^{I } }&=   m \ket{ \lambda^I}\,,  ~~~~ ~~
\bar{\bs{p}} \ket{ \lambda^I }=  m \Ket{ \tilde{\lambda}^{I }}\,,\\
\Bra{ \tilde{\lambda}^{I}} \bar{\bs{p}}&=  -  m \bra{\lambda^{I}}\,,  ~~~~
 \bra{{\lambda}^{I }} \bs{ p} =    -  m \Bra{\tilde{\lambda}^{I}} \,.
\eeq
In addition, the massive spinor bilinears satisfy,
\beq
\anb{\lambda^I \lambda_J} &=  m \delta^I_J \,,
~~~~\anb{\lambda^I \lambda^J} =  - m \epsilon^{IJ}  \,,
~~~~\anb{\lambda_I \lambda_J} =  m \epsilon_{IJ} \,,\\
\sqb{\tilde\lambda^I \tilde\lambda_J} &=   - m \delta^I_J \,,
~~~~\sqb{\tilde\lambda^I \tilde\lambda^J} = m {\epsilon}^{IJ} \,,
~~~~\sqb{\tilde\lambda_I \tilde\lambda_J} =- m \epsilon_{IJ}  \,,
\eeq
and
\beq
 \lambda^{I}_{ \alpha} \lambda^{\beta}_{I} = \ketbra{\lambda^{I}}{\lambda_{I}} & = -m \delta_{\alpha}^{\ \beta}\,,\qquad
 \tilde\lambda^{I \dot\alpha} \tilde\lambda_{I \dot\beta} = \KetBra{\tilde{\lambda}^I}{\tilde{\lambda}_I}=
m \delta^{\dot\alpha}_{\ \dot\beta}\,.
\label{eqn:mass_sumrule}
\eeq
For a real momentum, the dotted and undotted spinors are related by complex conjugation,
\beq
\left(\lambda^{I}_{\alpha}\right)^{\dag}&=  \tilde\lambda_{I \dot\alpha} \,,
\qquad
\left(\tilde\lambda^{I \dot\alpha}\right)^{\dag}= - \lambda_{I}^{\alpha} \,.
\label{eqn:dagger}
\eeq
For negative-energy spinors, the complex conjugation introduces an extra minus sign.

Dirac spinors are then given by,
\beq
u^I (p)& = 
\begin{pmatrix}
\lambda^I_{\alpha}\\
\tilde\lambda^{I \dot\alpha}
\end{pmatrix}\,, ~~~~
v^I (p) = 
\begin{pmatrix}
\lambda^I_{\alpha}\\
- \tilde\lambda^{I \dot\alpha}
\end{pmatrix}\,,\\
\overline{u}_I (p) &= 
\begin{pmatrix}
 - \lambda_I^{\alpha} & 
\tilde\lambda_{I \dot\alpha}
\end{pmatrix}\,, ~~~~ 
\overline{v}_I (p) = 
\begin{pmatrix}
\lambda_I^{\alpha} & 
\tilde\lambda_{I \dot\alpha}
\end{pmatrix}\,,
\eeq
and satisfy the equations of motion, 
\beq
\left(\Slash{p} - m \right) u^{I}( p) &
= 0\,,~~~~
\left(\Slash{p} + m \right) v^I (p) 
= 0\,,\\
\overline{u}_I (p) \left(\Slash{p} - m \right) & 
= 0\,,~~~~
\overline{v}_I (p)  \left(\Slash{p} + m \right)   =0\,,
\eeq
orthogonality conditions, 
\beq
\overline{u}_I (p) u^J(p) &= 2m \delta_{I}^{J}\,,\\
\overline{v}_I (p) v^J(p) &= - 2m \delta_{I}^{J}\,,\\
\overline{u}_I (p) v^J(p)  &= \overline{v}_I (p)u^J(p) = 0 \,,
\eeq
and spin sums,
\beq
\sum_{I} u^I (p) \overline{u}_I(p) = \Slash{p} + m\,,\\
\sum_{I} v^I (p) \overline{v}_I(p) =\Slash{p} - m \,.
\eeq

%%%%%%%%%%%%%%%%%%%%%%%%%%%%%%%%%%
\subsection{The high-energy limit} 
\label{sec:kandq}
%%%%%%%%%%%%%%%%%%%%%%%%%%%%%%%%%%

For a given massive momentum $\bs{p}$, different choices of $p^{I=1}$ are possible (with $p^{I=2}=\bs{p}-p^{I=1}$).
In order to make contact with helicity amplitudes, the spatial direction of the $p^I$'s can be chosen along the direction of motion of the particle.
It will be convenient to denote~\cite{Shadmi:2018xan},
\beq
 p_{\mu}= k_{\mu} + q_{\mu}\,,
 \label{eqn:kandq1}
\eeq
with
\beq
k_{\mu} = \frac{E+p}{2}(1,0,0,1)\,,~~~~q_{\mu} = \frac{E - p}{2}(1,0,0,-1)\,.
\label{eqn:momentum_decomposition}
\eeq
$k$  then scales with $E$ in the high-energy limit, while $q$ scales with $m^2/E$.
We  have,
\beq
\bs{p}_{\alpha \dot\alpha} &= \ket{k}\Bra{k} +  \ket{q}\Bra{q}\,,\\
\bar{\bs{p}}^{\dot\alpha \alpha} &= \Ket{k}\bra{k} +  \Ket{q}\bra{q}\,,
\eeq
with  $\anb{kk} = \anb{qq} = \sqb{kk} = \sqb{qq} = 0$, and
\beq
p^2 = 2 (k\cdot q)  = \anb{ k q} \sqb{ q k}= m^2\,.
\label{eqn:kandqlast}
\eeq
The spinors $\ket{k}$ and $\ket{q}$ are given by,
\beq
\ket{k} &=e^{i \phi_k}  \sqrt{E+p}  \begin{pmatrix} 1\\0 \end{pmatrix}\,,
~~~~ \Bra{k} =e^{- i \phi_k}  \sqrt{E+p} \begin{pmatrix}1 &0 \end{pmatrix}\,,\\
\ket{q} &= e^{i \phi_q} \sqrt{E-p} \begin{pmatrix} 0\\1 \end{pmatrix}\,,
~~~~ \Bra{q} = e^{- i \phi_q} \sqrt{E-p} \begin{pmatrix}0 &1\end{pmatrix}\,,
\eeq
where $\phi_k$ and $\phi_q$ are arbitrary phases.

Comparing
\eqs{eqn:kandq1}{eqn:kandqlast} and \autoref{eqn:massive_momenta_spinors}
we identify,
\begin{equation}\renewcommand{\arraystretch}{1.2}
\begin{array}{r@{\:}l@{\quad}r@{\:}l}
\tilde{\lambda}^{1 \dot\alpha}& =\Ket{\tilde{\lambda}^{1}}=  \Ket{k}\,,
&
\lambda_{1 \alpha} &= \ket{\lambda_{1}} = - \ket{k}\,, \\
\lambda^{ 2}_{\alpha}& =\ket{\lambda^{ 2 }} = \ket{k}\,,
&
\tilde{\lambda}_{2}^{ \dot\alpha} &=   \Ket{\tilde{\lambda}_{2}} = \Ket{k} \,,\\
\lambda^{1}_{\alpha} &=   \ket{ \lambda^{1}} =\ket{q}\,,
&
\tilde{\lambda}_{1}^{\dot\alpha} &= \Ket{\tilde{\lambda}_{1}}= \Ket{q}\,,\\
\tilde{\lambda}^{2 \dot\alpha} &= \Ket{\tilde{\lambda}^{2 }} = - \Ket{q}\,,
&
\lambda_{2\alpha} &= \ket{ \lambda_2} = \ket{q}\,,
\end{array}
\end{equation}
and
\begin{equation}\renewcommand{\arraystretch}{1.2}
\begin{array}{r@{\:}l@{\quad}r@{\:}l}
 \anb{k q } =&  m \,,&
 \sqb{q k} &=  m\,,\\
 p \Ket{k} =&  m \ket{q}\,,&
p \Ket{q}  &=   - m \ket{k}\,,\\
\bar{p} \ket{k} =& -m \Ket{q}\,,&
\bar{p} \ket{q} &= m \Ket{k}\,.
\end{array}
\end{equation}

Note that, choosing $\vec{k}$  parallel to $\vec{\bs{p}}$, the $| \bs p^I\rangle$ and $| \bs p^I]$ spinors reduce to massless spinors of definite helicity, with $I=1$ ($I=2$) corresponding to positive (negative) helicities.
Other choices are naturally possible, and give different spin quantization axes.%
\footnote{For general expressions for the $k$, $q$ spinors see ref.~\cite{Shadmi:2018xan}.}

\subsection{Massive particles of non-zero spin: bold notation} 
Obviously, the doublet of massless spinors can be used to span the polarizations of massive fermions.
Its utility is more general however, since any higher spin can be obtained from symmetric combinations of spin-$1/2$ representations.

Reference~\cite{Arkani-Hamed:2017jhn} introduced bold notation to implicitly indicate  symmetric combinations of the $SU(2)$ little-group indices of massive spinors.
To keep track of massive momenta, we will also bold massive fermion (or scalar) spinors and momenta, although a symmetrization is only relevant for particles of spin one or higher.
We find it convenient to use the standard Clebsch-Gordan $1/\sqrt2$ to define the symmetric combination of the two doublets,
\beq
\anb{\bs{3}1}\anb{\bs{3}2} \equiv
 \begin{cases}
    \anb{3^I1}\anb{3^I2}  & (I=J)\,, \\
\dfrac{1}{\sqrt{2}}\left( \anb{3^I1}\anb{3^J2} + \anb{3^J1}\anb{3^I2} \right) & (I\neq J)\,.
  \end{cases}
\label{eqn:bold_clebsch}
\eeq
With this definition, simple prescriptions are obtained for amplitudes $\mathcal{M}^{IJ}$ and polarization vectors $\varepsilon^{IJ}$ of definite helicities: $X^+ = X^{11}$, $X^-=X^{22}$ and $X^0=X^{12}=X^{21}$.\footnote{Instead, with \eg\ $\anb{\bs31} \anb{\bs32} \equiv (\anb{3^I1}\anb{3^J2} + \anb{3^J1}\anb{3^I2})/2$, one would have to define longitudinal amplitudes and polarizations vectors as $X^0=\sqrt{2}X^{12}=\sqrt{2}X^{21}$ to satisfy notably \autoref{eqn:pol_sum_mu} and \eqref{eqn:pol_sum_i}.}

%%%%%%%%%%%%%%%%%%%%%%%%%%%%%%%%%%
\subsection{Massive polarization vectors}
%%%%%%%%%%%%%%%%%%%%%%%%%%%%%%%%%%

The polarization vector of a massive vector boson of momentum $p$ and mass $m$ is~\cite{Chung:2018kqs}:
\beq
\varepsilon^{IJ}_{\mu} =  \frac{\bra{\bs{p}} \sigma_{\mu} \Ket{\bs{p}} }{\sqrt{2} m}\,,
\quad\text{or equivalently,}\quad
\varepsilon^{IJ}_{\alpha \dot\alpha} = \sqrt{2} \frac{\ket{\bs{p}} \Bra{\bs{p}}}{m}\,,
\label{eqn:massiveP}
\eeq
which corresponds to two transverse ($+,\,-$) and one longitudinal ($0$) modes:
\beq
\begin{gathered}
\bs\varepsilon^{+}_{\mu} 
	\equiv \varepsilon^{11}_{\mu}
	=  \frac{\bra{p^1} \sigma_{\mu} \Ket{p^1} }{\sqrt{2} m} \,,
\\
\bs\varepsilon^{-}_{\mu} 
	\equiv \varepsilon^{22}_{\mu}
	= \frac{\bra{p^2} \sigma_{\mu} \Ket{p^2} }{\sqrt{2} m} \,,
\\
\bs\varepsilon^{0}_{\mu} 
	\equiv \varepsilon^{12}_{\mu}
	=      \varepsilon^{21}_{\mu}
	=  \frac{	
		   \bra{p^1} \sigma_{\mu} \Ket{p^2}
		 + \bra{p^2} \sigma_{\mu} \Ket{p^1}
		}{2 m}\,\,.
\end{gathered}
\eeq
These massive polarization vectors satisfy the ordinary normalization for the vector boson,
\beq
\bs{\varepsilon}^{i}_{\mu} (\bs{\varepsilon}_j^{ \mu})^{\ast} &= - \delta^i_j\,,~~~\textrm{for}~i,j=+,-,0\,
\label{eqn:pol_sum_mu}
\eeq
The polarization sum gives 
\beq
 \sum_{i = +,-,0} \bs\varepsilon_{\mu}^{i} (\bs\varepsilon_{\nu}^{i})^{\ast} &= - \left(g_{\mu \nu} - \frac{p_{\mu} p_{\nu}}{m^2}\right)\,,
\label{eqn:pol_sum_i}
\eeq
which also corresponds to the numerator of  the massive vector propagator in the unitary gauge.

In the massless limit, the massive polarization vector matches the massless polarization vectors;
for $I=J=1$, 
\beq
\varepsilon^{11}_{\alpha \dot\alpha}
	= \sqrt{2} \frac{\ket{p^1} \Bra{p^1}}{m_3}
	= \sqrt{2} \frac{\ket{q} \Bra{k}}{\anb{k q}}
	\xrightarrow{m \to 0}  \sqrt{2} \frac{\ket{\zeta} \Bra{p}}{\anb{p \zeta}}
	\equiv \varepsilon^+_{\alpha \dot\alpha}
\,.
\label{eqn:polarimassless1}
\eeq
where the $k$ and $q$ spinor notation is given in \autoref{sec:kandq}.
For $I=J=2$, 
\beq
\varepsilon^{22}_{\alpha \dot\alpha}
	= \sqrt{2} \frac{\ket{p^2} \Bra{p^2}}{m}
	=-\sqrt{2} \frac{\ket{k} \Bra{q}}{\sqb{q k}}
	\xrightarrow{m \to 0} \sqrt{2} \frac{\ket{p} \Bra{\zeta}}{\sqb{p \zeta}}
	\equiv \varepsilon^-_{\alpha \dot\alpha}\,,
\label{eqn:polarimassless2}
\eeq
where $|\zeta]$, $|\zeta\rangle$, are reference spinors.
Note that the definition of $\varepsilon^+_{\alpha \dot\alpha}$ has a relative sign difference with respect to that of ref.~\cite{Shadmi:2018xan}.

%%%%%%%%%%%%%%%%%%%%%%%%%%%%%%%%%%%%%
\subsection{Useful identities}
\label{sec:identities}
%%%%%%%%%%%%%%%%%%%%%%%%%%%%%%%%%%%%%

If $\ket{i}$, $\ket{j}$ are independent spinors, any third spinor $\ket{k}$ can be written as,
\beq
\ket{k} = a \ket{i} + b \ket{j},
\eeq
where $a$, $b$ are complex coefficients. 
The coefficients $a$ and $b$ are determined by multiplying, for instance by $\bra{i}$. Then,
\beq
\label{eqn:general_Schouten}
\ket{i} \anb{jk} + \ket{j} \anb{ki} + \ket{k} \anb{ij} &= 0\,, \\
\Ket{i} \sqb{jk} + \Ket{j} \sqb{ki} + \Ket{k} \sqb{ij}& = 0\,.
\eeq

These relations are fundamentally equivalent to 
\beq
\varepsilon^{\alpha \beta}\varepsilon^{\gamma \delta}-\varepsilon^{\alpha \gamma }\varepsilon^{\beta \delta}+\varepsilon^{\alpha \delta}\varepsilon^{\beta \gamma} = 0\,, 
\label{eqn:Schouten1}\\
\varepsilon_{\dot\alpha \dot\beta}\varepsilon_{\dot\gamma \dot\delta}-\varepsilon_{\dot\alpha \dot\gamma }\varepsilon_{\dot\beta \dot\delta}+\varepsilon_{\dot\alpha \dot\delta}\varepsilon_{\dot\beta \dot\gamma} = 0\,,
\label{eqn:Schouten2}
\eeq
which are satisfied since a fully antisymmetric tensor with four indices taking only two different values must vanish.
These different identities are known as the Schouten identities.%
\footnote{
We also have 
$
\varepsilon^{I J}\varepsilon^{KL}-\varepsilon^{I K }\varepsilon^{JL}+\varepsilon^{I L}\varepsilon^{J K} = 0\,.
$
Note that $\epsilon^{IJ}  = \epsilon^{\alpha \beta}$ in our notation.}

%%%%%%%%%%%%%%%%%%%%%%%%%%%%%%%%%%%%%
\paragraph{Two spinor relations.}
%%%%%%%%%%%%%%%%%%%%%%%%%%%%%%%%%%%%%

The Schouten identity in \autoref{eqn:Schouten1} gives the following relation between two massive spinors of any spin,
\beq
0 
&= \left(  \varepsilon^{\alpha \beta}\varepsilon^{\gamma \delta}-\varepsilon^{\alpha \gamma }\varepsilon^{\beta \delta}+\varepsilon^{\alpha \delta}\varepsilon^{\beta \gamma} \right)
\lambda^{I}_{1, \gamma} \lambda^{J}_{2, \delta}\nn\\
& = - \varepsilon^{\alpha \beta} (\lambda_1^I \lambda_2^J) - \lambda_1^{I, \alpha} \lambda_2^{J, \beta}+ \lambda_1^{I, \beta} \lambda_2^{J, \alpha}\,,
\eeq
which corresponds to
\beq
\anb{\bs{1}^I \bs{2}^J} \delta_{\alpha}^{\beta} = \ket{\bs{2}^J}_{\alpha} \bra{\bs{1}^I}^{\beta} -  \ket{\bs{1}^I}_{\alpha} \bra{\bs{2}^J}^{\beta}\,.
\eeq
When $\ket{{1}}$ and $\ket{{2}}$ are massless non-collinear spinors, we obtain
\beq
\delta_{\alpha}^{\beta} = \frac{  \ket{{2}}_{\alpha} \bra{{1}}^{\beta} -  \ket{{1}}_{\alpha} \bra{{2}}^{\beta}}{\anb{{1} {2}} }\,.
\eeq

%%%%%%%%%%%%%%%%%%%%%%%%%%%%%%%%%%%%%
\paragraph{Application for two-massive one-massless amplitudes.}
%%%%%%%%%%%%%%%%%%%%%%%%%%%%%%%%%%%%%

Let us consider the special case of an amplitude involving three particles of arbitrary spins: two massive ($m_1, m_2\ne0$) and one massless ($m_3=0$).
Substituting $\ket{i} = \ket{\bs{1}^I}$, $\ket{j} =\ket{\bs{2}^J}$, and $\ket{k}=\ket{3}$ into \autoref{eqn:general_Schouten}, one obtains
\beq
\ket{3} \anb{\bs{1}^I \bs{2}^J} = - \ket{\bs{1}^I} \anb{\bs{2}^J 3} - \ket{\bs{2}^J} \anb{3 \bs{1}^I}\,.
\eeq
Multiplying by $\Bra{\zeta} \bar{p}_{\bs{1}}$
for some arbitrary spinor $\Bra{\zeta}$, one obtains 
\beq
[\zeta \bs{1}3 \rangle \anb{\bs{1}^I \bs{2}^J} = m_1 \sqb{\zeta \bs{1}^I} \anb{3 \bs{2}^J} + m_2 \sqb{\zeta \bs{2}^J} \anb{3 \bs{1}^I} + \sqb{ \zeta 3} \anb{3 \bs{1}^I} \anb{3 \bs{2}^J}\,,
\label{eqn:identity1}
\eeq
similarly,
\beq
\langle \zeta \bs{1}3] \sqb{\bs{1}^I \bs{2}^J} = m_1 \anb{\zeta \bs{1}^I} \sqb{3 \bs{2}^J} + m_2 \anb{\zeta \bs{2}^J} \sqb{3 \bs{1}^I} + \anb{\zeta  3} \sqb{3 \bs{1}^I} \sqb{3 \bs{2}^J}\,.
\eeq
The Schouten identities also lead to
\beq
\langle \zeta \bs{1} 3] \anb{\bs{1}^I \bs{2}^J}
&= m_1 \anb{\zeta \bs{2}^J} \sqb{3 \bs{1}^I} 
+ m_2 \anb{ \zeta \bs{1}^I} \sqb{3 \bs{2}^J}\,,\\
%%%
%%%
[\zeta \bs{1} 3\rangle \sqb{\bs{1}^I \bs{2}^J}
&= m_1 \sqb{\zeta \bs{2}^J} \anb{3 \bs{1}^I} 
+ m_2 \sqb{ \zeta \bs{1}^I} \anb{3 \bs{2}^J}\,.
\label{eqn:identity4}
\eeq
In the $m_1 = m_2$ limit, the above four equations imply:
\beq
\frac{\langle \zeta \bs{1} 3]}{ \anb{\zeta 3} } \left( \sqb{\bs{1}^I \bs{2}^J} -\anb{\bs{1}^I \bs{2}^J}\right)
&= \sqb{3 \bs{1}^I} \sqb{3 \bs{2}^J}\,,
\label{eqn:identity3132}\\
%%%
\frac{[\zeta  \bs{1} 3\rangle }{ \sqb{\zeta 3} } \left( \sqb{\bs{1}^I \bs{2}^J} - \anb{\bs{1}^I \bs{2}^J}  \right)
&= - \anb{3 \bs{1}^I} \anb{3 \bs{2}^J}\,.
\label{eqn:identityM3132}
\eeq

We can obtain similar identities starting from, 
\beq
m_1 \anb{\bs{1}^I 3} \Bra{3}& = - \Bra{\bs{1}^I} \bs{1} 3  = 
\Bra{\bs{1}^I}  3\bs{1} 
- 2 \left( p_{\bs{1}} \cdot p_3 \right)\Bra{\bs{1}^I} \nonumber \\
  & 
  = \sqb{\bs{1}^I 3} \bra{3}\bs{1} +(m_1^2 -m_2^2 )\Bra{\bs{1}^I} 
  \,.
\eeq
This gives,
\beq
\frac{[ \zeta  \bs{1} 3\rangle}
{\sqb{ \zeta 3}}
 \sqb{\bs{1}^I 3} = - m_1 \anb{\bs{1}^I 3}  + (m_1^2 -m_2^2 )\frac{\sqb{\zeta \bs{1}^I }}{\sqb{ \zeta 3}}\,,
\eeq
and, similarly,
\beq
%%%
\frac{[\zeta  \bs{1} 3\rangle}
{\sqb{ \zeta 3}}
 \sqb{\bs{2}^J 3} = m_2 \anb{\bs{2}^J 3}  + (m_1^2 -m_2^2 )\frac{\sqb{\zeta \bs{2}^J }}{\sqb{ \zeta 3} }\,.
\eeq
These two identities lead to
\beq
\frac{[\zeta  \bs{1} 3\rangle}
{\sqb{ \zeta 3}}
 \sqb{\bs{1}^I 3} 
  \sqb{\bs{2}^J 3}  &= \frac{1}{2} 
\biggl\{
 -  m_1 \anb{\bs{1}^I 3}  \sqb{\bs{2}^J 3} + m_2 \anb{\bs{2}^J 3} \sqb{\bs{1}^I 3}\nn\\
 & ~~~~~~~~   + (m_1^2 -m_2^2 )   \left[  \frac{\sqb{\zeta \bs{1}^I }\sqb{\bs{2}^J 3} }{\sqb{ \zeta 3} } +\frac{\sqb{\zeta \bs{2}^J }\sqb{\bs{1}^I 3}}{\sqb{ \zeta 3}}\right]
\biggr\} \nn\\
  & = 
 \frac{ m_1^2 + m_2^2 }{2}\sqb{\bs{1}^I \bs{2}^J} - m_1 m_2 \anb{\bs{1}^I \bs{2}^J} \nn\\
 &   ~~~~~~~~ 
 +\frac{m_1^2 -m_2^2 }{2}  \left[  \frac{\sqb{\zeta \bs{1}^I }\sqb{\bs{2}^J 3} }{\sqb{ \zeta 3} } +\frac{\sqb{\zeta \bs{2}^J }\sqb{\bs{1}^I 3}}{\sqb{ \zeta 3}}\right]\,.
\eeq
In the $m_1 = m_2$ limit, we find 
\beq
\frac{[\zeta \bs{1} 3\rangle}{ \sqb{\zeta 3}}\sqb{\bs{1}^I 3} \sqb{\bs{2}^J 3} &= 
m^2 \left( \sqb{\bs{1}^I \bs{2}^J}- \anb{\bs{1}^I \bs{2}^J}\right) \,,
\eeq
and similarly,
\beq
\frac{\langle \zeta\bs{1} 3]}{ \anb{\zeta 3}} \anb{\bs{1}^I 3} \anb{\bs{2}^J 3}&= 
- m^2 \left( \sqb{\bs{1}^I \bs{2}^J}  - \anb{\bs{1}^I \bs{2}^J} \right)\,.
\eeq

%%%%%%%%%%%%%%%%%%%%%%%%%%%%%%%%%%
\subsection{Massive gluing with non-zero spin}
\label{sec:massive_gluing}
%%%%%%%%%%%%%%%%%%%%%%%%%%%%%%%%%%

When gluing two amplitudes to obtain higher-point amplitudes, one (or more) external leg has to be treated as outgoing.
The spinors associated with this outgoing leg must be contracted with the spinors associated with the incoming leg on the ``other side''.
Here we clarify our conventions for the polarization associated with the external outgoing leg and the resulting contractions.
To derive these, we consider the product of two amplitudes with a vector propagator, and require that the correct propagator is reproduced.
For a propagating vector of mass $m$, we have:
\begin{equation}
-
\frac{1}{p^2-m^2}
\left[
\sum_{I,J,\tilde{I},\tilde{J}=1,2}
	\frac{\mathcal{M}^{IJ}_1(p) + \mathcal{M}^{JI}_1(p)}{2}\;
	\epsilon_{I\tilde I}\;\epsilon_{J\tilde J}
	\frac{\mathcal{M}^{\tilde I\tilde J}_2(\tilde p) + \mathcal{M}^{\tilde J\tilde I}_2(\tilde p)}{2}
\right]_{p^2=m^2}\,.
\label{eqn:gluing_prescription_app}
\end{equation}
There are no $J,\tilde{J}$ indices for a propagating massive fermion, and no indices at all for a propagating scalar.
The overall minus sign corresponds to the two factors of $i$ that appear in each of the vertices in a Feynman-rule computation.
In the second amplitude $\mathcal{M}_2$, the propagating momentum $p$ is flipped from \emph{incoming} to \emph{outgoing}.
(Our convention is that all momenta are otherwise incoming.)
We denote this outgoing momentum as $\tilde{p}$ and the corresponding spinors as $\bra{\tilde{p}}$ and $\Bra{\tilde{p}}$.
Special contraction rules apply for such flipped-momenta spinors:
\begin{equation}
\begin{aligned}
| p^I]\langle \tilde p_I |	&= + \bar{\bs{p}}\,,\\
|p^I\rangle [\tilde p_I	| &= + \bs p\,,
\end{aligned}
\hspace{2cm}
\begin{aligned}
| p^I\rangle\langle \tilde p_I|	&=+m\,,\\
| p^I][\tilde p_I |		&=+m\,,
\end{aligned}
\end{equation}
where the summation over $I$ is implicit.
A plus sign appears in the right-hand side of all these sum rules, in contrast the usual (see \autoref{eqn:spinor_momenta} and \eqref{eqn:mass_sumrule}):
\begin{equation}
\begin{aligned}
| p^I]\langle  p_I |	&= - \bar{\bs{p}}\,,\\
|p^I\rangle [ p_I	| &= + \bs p\,,
\end{aligned}
\hspace{2cm}
\begin{aligned}
| p^I\rangle\langle  p_I|	&=-m\,,\\
| p^I][ p_I |		&=+m\,.
\end{aligned}
\end{equation}
For a propagating vector, flipping the momentum in one amplitude or the other (exchanging ${}_1$ and ${}_2$ labels above) yields the same final result.
For fermion propagators (in fermion-number conserving interactions), the fermion flow has to be followed.
The flipped momentum should be taken at the vertex with outgoing propagating fermion flow.

If all $IJ$ indices are carried by polarization vectors, the prescription above consistently reproduces the unitary gauge propagator:
\begin{equation}
\frac{1}{p^2-m^2}
\left[
\sum_{I,J,\tilde{I},\tilde{J}=1,2}
	\frac{\varepsilon^{IJ}_\mu(p) + \varepsilon^{JI}_\mu(p)}{2}\;
	\epsilon_{I\tilde I}\; \epsilon_{J\tilde J}\;
	\frac{\varepsilon^{\tilde I\tilde J}_\nu(\tilde p) + \varepsilon^{\tilde J\tilde I}_\nu(\tilde p)}{2}
\right]_{p^2=m^2}
= -\frac{g_{\mu\nu} - \dfrac{p_\mu p_\nu}{m^2}}{p^2-m^2}
\,.
\end{equation}
For a propagating fermion, one also obtains the desired expression:
\begin{equation}
\frac{1}{p^2-m^2}
\left[
\sum_{I,\tilde{I}=1,2}
	\begin{pmatrix}| p^I\rangle \\ |p^I]\end{pmatrix}\;
	\epsilon_{I\tilde I}\;
	\begin{pmatrix} \langle \tilde p^{\tilde{I}}| & [ \tilde p^{\tilde{I}}| \end{pmatrix}\;
\right]_{p^2=m^2}
= \frac{1}{p^2-m^2}\begin{pmatrix} m & \bs p \\ \bar{\bs{p}} & m\end{pmatrix}
= \frac{1}{\slashed{p}-m}
\,.
\end{equation}

%%%%%%%%%%%%%%%%%%%%%%%%%%%%%%%%%%%%%%%%%%%%%%%%%%%%%%%%%%%%%%%%%%%%%%%%%%%%%%%%%%%%%
\subsection{Spin magnetic dipole moment (\texorpdfstring{$g$}{g}-factor)}
\label{sec:g2}
%%%%%%%%%%%%%%%%%%%%%%%%%%%%%%%%%%%%%%%%%%%%%%%%%%%%%%%%%
%
To demonstrate the application of the massive spinor techniques, we give a detailed derivation of the fermion $g$-factor.
Our derivation closely follows ref.~\cite{Chung:2018kqs}, although we differ by factors of two along the
way.
In the process, we also derive the Gordon identity using the massive spinor formalism.
We start from the amplitudes of \autoref{eqn:mffa}, whose terms are separately well-behaved in the high-energy limit.
The first and third (and also the second and fourth) identities  in \autoref{eqn:x_factor_equalities} give the following relations:
\begin{equation}
\begin{aligned}
(\sqb{\bs1\bs2} + \anb{\bs1\bs2})\langle\zeta\bs13]/\anb{\zeta3}
&=2 m
(\anb{\zeta\bs1}\sqb{3\bs2}
+\anb{\zeta\bs2}\sqb{3\bs1})/\anb{\zeta3}
+\sqb{\bs13}\sqb{\bs23}\,,\\
(\sqb{\bs1\bs2} + \anb{\bs1\bs2})[\zeta\bs13\rangle/\sqb{\zeta3}
&=2 m
(\sqb{\zeta\bs1}\anb{3\bs2}
+\sqb{\zeta\bs2}\anb{3\bs1})/\sqb{\zeta3}
+\anb{\bs13}\anb{\bs23}\,,
\end{aligned}
\end{equation}
or equivalently,
\begin{equation}
\begin{aligned}
\langle \bs1 \varepsilon_3^{+} \bs2 ] + \langle \bs2 \varepsilon_3^{+} \bs1 ]
&=
\frac{1}{\sqrt{2} m}\frac{\langle\zeta\bs13]}{\anb{\zeta3}} \left(\sqb{\bs1\bs2} + \anb{\bs1\bs2}\right) - \frac{1}{\sqrt{2} m} \sqb{\bs13}\sqb{\bs23}
\,,\\
\langle \bs1 \varepsilon_3^{-} \bs2 ] + \langle \bs2 \varepsilon_3^{-} \bs1 ]
& = 
\frac{1}{\sqrt{2} m}
\frac{[\zeta\bs1 3\rangle}{\sqb{\zeta 3}}
\left(\sqb{\bs1\bs2} + \anb{\bs1\bs2}\right)  - \frac{1}{\sqrt{2} m} \anb{\bs13}\anb{\bs23}
\,.
\end{aligned}
\label{eqn:identity_for_g}
\end{equation}
We can therefore also re-express \autoref{eqn:mffa} as:
\begin{equation}
\begin{aligned}
\mathcal{M}(\bs1_\psic,\bs2_\psi,3_\gamma^+)
	&=
	 \frac{c_{\psic\psi\gamma}}{\sqrt{2} m_\psi} \frac{\langle \zeta \bs{1} 3 ]}{\anb{\zeta 3}}\left(\sqb{\bs1\bs2} + \anb{\bs1\bs2}\right)
	-   \frac{c_{\psic\psi\gamma}}{\sqrt{2} m_\psi}\sqb{\bs13}\sqb{\bs23}
	+ \frac{c^{RRR}_{\psic\psi\gamma}}{\Lam}
	\sqb{\bs13}\sqb{\bs23}\,,
\\
\mathcal{M}(\bs1_\psic,\bs2_\psi,3_\gamma^-)
	&= 
	\frac{c_{\psic\psi\gamma}}{ \sqrt{2} m_\psi} \frac{[\zeta \bs{1} 3 \rangle}{\sqb{\zeta 3}}\left(\sqb{\bs1\bs2} + \anb{\bs1\bs2}\right)
- 
	\frac{c_{\psic\psi\gamma}}{\sqrt{2} m_\psi} \anb{\bs13}\anb{\bs23}
	+ \frac{c^{LLL}_{\psic\psi\gamma}}{\Lam}
	\anb{\bs13}\anb{\bs23}\,.
\end{aligned}
	\label{eqn:ffA_dipole}
\end{equation}
Note that in the high-energy limit, two $1/m_\psi$ singularities are canceled between the first and second terms, so that one still obtains  good high-energy behavior.
The relations in \autoref{eqn:identity_for_g} actually correspond to the Gordon identity~\cite{Gordon1928}:%
\footnote{One can easily check the Gordon identity itself in the massive spinor formalism:
\beq
 \langle \bs1  \sigma^{\mu} \bs2 ]   +  \langle \bs2  \sigma^{\mu} \bs1 ]  =&
 \frac{1}{2m} (- p_{\bs1} + p_{\bs2})^{\mu} \left( \sqb{\bs{12}} + \anb{\bs{12}}\right) \nonumber \\
  &+ \frac{1}{4 m} \left(  \langle \bs1 \left\{\sigma^{\mu} ( \bar{\bs{1}} + \bar{\bs{2}})
  - ({\bs{1}} + {\bs{2}})\bar \sigma^{\mu} \right\}\bs2 \rangle
+  [ \bs1 \left\{\bar \sigma^{\mu} ( {\bs{1}} + {\bs{2}})
  - (\bar{\bs{1}} + \bar{\bs{2}}) \sigma^{\mu} \right\}\bs2 ]\right)\,.
\eeq
}
\begin{equation}
\bar{v}_{\bs{1}}  \gamma^{\mu} u_{\bs{2}}
= \frac{1}{2 m}\bar{v}_{\bs{1}}(-p_{\bs{1}} + p_{\bs{2}})^{\mu}u_{\bs{2}}
-\frac{i}{2 m}\bar{v}_{\bs{1}} \sigma^{\mu \nu}(p_{\bs{1}} + p_{\bs{2}})_{\nu} u_{\bs{2}}\,,
\label{eqn:gordon}
\end{equation}
where $p_{\bs{1}}$ and $p_{\bs{2}}$ are incoming momenta for $\bs1_\psic$ and $\bs2_\psi$, respectively, and $\sigma^{\mu \nu}\equiv i[\gamma^{\mu}, \gamma^{\nu}]/2$.
Multiplying \autoref{eqn:gordon} by $\varepsilon_3^+$ or $\varepsilon_3^-$ and using three-point kinematics $p_{\bs{1}} + p_{\bs{2}} = - p_3$,
one immediately finds that the first and second terms of the right-hand side coincide with that of \autoref{eqn:identity_for_g}. 
In the non-relativistic limit, the first term corresponds to an interaction independent of $\bs1,\bs2$ spins
because all the coefficients of $\sqb{\bs{1}^I \bs{2}^J}$ and $\anb{\bs{1}^I \bs{2}^J}$ with any $I,\,J$ are the same, and the coefficient $\langle \zeta \bs{1} 3 ] / \anb{ \zeta 3}$ (or $[ \zeta \bs{1} 3\rangle /\sqb{\zeta 3}$) corresponds to the spinor structure of a $\phi^+\phi^-\gamma$ amplitude  (see \autoref{sec:hhZhhA}), namely the scalar QED interaction.
On the other hand, the second term is a spin-dependent magnetic dipole moment~\cite{Schwartz:2013pla}.
This can be seen by noting that the second term of \autoref{eqn:identity_for_g} vanishes when the $\Bra{\bs{1}}$ and $\Ket{\bs{2}}$ spinors (or $\bra{\bs{1}}$ and $\ket{\bs{2}}$ ones) are factored out and the resulting tensor contracted with $\epsilon_{\dot\alpha\dot\beta}$ (or $\epsilon_{\alpha\beta}$) to obtain the amplitude for a scalar from that of a fermion.

Using the relations (see derivation in \autoref{eqn:dipole_identity}),
\beq
[\bs1 3][\bs2 3] = - \frac{i}{\sqrt{2}}
	\;\bar{v}_{\bs1} \sigma^{\mu \nu}  u_{\bs2}
	\;\varepsilon_{3, \mu}^+p_{3,\nu}\,,\qquad
\anb{\bs1 3}\anb{\bs2 3} = - \frac{i}{\sqrt{2}}
	\;\bar{v}_{\bs1}   \sigma^{\mu \nu}  u_{\bs2}
	\;\varepsilon_{3, \mu}^-p_{3,\nu} \,,
\eeq
the second terms of \autoref{eqn:ffA_dipole} become
\begin{equation}
\frac{c_{\psic\psi\gamma}}{2 m_\psi} i
	\;\bar{v}_{\bs1} \sigma^{\mu \nu}  u_{\bs2}
	\;\varepsilon_{3, \mu}^+ p_{3,\nu}\,,
\quad 
\frac{c_{\psic\psi\gamma}}{2 m_\psi} i
	\;\bar{v}_{\bs1} \sigma^{\mu \nu} u_{\bs2}
	\; \varepsilon_{3, \mu}^- p_{3,\nu}\,.
\end{equation}
Furthermore, matching onto the $U(1)_{\text{EM}}$ gauge theory, with the $D_{\mu} \psi = (\partial_{\mu} + i Q_{\psi} e {A}_{\mu})\psi $ sign convention for the covariant derivative,
provides $c_{\psic\psi\gamma} = - Q_\psi e$ (see \autoref{eqn:ffA_tree}).
Eventually, we obtain
\begin{equation}
-  \frac{Q_\psi e}{2 m_\psi} i
	\;\bar{v}_{\bs1} \sigma^{\mu \nu}  u_{\bs2}
	\; \varepsilon_{3, \mu}^{\pm}  p_{3,\nu}\,,
\end{equation}
which corresponds to the factor $g=2$.%
\footnote{
To see this, note that the magnetic field four-vector and spin operator are given by,
\beq
B_{\mu}^{h} &= - \frac{1}{2m} \epsilon_{\mu \nu \rho \sigma}p_{\bs2}^{\nu} F^{h, \rho \sigma}\,, \quad \textrm{with} \quad F^{h, \rho \sigma} = -  i (p_3^{\rho} \varepsilon_3^{h, \sigma }- p_3^{\sigma} \varepsilon_3^{h, \rho })\,,\\
S_{\mu} &= \frac{1}{2m} \epsilon_{\mu \nu \rho \sigma} p_{\bs2}^{\nu} J^{\rho \sigma}\,,
\quad \textrm{with} \quad J^{\rho\sigma}  = \frac{i}{4}[\gamma^{\rho},\gamma^{\sigma}]\,,
\eeq
where $h$ is the photon helicity (for $\epsilon_{0123}=+1 = - \epsilon^{0123}$).
One then for instance obtains: 
\begin{equation}
    S\cdot B^{+}
     = \frac{1}{2 \sqrt{2} m^2}
     \begin{pmatrix}
     \bs2 \Ket{3}\Bra{3} \bar{\bs2} & 0\\
     0 & (m^2) \Ket{3}\Bra{3}
     \end{pmatrix}\,,
\end{equation}
and 
\begin{equation}
   \bar{v}_{\bs1} S\cdot B^{+} u_{\bs2}= - \frac{1}{\sqrt{2}} [\bs1 3] [\bs2 3]\,,
   \quad   \bar{v}_{\bs1} S\cdot B^{-} u_{\bs2}= - \frac{1}{\sqrt{2}} \anb{\bs1 3} \anb{\bs2 3}\,.
\end{equation}
The second terms of \autoref{eqn:ffA_dipole} amplitudes are then,
\begin{equation}
+ c_{\psic\psi\gamma}\,  \bar{v}_{\bs1} S\cdot B^{h} u_{\bs2}/ m_\psi \,.
\end{equation}
In the rest frame of $\psi_{\bs2}$ where $p_{\bs2} = (m,0,0,0)$, $S\cdot B^h = - \vec{S} \cdot \vec{B}^h$ holds.
Using $c_{\psic\psi\gamma} = - Q_\psi e$, we obtain the Hamiltonian
\begin{equation}
    \mathcal{H}  = - Q_{\psi} e  \vec{S} \cdot \vec{B}^h  /m_{\psi}\,.
        \label{eqn:dipole_direct}
\end{equation}
On the other hand, the spin magnetic dipole moment (and $g$-factor) is defined by 
\begin{equation}
    \mathcal{H} = - \vec{\mu}_{\psi} \cdot \vec{B}\,,\quad \textrm{with} \quad \vec{\mu}_{\psi} = g \frac{ Q_\psi e}{2 m_{\psi}} \vec{S}\,,
\end{equation}
where the covariant derivative is 
$
D_{\mu} \psi = (\partial_{\mu} + i Q_{\psi} e {A}_{\mu})\psi 
$ \cite{Roberts:2010zz}.
Compared to \autoref{eqn:dipole_direct}, we obtain $g=2$.}

%%%%%%%%%%%%%%%%%%%%%%%%%%%%%%%%%%%%%%
\section{Tree-level matching to the  broken-phase SMEFT}
\label{sec:matching}
%%%%%%%%%%%%%%%%%%%%%%%%%%%%%%%%%%%%%%
In this section, we match the coefficients of the massive three-point amplitudes of \autoref{sec:3pt}, and the four-point amplitude $\psic \psi Z h$ of \autoref{sec:4-pt}, onto the SMEFT in the broken electroweak phase~\cite{Dedes:2017zog}.
The Lagrangian of ref.~\cite{Dedes:2017zog}, which includes operators of dimension $\leq 6$, is based on the Warsaw basis~\cite{Grzadkowski:2010es}.
All fields are canonically normalized (see also refs.~\cite{Helset:2018fgq,Misiak:2018gvl}).
The matching relations we provide extend the work of ref.~\cite{Aoude:2019tzn} to include fermionic three-point amplitudes, as well as the $\psic\psi Zh$ amplitude.
Since our toy model has a single generation, the CKM matrix is set to the unit matrix.

Working up to dimension-six, the SMEFT Lagrangian in the symmetric electroweak phase is given by
\beq
\mathcal{L}_{\rm SMEFT}
= \mathcal{L}_{\rm SM} 
+ \sum_i \frac{C_i}{\Lambda^2} \mathcal{O}_i\,.
\label{eqn:SMEFT}
\eeq
In the broken electroweak phase, the covariant derivative becomes 
\beq
D_{\mu} = \partial_{\mu} + i \bar{g}' \bar{B}_{\mu} Y + i \bar{g} \bar{W}_{\mu}^I T^I + i \bar{g}_s \bar{G}_{\mu}^{A} T^A\,,   
\eeq
where 
\beq
\bar{g}' = Z_{g'}^{-1} g'\,,~~~~ \bar{g} = Z_{g}^{-1} g\,,~~~~ \bar{g}_s = Z_{g_s}^{-1} g_s\,,
\eeq
and
\beq
Z_{g'} = 1 - \frac{v^2}{\Lambda^2} C_{\varphi B}\,,~~~~
Z_{g} = 1 - \frac{v^2}{\Lambda^2} C_{\varphi W}\,,~~~~
Z_{g_s} = 1 - \frac{v^2}{\Lambda^2} C_{\varphi G}\,,
 \eeq
with  $v \simeq 246 $ GeV.
$\bar{G}_{\mu}^{A}$ is the canonically normalized gluon field, and the canonically normalized $W^{\pm}_{\mu}$ bosons are
\beq
W_{\mu}^{\pm} = \frac{1}{\sqrt{2}} \left( \bar{W}^1_{\mu} \mp i \bar{W}_{\mu}^2 \right)\,,
\quad\text{with mass }\quad
m_W = \frac{1}{2} \bar{g} v\,.
\eeq
The canonically normalized $Z_{\mu}$ and $A_{\mu}$ are a linear combination of $\bar{B}_{\mu}$ and $\bar{W}^3_{\mu}$~\cite{Dedes:2017zog}.
The $Z$-boson mass is 
\beq
m_Z = \frac{1}{2} \sqrt{\bar{g}^2 + \bar{g}^{\prime 2}} v + \frac{1}{2} \frac{v^3}{\Lambda^2}
\frac{\bar{g}\bar{g}'}{ \sqrt{\bar{g}^2 + \bar{g}^{\prime 2}}} C_{\varphi W B} 
+ \frac{1}{8}  \frac{v^3}{\Lambda^2}\sqrt{\bar{g}^2 + \bar{g}^{\prime 2}} C_{\varphi D}\,.
\label{eqn:mZ}
\eeq
Finally, the canonically normalized Higgs field has mass:
\begin{equation}
m_h^2 = \lambda v^2-(3C^\varphi - 2\lambda C^{\varphi\square} + \lambda C^{\varphi D}/2)\frac{v^4}{\Lambda^2}.
\end{equation}
 
In order to fix $\bar{g},\,\bar{g}^{\prime},\,\lambda$, and $v$ from measurements, one has to choose an input scheme.
In an on-shell amplitude context, it would be appropriate to use pole masses as input parameters, necessarily complemented by an additional measurement like that of $G_F$ or $\alpha_\text{EM}$.

In order to treat fermion couplings generally, we define $I_\psi=\pm1/2$ for up-type quarks or neutrinos, and down-type quarks or charged leptons, respectively.
We also take $Y_{u,d,e,\nu}=1/6,1/6,-1/2,-1/2$.
$Q_\psi$ is their electric charge in units of $e$.

\subsection{\texorpdfstring{$\psic\psi Z$}{ffZ}}
\label{sec:ffZ}

The general form of the massive $\psic\psi Z$ amplitude is
\begin{equation}
\mathcal{M}(\bs{1}_\psic, \bs{2}_{\psi},\bs{3}_{Z})=
\frac{c^{LR0}_{\psic\psi Z}}{m_Z}	\anb{\bs{13}}\sqb{\bs{23}}
+ \frac{c^{RL0}_{\psic\psi Z}}{m_Z}	\sqb{\bs{13}}\anb{\bs{23}}
+   \frac{c_{\psic\psi Z}^{RRR}}{\Lam}	\sqb{\bs{13}}\sqb{\bs{23}}
+ \frac{c_{\psic\psi Z}^{LLL}}{\Lam}	\anb{\bs{13}}\anb{\bs{23}}\,.
\end{equation}
Matching to the broken-phase SMEFT~\cite{Dedes:2017zog}, we find 
\beq
c^{LR0}_{\psic\psi Z}
= & 
 -\sqrt{2} Q_{\psi} \frac{\bar{g}^{\prime 2}}{\sqrt{\bar{g}^2 + \bar{g}^{\prime 2}}}\nn\\
 & + \frac{v^2}{\Lambda^2}  \left[
 - \sqrt{2} \frac{\bar{g}^3 \bar{g}'}{ \left(\bar{g}^2 +  \bar{g}^{\prime 2}\right)^{3/2}} Q_{\psi}
 C_{\varphi W B} 
 - \frac{1}{\sqrt{2}}  \sqrt{\bar{g}^2 + \bar{g}^{\prime 2}} C_{\varphi \psi_R}
 \right]\,,\\
c^{RL0}_{\psic\psi Z}
 =& \sqrt{2}  I_{\psi} \frac{\bar{g}^2}{\sqrt{\bar{g}^2 + \bar{g}^{\prime 2}}}
- \sqrt{2} Y_{\psi} \frac{\bar{g}^{\prime 2}}{\sqrt{\bar{g}^2 + \bar{g}^{\prime 2}}}\nn\\
&+ \frac{v^2}{\Lambda^2}  \left[ \sqrt{2} \frac{\bar{g}\bar{g}'}{ \left(\bar{g}^2 +  \bar{g}^{\prime 2}\right)^{3/2}}
\left( - Y_{\psi} \bar{g}^2 + I_{\psi} \bar{g}^{\prime 2} \right)C_{\varphi WB}
- \frac{1}{\sqrt{2}}  \sqrt{\bar{g}^2 + \bar{g}^{\prime 2}}\left( C_{\varphi \psi_L}^1- 2 I_{\psi} C_{\varphi \psi_L}^3 \right)\right]\,,\\
\frac{c_{\psic\psi Z}^{RRR}}{\bar\Lambda}
= &
  \frac{v}{\Lambda^2}\left( - 4 I_{\psi}   \frac{\bar{g}}{ \sqrt{\bar{g}^2 +  \bar{g}^{\prime 2}}} C_{\psi W} 
  +  2 \frac{ \bar{g}'}{ \sqrt{\bar{g}^2 +  \bar{g}^{\prime 2}}} C_{\psi B}\right)\,,\\
  c_{\psic\psi Z}^{LLL}
  = & \left(c_{\psic\psi Z}^{RRR}\right)^{\ast}\,.
\eeq
Here, we used
\beq
\bar{\psi}_{\bs{1}} \gamma^{\mu} P_R \psi_{\bs{2}} \bs{\varepsilon}_{\bs{3}, \mu} &=
\bar{v}_{\bs{1}}\gamma^{\mu} P_R u_{\bs{2}} \bs{\varepsilon}_{\bs{3}, \mu} = \langle \bs{1} \bs{\varepsilon}_{\bs{3}} \bs{2}]
= - \frac{\sqrt{2}}{m_3} \anb{\bs{1} \bs{3}} \sqb{\bs{2} \bs{3}}\,,\\
\bar{\psi}_{\bs{1}} \gamma^{\mu} P_L \psi_{\bs{2}} \bs{\varepsilon}_{\bs{3}, \mu} &=
- \frac{\sqrt{2}}{m_3} \sqb{\bs{1} \bs{3}} \anb{\bs{2} \bs{3}}\,,
\eeq
and
\beq
\bar{\psi}_{\bs{1}} \sigma^{\mu \nu} P_R \psi_{\bs{2}} \bs{\varepsilon}_{\bs{3}, \mu} {p}_{\bs{3}, \nu}
&= \frac{i}{2} \bar{v}_{\bs{1}} [\gamma^{\mu}, \gamma^{\nu}] P_R u_{\bs{2}}  \bs{\varepsilon}_{\bs{3}, \mu} {p}_{\bs{3}, \nu}
= \frac{i}{2} [ \bs{1} \left( \bar\sigma^{\mu} \bs3 - \bar{\bs{3}} \sigma^{\mu}\right) \bs{2}] \bs{\varepsilon}_{\bs{3}, \mu}\nn \\
& =-  \frac{i}{2}   \left( [ \bs{1} \bar{\bs{3}} \bs{\varepsilon}_{\bs{3}} \bs{2}] + [ \bs{2}\bar{\bs{3}}\bs{\varepsilon}_{\bs{3}} \bs{1}]\right) 
= - \frac{i}{2}  \frac{\sqrt{2}}{m_3}  \left( [ \bs{1} \bar{\bs{3}} \bs{3} \rangle \sqb{\bs{3 2}} + [ \bs{2} \bar{\bs{3}} \bs{3} \rangle \sqb{\bs{3 1}} \right)\nn\\
&= \sqrt{2} i \sqb{\bs{1 3}} \sqb{\bs{2 3}}\,,\\
\bar{\psi}_{\bs{1}} \sigma^{\mu \nu} P_L \psi_{\bs{2}} \bs{\varepsilon}_{\bs{3}, \mu} {p}_{\bs{3}, \nu}
&
= \sqrt{2} i \anb{\bs{1 3}} \anb{\bs{2 3}}\,.
\eeq

\subsection{\texorpdfstring{$\psic\psi \gamma$}{ffA}}

General form:
\begin{align}
\mathcal{M}(\bs1_\psic,\bs2_\psi,3_\gamma^+)
	&= c_{\psic\psi\gamma} (\langle\bs1\varepsilon_3^+\bs2]+\langle\bs2\varepsilon_3^+\bs1])
	+ \frac{c^{RRR}_{\psic\psi\gamma}}{\Lam}
	\sqb{\bs13}\sqb{\bs23}\,,
\\
\mathcal{M}(\bs1_\psic,\bs2_\psi,3_\gamma^-)
	&= c_{\psic\psi\gamma} (\langle\bs1\varepsilon_3^-\bs2]+\langle\bs2\varepsilon_3^-\bs1])
	+ \frac{c^{LLL}_{\psic\psi\gamma}}{\Lam}
	\anb{\bs13}\anb{\bs23}\,,
\end{align}
or equivalently (see the last two equalities of \autoref{eqn:x_factor_equalities}), 
\begin{align}
\mathcal{M}(\bs1_\psic,\bs2_\psi,3^+)
	&= c_{\psic\psi\gamma} \frac{\sqrt{2}}{m_\psi} \frac{\langle \zeta \bs{1} 3 ]}{\anb{\zeta 3}}\anb{\bs{12}}
	+ \frac{c^{RRR}_{\psic\psi\gamma}}{\Lam}
	\sqb{\bs13}\sqb{\bs23}\,,
\\
\mathcal{M}(\bs1_\psic,\bs2_\psi,3^-)
	&= c_{\psic\psi\gamma}
	\frac{\sqrt{2}}{m_\psi} \frac{[\zeta \bs{1} 3 \rangle}{\sqb{\zeta 3}}\sqb{\bs{12}}
	+ \frac{c^{LLL}_{\psic\psi\gamma}}{\Lam}
	\anb{\bs13}\anb{\bs23}\,.
\end{align}

We obtain
\beq
\label{eqn:ffA_tree}
c_{\psic\psi\gamma} &= - Q_\psi \left[ 
\frac{\bar{g} \bar{g}'}{\sqrt{\bar{g}^2 +  \bar{g}^{\prime 2}}}
- \frac{v^2}{\Lambda^2} \frac{ \bar{g}^2 \bar{g}^{\prime 2}}{\left(\bar{g}^2 +  \bar{g}^{\prime 2}\right)^{3/2}} 
C_{\varphi WB}
\right]\,,\\
\frac{c^{RRR}_{\psic\psi\gamma}}{\bar\Lambda}& = \frac{v}{\Lambda^2}\left( - 2\frac{\bar{g}}{ \sqrt{\bar{g}^2 +  \bar{g}^{\prime 2}}} C_{\psi B}- 4 I_{\psi} \frac{ \bar{g}'}{ \sqrt{\bar{g}^2 +  \bar{g}^{\prime 2}}} C_{\psi W}\right)\,,
\\
c^{LLL}_{\psic\psi\gamma}& = \left(c^{RRR}_{\psic\psi\gamma}\right)^{\ast}\,,
\eeq
where we used
\beq
\bar{\psi}_{\bs{1}} \sigma^{\mu \nu} P_R \psi_{\bs{2}} \varepsilon^+_{3, \mu} p_{3, \nu}
&= \frac{i}{2} \bar{v}_{\bs{1}} [\gamma^{\mu}, \gamma^{\nu}] P_R u_{\bs{2}} \varepsilon^+_{3, \mu} p_{3, \nu}
=-  \frac{i}{2}   \left( [ \bs{1} \bar{3} \varepsilon^+_3\bs{2}] + [ \bs{2} \bar{3} \varepsilon^+_3\bs{1}]\right) \nn\\
&= - \frac{i}{2}   \left( [ \bs{1} 3] \langle 3 \varepsilon^+_3\bs{2}] + [ \bs{2} 3] \langle 3  \varepsilon^+_3\bs{1}]\right)\nn\\
&= \sqrt{2} i \sqb{\bs{1} 3} \sqb{\bs{2} 3}\,,
\label{eqn:dipole_identity}
\eeq
and similarly,
\beq
\bar{\psi}_{\bs{1}} \sigma^{\mu \nu} P_L \psi_{\bs{2}} \varepsilon^-_{3,\mu} p_{3, \nu}
& = \sqrt{2} i \anb{\bs{1} 3} \anb{\bs{2} 3}\,,\\
\bar{\psi}_{\bs{1}} \sigma^{\mu \nu} P_R \psi_{\bs{2}} \varepsilon^-_{3,\mu} p_{3, \nu}
& = \bar{\psi}_{\bs{1}} \sigma^{\mu \nu} P_L \psi_{\bs{2}} \varepsilon^+_{3,\mu} p_{3, \nu} 
=0\,.
\eeq

\subsection{\texorpdfstring{$\psic\psi' W$}{ff'W}}
\label{sec:ffWmatch}

General form:
\begin{equation}
\mathcal{M}(\bs{1}_\psic, \bs{2}_{\psi'},\bs{3}_{W})=
 \frac{c^{LR0}_{\psic\psi' W}}{m_W}	\anb{\bs{13}}\sqb{\bs{23}}
+ \frac{c^{RL0}_{\psic\psi' W}}{m_W}	\sqb{\bs{13}}\anb{\bs{23}}
+   \frac{c_{\psic\psi' W}^{RRR}}{\Lam}	\sqb{\bs{13}}\sqb{\bs{23}}
+ \frac{c_{\psic\psi' W}^{LLL}}{\Lam}	\anb{\bs{13}}\anb{\bs{23}}
\,.
\end{equation}

We obtain (incoming $W^-$)
\beq
c^{LR0}_{\psic\psi' W} & =  \frac{\bar{g}}{2} \frac{v^2}{\Lambda^2} C_{\varphi u d}^{\ast} \,,\\
c^{RL0}_{\psic\psi' W} & = \bar{g} \left( 1 + \frac{v^2}{\Lambda^2} C_{\varphi \psi_L}^3 \right)\,,\\
\frac{c_{\psic\psi' W}^{RRR}}{\bar\Lambda} & =- 2 \sqrt{2}\frac{v}{\Lambda^2}  C_{\psi' W}\,,\\
\frac{c_{\psic\psi' W}^{LLL}}{\bar\Lambda} & = - 2 \sqrt{2}\frac{v}{\Lambda^2}  C_{\psi W}^{\ast}\,,
\eeq
where $C_{\varphi u d}$ is absent for the lepton case because there is no right-handed neutrino in the SMEFT.

\subsection{\texorpdfstring{$\psic\psi h$}{ffh}}

 General form:
\beq
\mathcal{M}(\bs{1}_{\psi^c},\bs{2}_{\psi}, \bs{3}_{h})=
	c^{RR}_{\psic\psi h} \sqb{\bs{12}} + c^{LL}_{\psic\psi h}\anb{\bs{12}}\,.
	\label{eqn:pph3pt}
\eeq
In addition, when $h$ is a neutral scalar, 
the Hermitian condition for the effective Lagrangian gives 
\beq
c^{LL}_{\psic\psi h} = \left(c^{RR}_{\psic\psi h}\right)^{\ast}\,.
\eeq

We obtain
\beq
\label{eqn:matching_ffh}
\frac12(c^{RR}_{\psic\psi h}  +c^{LL}_{\psic\psi h}) &
= \textrm{Re}c^{RR}_{\psic\psi h}=   - \frac{m_\psi}{v} 
+ \frac{v}{\Lambda^2} \left(- m_\psi C_{\varphi \square } + \frac{1}{4} m_\psi C_{\varphi D}
+ \frac{v}{\sqrt{2}} \textrm{Re} C_{\psi \varphi}\right) \,, \\ 
\frac12(c^{RR}_{\psic\psi h}  -c^{LL}_{\psic\psi h})  &= i \textrm{Im}c^{RR}_{\psic\psi h} = i \frac{1}{\sqrt{2}}  \frac{v^2}{\Lambda^2} \textrm{Im} C_{\psi \varphi}  \,,
\eeq
where  $m_\psi$ is the fermion mass and it can be written by the Lagrangian parameters as
\beq
\frac{m_\psi}{v} = \frac{1}{\sqrt{2}} \left( y_\psi - \frac{1}{2}\frac{v^2}{\Lambda^2}  C_{\psi \varphi}\right)\,, \textrm{~~~for~}\psi=e,u,d.
\label{eqn:fermionmass}
\eeq

\subsection{\texorpdfstring{$Z Z h$}{ZZh}}
General form:
\begin{equation}
\mathcal{M}(\bs1_Z, \bs2_Z, \bs3_h)
	= 
	\frac{c^{00}_{ZZh}}{m_Z}	\sqb{\bs1\bs2}\anb{\bs1\bs2}
	+\frac{c^{RR}_{ZZh}}{\Lam}	\sqb{\bs1\bs2}^2
	+ \frac{c^{LL}_{ZZh}}{\Lam}	\anb{\bs1\bs2}^2\,.
\end{equation}

We obtain
\beq
\label{eqn:matching_zzh}
c^{00}_{ZZh} & = - \sqrt{\bar{g}^2 + \bar{g}^{\prime 2}} \left[1  + \frac{v^2}{\Lambda^2}  \left(  C_{\varphi \square}
+ \frac{1}{2} C_{\varphi D}  + \frac{\bar{g}\bar{g}'}{\bar{g}^2 + \bar{g}^{\prime 2}} C_{\varphi WB} \right) \right]\,,\\
\frac{c^{RR}_{ZZh}}{\bar\Lambda}
& = - \frac{2}{\bar{g}^2 + \bar{g}^{\prime 2}}  \frac{v}{\Lambda^2} 
\left[ \bar{g}^2 \left( C_{\varphi W} + i C_{\varphi \tilde{W}} \right) +
\bar{g}^{\prime 2} \left( C_{\varphi B} + i C_{\varphi \tilde{B}} \right)  
+ \bar{g} \bar{g}' \left( C_{\varphi W B} + i C_{\varphi \tilde{W} B} \right) 
\right]\,,\\
c^{LL}_{ZZh} & = \left( c^{RR}_{ZZh}\right)^{\ast}\,.
\eeq

Above we used,
\beq
\left(p_{\bs{1}}^{\nu} p_{\bs{2}}^{\mu} - \left( p_{\bs{1}} \cdot p_{\bs{2}} \right) \eta^{\mu\nu}\right)
\bs{\varepsilon}_{\bs{1},\mu} \bs{\varepsilon}_{\bs{2},\nu} 
& = - \frac{1}{2} \left(  \sqb{\bs{12}}^2 + \anb{\bs{12}}^2\right)\,,\\
p_{\bs{1}}^{\rho} p_{\bs{2}}^{\sigma} \epsilon_{\mu \nu \rho \sigma} \bs{\varepsilon}_{\bs{1}}^{\mu} \bs{\varepsilon}_{\bs{2}}^{\nu}
& =  - \frac{i }{2} \left( \sqb{\bs{12}}^2 - \anb{\bs{12}}^2\right)\,,
\eeq
which follow from the Schouten identity and,%
\footnote{An alternative form is $\;
4 i \epsilon_{\mu\nu\rho\sigma} = \textrm{Tr}[
\bar\sigma_{\mu}\sigma_{\nu}\bar\sigma_{\rho}\sigma_{\sigma}]
-  \textrm{Tr}[\sigma_{\mu}\bar\sigma_{\nu}\sigma_{\rho}\bar{\sigma}_{\sigma }] \,.
$} 
\beq
\epsilon_{\mu \nu \rho \sigma} \sigma^{[\mu}_{\alpha \dot \alpha} \sigma^{\nu]}_{\beta \dot \beta} 
& = - {i}\left( \sigma_{[\rho} \bar{\sigma}_{\sigma]}\right)_{\alpha}^{~\gamma} \epsilon_{\gamma \beta} 
\epsilon_{\dot \alpha \dot\beta}
+  {i}\left( \bar{\sigma}_{[\rho} {\sigma}_{\sigma]}\right)^{\dot \gamma}_{~\dot\beta} \epsilon_{\alpha \beta} 
\epsilon_{\dot \alpha \dot\gamma}\,.
\label{eqn:identityepsilon}
\eeq
Note that in the Warsaw basis, the dual tensors are defined by $\tilde{X}_{\mu \nu} = \frac{1}{2} \epsilon_{\mu \nu \rho \sigma} X^{\rho \sigma} $ with $ \epsilon_{0123}= +1$ \cite{Grzadkowski:2010es}, so that the coefficients $C_{\varphi \tilde{B}}$,  $C_{\varphi \tilde{W}}$, and $C_{\varphi \tilde{W}B}$ are purely real.

\subsection{\texorpdfstring{$WWh$}{WWh}}
General form:
\begin{equation}
\mathcal{M}(\bs1_W, \bs2_W, \bs3_h)
	= 
	\frac{c^{00}_{WWh}}{m_W}	\sqb{\bs1\bs2}\anb{\bs1\bs2}
	+\frac{c^{RR}_{WWh}}{\Lam}	\sqb{\bs1\bs2}^2
	+ \frac{c^{LL}_{WWh}}{\Lam}	\anb{\bs1\bs2}^2\,.
\end{equation}

We obtain
\beq
c^{00}_{WWh} & = 
-  \bar{g} \left[ 1  +  \frac{v^2}{\Lambda^2} \left( C_{\varphi \square} - \frac{1}{4}    C_{\varphi D}\right)\right]\,,\\
\frac{c^{RR}_{WWh}}{\bar\Lambda} &=
- 2 \frac{v}{\Lambda^2} \left( C_{\varphi W} + i  C_{\varphi \tilde{W}}\right)\,,\\
c^{LL}_{WWh}& = \left( c^{RR}_{WWh} \right)^{\ast}\,.
\eeq

\subsection{\texorpdfstring{$\gamma \gamma h$}{AAh}}
General form:
\begin{equation}
\mathcal{M}(1_\gamma^+,2_\gamma^+,\bs3_h) = \frac{c^{RR}_{\gamma\gamma}}{\Lam} \sqb{12}^2\,,
\qquad\text{and}\qquad
\mathcal{M}(1_\gamma^-,2_\gamma^-,\bs3_h) = \frac{c^{LL}_{\gamma\gamma}}{\Lam} \anb{12}^2\,.
\end{equation}

We obtain
\beq
\frac{c^{RR}_{\gamma\gamma} }{\bar\Lambda}& = 
- \frac{2}{\bar{g}^2 + \bar{g}^{\prime 2}}  \frac{v}{\Lambda^2}  \left[ \bar{g}^{\prime 2} \left( C_{\varphi W} + i C_{\varphi \tilde{W} }\right)
+ \bar{g}^2 \left( C_{\varphi B} + i C_{\varphi \tilde{B}}\right) - \bar{g} \bar{g}' \left( C_{\varphi W B} + i C_{\varphi \tilde{W} B }\right) \right] \,,\\
c^{LL}_{\gamma\gamma}& = \left( c^{RR}_{\gamma\gamma} \right)^{\ast}\,.
\eeq

\subsection{\texorpdfstring{$\gamma Z h$}{AZh}}

General form:
\begin{equation}
\mathcal{M}(1_\gamma^+,\bs2_Z,\bs3_h) = \frac{c^{RR}_{\gamma Z}}{\Lam} \sqb{1\bs2}^2\,,
\qquad\text{and}\qquad
\mathcal{M}(1_\gamma^-,\bs2_Z,\bs3_h) = \frac{c^{LL}_{\gamma Z}}{\Lam} \anb{1\bs2}^2\,.
\end{equation}

We obtain
\beq
\frac{c^{RR}_{\gamma Z}}{\bar\Lambda} & =
-  \frac{1}{\bar{g}^2 + \bar{g}^{\prime 2}}\frac{v}{\Lambda^2}  \left[ 
2 \bar{g} \bar{g}' \left( C_{\varphi W} + i C_{\varphi \tilde{W}}\right) - 2 \bar{g} \bar{g}' \left( C_{\varphi B} + i C_{\varphi \tilde{B}} \right) + \left(\bar{g}^{\prime 2} - \bar{g}^2 \right) \left( C_{\varphi W B} + i C_{\varphi \tilde{W} B}\right) 
\right] 
 \,,\\
c^{LL}_{\gamma Z}& = \left( c^{RR}_{\gamma Z} \right)^{\ast}\,.
\eeq

\subsection{\texorpdfstring{$hhh$}{hhh}}

General form:
\beq
\mathcal{M}(\bs{1}_h,\bs{2}_h, \bs{3}_h) = m_h c_{hhh}\,.
\eeq

We obtain
\beq
m_h c_{hhh} & = -3  v \left[ \lambda + \frac{v^2}{\Lambda^2}\left(
- 5  C_{\varphi} 
+  5 \lambda  C_{\varphi \square} 
- \frac{5}{4} \lambda   C_{\varphi D} \right) \right]\,.
\eeq

\subsection{\texorpdfstring{$WWZ$}{WWZ}}
\label{sec:matching_wwz}

General form:
\begin{equation}
\begin{aligned}
\mathcal{M}(\bs 1_W,\bs 2_W,\bs 3_Z)
	=&2\frac{c_{WWZ}}{m_Z m_W } \left( \frac{m_Z}{m_W}  \anb{\bs{12}} \sqb{\bs{13}} \sqb{\bs{23}} 
	   +\sqb{\bs{12}} \anb{\bs{13}} \sqb{\bs{23}} \;
	   +\sqb{\bs{12}} \sqb{\bs{13}} \anb{\bs{23}}\right)
\\	&+ \frac{\cwwzLA}{m_Z \bar\Lambda}  \anb{\bs{12}} \left( \anb{\bs{13}} \sqb{\bs{23}} 
		- \sqb{\bs{13}} \anb{\bs{23}}	\right)	
%\\	&
+ \frac{\cwwzLS}{m_Z \bar\Lambda}  \anb{\bs{12}} \left(\anb{\bs{13}} \sqb{\bs{23}} 
		+ \sqb{\bs{13}} \anb{\bs{23}}	\right)	
\\	&+ \frac{\cwwzRA}{m_Z \bar\Lambda}  \sqb{\bs{12}} \left(\anb{\bs{13}} \sqb{\bs{23}}
		- \sqb{\bs{13}} \anb{\bs{23}}	\right)	
%\\	&
+ \frac{\cwwzRS}{m_Z \bar\Lambda}  \sqb{\bs{12}} \left(\anb{\bs{13}} \sqb{\bs{23}}
		+ \sqb{\bs{13}} \anb{\bs{23}}	\right)	
\\	&+  \frac{ c_{WWZ}^{RRR}}{\Lam^2} \sqb{\bs{12}} \sqb{\bs{13}} \sqb{\bs{23}} + \frac{c_{WWZ}^{LLL}}{\Lam^2} \anb{\bs{12}} \anb{\bs{13}} \anb{\bs{23}} \,.
\end{aligned}
\end{equation}
We obtain (with incoming $W^+$ ($\bs1_W $) and  incoming $W^-$ ($\bs2_W $)),
\beq
\label{eqn:WWZrenormalizable}
2 c_{WWZ} & = - \sqrt{2} \frac{\bar{g}^2}{\sqrt{ \bar{g}^2 + \bar{g}^{\prime 2}}}
+ \sqrt{2}  \frac{\bar{g}^3 \bar{g}'}{(\bar{g}^2 + \bar{g}^{\prime 2})^{3/2}} \frac{v^2}{\Lambda^2} C_{\varphi W B}\,,\\
\label{eqn:WWZA}
\cwwzLA & = \cwwzRA =0\,,\\
\label{eqn:WWZS}
\frac{\cwwzRS}{m_Z \bar\Lambda} & =  - \frac{1}{\sqrt{2}m_W m_Z} \frac{\bar{g} \bar{g}'}{ \sqrt{\bar{g}^2 + \bar{g}^{\prime 2}}} \frac{v^2}{\Lambda^2} \left( C_{\varphi W B} + i C_{\varphi \tilde{W} B}\right)\,,\\
\cwwzLS & = \left( \cwwzRS\right)^{\ast}\,,\\
\frac{\cwwz+++}{\bar{\Lambda}^2} & = - 3 \sqrt{2} \frac{\bar{g }}{\sqrt{\bar{g}^2 + \bar{g}^{\prime 2}}}\frac{1}{\Lambda^2}(C_{W} + i C_{\tilde{W}})\,, \\
\cwwz---& = \left( \cwwz+++\right)^{\ast}\,.
\eeq
Note that the renormalizable term in \autoref{eqn:WWZrenormalizable} has the opposite sign from ref.~\cite{Aoude:2019tzn}.

In order to obtain these relations, we used the following identities: 
\beq
&\left[ \eta_{\mu \nu} (p_{\bs{1}} - p_{\bs{2}})_{\rho} + \eta_{ \nu \rho} (p_{\bs{2}} - p_{\bs{3}})_{\mu} +\eta_{\rho \mu} (p_{\bs{3}} - p_{\bs{1}})_{\nu} \right]\bs{\varepsilon}^{\mu}_{\bs{1}}\bs{\varepsilon}^{\nu}_{\bs{2}}
\bs{\varepsilon}^{\rho}_{\bs{3}}\nn\\
& \quad =
- \frac{\sqrt{2}}{m_1 m_2 m_3} \left( m_1 \sqb{\bs{12}} \sqb{\bs{13}}  \anb{\bs{23}}+ m_2 \sqb{\bs{12}} \anb{\bs{13}}  \sqb{\bs{23}} + m_3  \anb{\bs{12}} \sqb{\bs{13}}  \sqb{\bs{23}} \right) \nn \\
& \quad =
- \frac{\sqrt{2}}{m_1 m_2 m_3} \left( m_1 \anb{\bs{12}} \anb{\bs{13}}  \sqb{\bs{23}}+ m_2 \anb{\bs{12}} \sqb{\bs{13}}  \anb{\bs{23}} + m_3  \sqb{\bs{12}} \anb{\bs{13}}  \anb{\bs{23}} \right)\,,
\nn\\& \quad = 
- \frac{1}{\sqrt{2} m_1 m_2 m_3} \big( m_1 \anb{\bs{12}} \anb{\bs{13}}  \sqb{\bs{23}}+ m_2 \anb{\bs{12}} \sqb{\bs{13}}  \anb{\bs{23}} + m_3  \sqb{\bs{12}} \anb{\bs{13}}  \anb{\bs{23}}
\nn\\&\qquad\qquad\qquad\qquad +m_1 \sqb{\bs{12}} \sqb{\bs{13}}  \anb{\bs{23}}+ m_2 \sqb{\bs{12}} \anb{\bs{13}}  \sqb{\bs{23}} + m_3  \anb{\bs{12}} \sqb{\bs{13}}  \sqb{\bs{23}}
 \big)\,,
\eeq
\beq
\left( p_{\bs{3}, \mu} \eta_{\nu \rho } - p_{\bs{3}, \nu} \eta_{\mu \rho } \right) \bs{\varepsilon}^{\mu}_{\bs{1}}\bs{\varepsilon}^{\nu}_{\bs{2}}
\bs{\varepsilon}^{\rho}_{\bs{3}}& = 
\frac{1}{\sqrt{2} m_1 m_2 } \left(\sqb{\bs{12}}\anb{\bs{13}}\anb{\bs{23}} + \anb{\bs{12}}\sqb{\bs{13}}\sqb{\bs{23}}\right)\,,\\
\epsilon_{\mu \nu \rho \sigma} p_{\bs{3}}^{\sigma} \bs{\varepsilon}^{\mu}_{\bs{1}}\bs{\varepsilon}^{\nu}_{\bs{2}}
\bs{\varepsilon}^{\rho}_{\bs{3}} &= 
\frac{i }{\sqrt{2} m_1 m_2 } \left(\sqb{\bs{12}}\anb{\bs{13}}\anb{\bs{23}} - \anb{\bs{12}}\sqb{\bs{13}}\sqb{\bs{23}}\right)\,.
\eeq
%%%
\beq
&\Bigl\{
p_{\bs3,\mu} p_{\bs1,\nu} p_{\bs2,\rho} - p_{\bs2,\mu} p_{\bs3,\nu} p_{\bs1,\rho} 
+ \eta_{\mu \nu} [p_{\bs1,\rho} (p_\bs2 \cdot p_\bs3 ) - p_{\bs2,\rho} (p_\bs1 \cdot p_\bs3 )]\nonumber \\
& +  \eta_{ \nu \rho} [p_{\bs2,\mu} (p_\bs1 \cdot p_\bs3 ) - p_{\bs3,\mu} (p_\bs1 \cdot p_\bs2 )]
+  \eta_{ \rho \mu} [p_{\bs3,\nu} (p_\bs1 \cdot p_\bs2 ) - p_{\bs1,\nu} (p_\bs2 \cdot p_\bs3 )]\Bigr\} \bs{\varepsilon}^{\mu}_{\bs{1}}\bs{\varepsilon}^{\nu}_{\bs{2}}
\bs{\varepsilon}^{\rho}_{\bs{3}}  \nonumber \\
&=   \frac{1}{\sqrt{2}} ([\bs{12}] [\bs{13}][\bs{23}] + \langle \bs{12}\rangle \langle \bs{13}\rangle\langle \bs{23}\rangle)\,.\\
%%%
&
\Bigl\{\epsilon_{\mu \nu \rho \sigma} \left[ p_{\bs1}^{\sigma} (p_\bs2 \cdot p_\bs3 ) +p_{\bs2}^{\sigma} (p_\bs1 \cdot p_\bs3 )+p_{\bs3}^{\sigma} (p_\bs1 \cdot p_\bs2 )  \right]
+ \epsilon_{\mu \nu \sigma \delta} (p_\bs1 - p_\bs2 )_{\rho} p_\bs1^{\sigma} p_\bs2^\delta \nonumber \\
&
+ \epsilon_{ \nu \rho \sigma \delta} (p_\bs2 - p_\bs3 )_{\mu} p_\bs2^{\sigma} p_\bs3^\delta 
+ \epsilon_{ \rho \mu \sigma \delta} (p_\bs3 - p_\bs1 )_{\nu} p_\bs3^{\sigma} p_\bs1^\delta  \Bigr\} \bs{\varepsilon}^{\mu}_{\bs{1}}\bs{\varepsilon}^{\nu}_{\bs{2}}
\bs{\varepsilon}^{\rho}_{\bs{3}}\nonumber \\
&
=  \frac{3 i}{\sqrt{2}} ([\bs{12}] [\bs{13}][\bs{23}] - \langle \bs{12}\rangle \langle \bs{13}\rangle\langle \bs{23}\rangle)
\,.
\eeq

\subsection{\texorpdfstring{$WW\gamma$}{WWA}}

General form:
\begin{equation}
\begin{aligned}
\mathcal{M}(\bs 1_W,\bs 2_W, 3_\gamma^+) =
	&\frac{\cwwa{}{}{}}{m_W} \anb{\bs{12}} \left(\langle\bs1\varepsilon^+_3\bs2]+\langle\bs2\varepsilon^+_3\bs1]\right) 
+\frac{\cwwa00+}{m_W\Lam} \anb{\bs{12}} \sqb{\bs1 3} \sqb{\bs2 3}	\\
	&	+\frac{\cwwa+++}{\Lam^2} \sqb{\bs{12}} \sqb{\bs1 3} \sqb{\bs2 3}	\,,\\
\mathcal{M}(\bs 1_W,\bs 2_W, 3_\gamma^-) =
	&\frac{\cwwa{}{}{}}{m_W} \sqb{\bs{12}} \left(\langle\bs1\varepsilon^-_3\bs2]+\langle\bs2\varepsilon^-_3\bs1]\right)
	+\frac{\cwwa00-}{m_W\Lam } \sqb{\bs{12}} \anb{\bs1 3} \anb{\bs2 3} \\
	&+\frac{\cwwa---}{\Lam^2} \anb{\bs{12}} \anb{\bs1 3} \anb{\bs2 3}	\,,
\end{aligned}
\end{equation}
or equivalently (see the last two equalities of \autoref{eqn:x_factor_equalities}), 
\begin{equation}
\begin{aligned}
\mathcal{M}(\bs 1_W,\bs 2_W, 3_\gamma^+) =
	&\frac{\sqrt{2} \cwwa{}{}{}}{m_W^2 } \frac{ \langle \zeta \bs{1} 3]}{\anb{\zeta 3}}  \anb{\bs{12}}^2
+\frac{\cwwa00+}{m_W\Lam} \anb{\bs{12}} \sqb{\bs1 3} \sqb{\bs2 3}	+\frac{\cwwa+++}{\Lam^2} \sqb{\bs{12}} \sqb{\bs1 3} \sqb{\bs2 3}	\,,\\
\mathcal{M}(\bs 1_W,\bs 2_W, 3_\gamma^-) =
	&\frac{\sqrt{2} \cwwa{}{}{}}{m_W^2} 
	\frac{[ \zeta \bs{1} 3\rangle}{\sqb{\zeta 3}} \sqb{\bs{12}}^2
	+\frac{\cwwa00-}{m_W\Lam } \sqb{\bs{12}} \anb{\bs1 3} \anb{\bs2 3} +\frac{\cwwa---}{\Lam^2} \anb{\bs{12}} \anb{\bs1 3} \anb{\bs2 3}	\,.
\end{aligned}
\end{equation}

We obtain (with incoming $W^+$ ($\bs1_W $) and  $W^-$ ($\bs2_W $)),
\beq
{\cwwa{}{}{}} = & 
 \frac{\bar{g}\bar{g}'}{\sqrt{\bar{g}^2 + \bar{g}^{\prime 2}} } - 
 \frac{\bar{g}^2 \bar{g}^{\prime 2}  }{(\bar{g}^2 + \bar{g}^{\prime 2})^{3/2}}
\frac{v^2}{\Lambda^2}  C_{\varphi W B} \,,
\\
\frac{\cwwa{0}{0}{R}}{m_W \bar\Lambda } = &
 - \frac{\sqrt{2} }{m_W  }\frac{\bar{g}}{\sqrt{\bar{g}^2 + \bar{g}^{\prime 2}}}
\frac{v}{\Lambda^2} \left( C_{\varphi W B} + i C_{\varphi \tilde{W} B}\right)\,,
\\
{\cwwa{0}{0}{L}} = & \left( {\cwwa{0}{0}{R}}\right)^{\ast}\,,\\
 \frac{\cwwa+++}{\Lam^2} = &
 - 3 \sqrt{2} \frac{\bar{g}'}{\sqrt{\bar{g}^2 + \bar{g}^{\prime 2}}}\frac{1}{\Lambda^2} \left( C_W + i C_{\tilde{W}} \right)\,,\\
 {\cwwa---} = & \left( {\cwwa+++} \right)^{\ast}\,.
\eeq

\subsection{\texorpdfstring{$\psic \psi Z h$}{ffZh}}
\label{sec:matching_ffzh}

General form (see \autoref{eqn:mffzh}):
\begin{equation}
\begin{aligned}
\mathcal{M}\nf(\bs1_\psic,\bs2_\psi,\bs3_Z,\bs4_h)
&=	\frac{\cffzh+++}{\Lam^2}	\sqb{\bs{13}}\sqb{\bs{23}}
+	\frac{\sqb{\bs{12}}}{\Lam^3} \langle\bs3\;\{
		 \cffzh[_A]++0 (\bs1+\bs2)
		+\cffzh[_S]++0 (\bs1-\bs2)
		\}\;\bs3]
\\&+	\frac{\cffzh+-0}{\Lam^2}\sqb{\bs{13}}\anb{\bs{23}}
+	\frac{\cffzh+-+}{\Lam^3}[\bs{312}\rangle \sqb{\bs{13}}
+	\frac{\cffzh+--}{\Lam^3}\langle\bs{321}] \anb{\bs{23}}
\\&+	\frac{\cffzh-+0}{\Lam^2}\anb{\bs{13}}\sqb{\bs{23}}
+	\frac{\cffzh-++}{\Lam^3}[\bs{321}\rangle \sqb{\bs{23}}
+	\frac{\cffzh-+-}{\Lam^3}\langle\bs{312}] \anb{\bs{13}}
\\&+	\frac{\cffzh---}{\Lam^2}	\anb{\bs{13}}\anb{\bs{23}}
+	\frac{\anb{\bs{12}}}{\Lam^3} \langle\bs3\;\{
		 \cffzh[_A]--0 (\bs1+\bs2)
		+\cffzh[_S]--0 (\bs1-\bs2)
		\}\;\bs3]
\,.
\end{aligned}
\end{equation}

At the level of dimension-six operators, we have,
\beq
\frac{\cffzh+-0}{\Lam^2} & = - \frac{2 \sqrt{2}}{\Lambda^2} \left( C_{\varphi \psi_L}^1 - 2 I_\psi C_{\varphi \psi_L}^3 \right)\,,\\
\frac{\cffzh-+0}{\Lam^2} & = -\frac{2 \sqrt{2}}{\Lambda^2} C_{\varphi \psi_R}\,,\\
\frac{\cffzh+++}{\Lam^2} & = \frac{2}{\Lambda^2}\frac{  - 2 I_\psi \bar{g} C_{\psi W} + \bar{g}' C_{\psi B}}{\sqrt{\bar{g}^2 + \bar{g}^{\prime 2}}}\,,\\
\cffzh--- & = \left( \cffzh+++ \right)^{\ast}\,,
\eeq
or equivalently 
\beq
\mathcal{M}\nf\left( 
\bs{1}_\psic, \bs{2}_\psi, \bs{3}_Z, \bs{4}_h \right) =&
 \frac{2}{\Lambda^2}\frac{  - 2 I_\psi \bar{g} C_{\psi W} + \bar{g}' C_{\psi B}}{\sqrt{\bar{g}^2 + \bar{g}^{\prime 2}}}
\sqb{\bs{13}} \sqb{\bs{23}}
- \frac{2 \sqrt{2}}{\Lambda^2} \left( C_{\varphi \psi_L}^1 - 2 I_\psi C_{\varphi \psi_L}^3 \right) \sqb{\bs{13}}\anb{\bs{23}}\nn\\
& -\frac{2 \sqrt{2}}{\Lambda^2} C_{\varphi \psi_R}
\anb{\bs{13}}\sqb{\bs{23}}
 +\frac{2}{\Lambda^2}\frac{  - 2 I_\psi \bar{g} C_{\psi W}^{\ast} + \bar{g}' C_{\psi B}^{\ast}}{\sqrt{\bar{g}^2 + \bar{g}^{\prime 2}}} \anb{\bs{13}}\anb{\bs{23}}\,. 
\eeq
Unbolding $\bs{1}$ and $\bs{2}$ corresponds to the massless fermion amplitudes.
Here, there is a non-trivial cancellation of the vacuum expectation value in the second and third terms which involve $v/\Lambda^2 m_Z$. The $m_Z$ comes from the denominator of the massive polarization vector in \autoref{eqn:massiveP}.

The matching onto the symmetric-phase SMEFT at dimension-six level provides \cite{Ma:2019gtx},
\begin{equation}
\begin{aligned}
\mathcal{M}\nf\left( 
{1}_\psic^+, {2}_\psi^+, {3}_Z^+, {4}_H \right) =&
 \frac{2 \sqrt{2} }{\Lambda^2}\frac{  - 2 I_\psi {g} C_{\psi W} + {g}' C_{\psi B}}{\sqrt{{g}^2 + {g}^{\prime 2}}}
\sqb{{13}} \sqb{{23}}\,,
\\
\mathcal{M}\nf\left( 
{1}_\psic^-, {2}_\psi^-, {3}_Z^-, {4}_H \right) =&
 \frac{2 \sqrt{2}}{\Lambda^2}\frac{  - 2 I_\psi {g} C_{\psi W}^{\ast} + {g}' C_{\psi B}^{\ast}}{\sqrt{{g}^2 + {g}^{\prime 2}}} \anb{{13}}\anb{{23}}\,,\\
  \mathcal{M}\nf\left( 
{1}_\psic^+, {2}_\psi^+, {3}_Z^{-}, {4}_H \right)
 & = \mathcal{M}\nf\left( 
{1}_\psic^-, {2}_\psi^-, {3}_Z^{+}, {4}_H \right)= 0\,,\\
 \mathcal{M}\nf\left( 
{1}_\psic^+, {2}_\psi^-, {3}_Z^{\pm}, {4}_H \right)
 & = \mathcal{M}\nf\left( 
{1}_\psic^-, {2}_\psi^+, {3}_Z^{\pm}, {4}_H \right)= 0\,,
\label{eqn:matching_massless_ffzh}
\end{aligned}
\end{equation}
and 
\begin{equation}
\begin{aligned}
\mathcal{M}\nf\left( 
{1}_\psic^+, {2}_\psi^-, {3}_{H}, {4}_{H^c} \right) =&
\frac{1}{\Lambda^2} \left( C_{\varphi \psi_L}^1 - 2 I_\psi C_{\varphi \psi_L}^3 \right) [1 (\bar3 - \bar{4}) 2 \rangle\,,\\
\mathcal{M}\nf\left( 
{1}_\psic^-, {2}_\psi^+, {3}_{H}, {4}_{H^c} \right) =&
\frac{1}{\Lambda^2} C_{\varphi \psi_R}
\langle 1 (3-4) 2]\,,\\
\mathcal{M}\nf\left( 
{1}_\psic^+, {2}_\psi^{+}, {3}_{H}, {4}_{H^c} \right) =&\mathcal{M}\nf\left( 
{1}_\psic^-, {2}_\psi^{-}, {3}_{H}, {4}_{H^c} \right) = 0\,,
\end{aligned}
\end{equation}
where $H$ is a complex scalar field.
Thus, the $\psi^2 X \varphi$ (dipole) operators contribute to the transverse modes of $Z$ boson, while the $\psi^2 \varphi^2 D$ operators contribute to the longitudinal one in the high-energy limit.

\subsection{The equivalence theorem}
\label{sec:equivalence}
In the high-energy limit of the $\psic \psi Z$ amplitude, the leading longitudinal amplitudes are (see \autoref{sec:ppZ})
\beq
\mffz++0 &\to -\frac{m_{\psi}}{\sqrt{2} m_Z}  (c^{LR0}_{\psic\psi Z}-c^{RL0}_{\psic\psi Z})\;  \sqb{12}\,,
\label{eqn:highE:ffZ1}
\\
  \mffz--0 &\to +\frac{m_{\psi}}{\sqrt{2} m_Z}  (c^{LR0}_{\psic\psi Z}-c^{RL0}_{\psic\psi Z}) \;\anb{12}\,.
  \label{eqn:highE:ffZ2}
\eeq
From the matching onto the broken-phase SMEFT in \autoref{sec:ffZ}, we find 
\begin{multline}
- \frac{m_{\psi}}{\sqrt{2} m_Z} \left( c^{LR0}_{\psic\psi Z}-c^{RL0}_{\psic\psi Z}\right)
=   I_{\psi}  \sqrt{\bar{g}^2 + \bar{g}^{\prime 2}} \frac{m_{\psi}}{m_Z}\\
  + \frac{v^2}{\Lambda^2}  \frac{m_{\psi}}{m_Z} \left[
 \frac{\bar{g} \bar{g}'}{ \sqrt{\bar{g}^2 + \bar{g}^{\prime 2}}}I_{\psi} C_{\varphi WB}
 - \frac{1}{2} \sqrt{\bar{g}^2 + \bar{g}^{\prime 2}} \left( C_{\varphi \psi_L}^1- 2 I_{\psi} C_{\varphi \psi_L}^3  - C_{\varphi \psi_R} \right)\right]\\
 =   2 I_{\psi}  \frac{m_{\psi}}{v}   
 + \frac{ m_{\psi} v}{\Lambda^2}  \left(-   \frac{1}{2}I_{\psi}   C_{\varphi D} 
 -  C_{\varphi \psi_L}^1 + 2 I_{\psi} C_{\varphi \psi_L}^3  +  C_{\varphi \psi_R} \right)\,,
\end{multline}
where \autoref{eqn:mZ} is used for $m_Z$.

On the other hand, matching the Goldstone-boson amplitude $\psic\psi G^0$, which has the structure \autoref{eqn:pph3pt}, we obtain
\beq
c^{RR}_{\psic\psi G^0}    
&= i  2 I_3 \frac{m_\psi}{v}   
+ \frac{m_\psi v}{\Lambda^2} \left(- i \frac{1}{2} I_3   C_{\varphi D}
- i  C_{\varphi \psi_L}^1 
+ i  2 I_3   C_{\varphi \psi_L}^3 
+ i   C_{\varphi \psi_R} \right)
\,,\label{eqn:NGcoupling}\\
c^{LL}_{\psic\psi G^0}
&=- c^{RR}_{\psic\psi G^0}  
\,,
\eeq 
which correspond to the leading longitudinal $\psic\psi Z$ couplings of \eqs[,\,]{eqn:highE:ffZ1}{eqn:highE:ffZ2} including non-renormalizable terms. 
This is exactly the Goldstone boson equivalence theorem proved by using renormalizable massive vector-boson theories at the amplitude level up to $\mathcal{O}(m_Z/E)$ corrections and up to a factor $i^n$, where $n$ is the number of longitudinally polarized vector bosons~\cite{Cornwall:1974km, Vayonakis:1976vz, Lee:1977eg, Chanowitz:1985hj}.

Similarly, in the high-energy limit of the $\psic \psi' W$ amplitude, the leading amplitudes for the longitudinal mode are (see \autoref{sec:ffW})
\begin{align}
\mffw++0 &\to -  \frac{1}{\sqrt{2}m_W} (m_\psi\, c^{LR0}_{\psic\psi' W} - m_{\psi'}\,c^{RL0}_{\psic\psi' W}) \; \sqb{12} \,,\\
  \mffw--0 &\to + \frac{1}{\sqrt{2}m_W}  (m_{\psi'}\, c^{LR0}_{\psic\psi' W} - m_{\psi}\,c^{RL0}_{\psic\psi' W})\;  \anb{12}\,.
\end{align}
The matching onto the broken-phase SMEFT in \autoref{sec:ffWmatch} gives 
\beq
 -  \frac{1}{\sqrt{2}m_W} (m_\psi\, c^{LR0}_{\psic\psi' W} - m_{\psi'}\,c^{RL0}_{\psic\psi' W}) 
 = &  
 -  \frac{1}{\sqrt{2}m_W} 
\left[ m_\psi\,   \frac{\bar{g}}{2} \frac{v^2}{\Lambda^2} C_{\varphi u d}^{\ast}  
- m_{\psi'} \bar{g} \left( 1 + \frac{v^2}{\Lambda^2} C_{\varphi \psi_L}^3 \right) \right]\nn\\
= & \sqrt{2} \frac{m_{\psi'}}{v} + \frac{v}{\Lambda^2} \left(
 \sqrt{2}  m_{\psi'} C_{\varphi \psi_L}^3 -  \frac{1}{\sqrt{2}} m_{\psi} C_{\varphi u d}^{\ast} \right)\,.
\eeq

Also, 
for a charged-Goldstone boson (incoming $G^-$), from the SMEFT matching 
we obtain
\beq
c^{RR}_{\psic\psi' G^-} & = \sqrt{2} \frac{m_{\psi'}}{v} + \frac{v}{\Lambda^2} \left(
 \sqrt{2}  m_{\psi'} C_{\varphi \psi_L}^3 -  \frac{1}{\sqrt{2}} m_{\psi} C_{\varphi u d}^{\ast} \right)\,,
 \label{eqn:ffG-1}\\
 c^{LL}_{\psic\psi' G^-} & = -\sqrt{2} \frac{m_{\psi}}{v} + \frac{v}{\Lambda^2} \left(
 -\sqrt{2} m_{\psi} C_{\varphi \psi_L}^3 + \frac{1}{\sqrt{2}}m_{\psi'} C_{\varphi u d}^{\ast} \right)\,.
  \label{eqn:ffG-2}
\eeq
Note that $C_{\varphi u d}$ is absent in the lepton sector.
Again, the longitudinal modes 
are consistent with  $ c^{RR}_{\psic\psi' G^{-}}$ 
and  $ c^{LL}_{\psic\psi' G^{-}}$ including dimension-six contributions, and we hence confirm the Goldstone boson equivalence theorem.

\subsection{\texorpdfstring{$\psic\psi Zh$}{ffZh} unitarity relation}
\label{sec:four-point-relation-matching}
One can explicitly verify that the restriction of \autoref{eqn:four-point-relation} required to preserve perturbative unitarity in the $\psic\psi Zh$ four-point amplitude is satisfied in SMEFT.
At the renormalizable level, it leads to (see \autoref{eqn:matching_ffh} and \eqref{eqn:matching_zzh})
\begin{equation}
\cffh++ = \cffh--  = - \frac{m_\psi}{v}\,,
\qquad\text{and}\qquad
\czzh00 = - \sqrt{\bar{g}^2 + \bar{g}^{\prime 2}} \,,
\end{equation}
so that, including also dimension-six contributions,
\begin{equation}
\czzh00\frac{ m_\psi}{2m_Z} - c_{\psic \psi h}^{RR(LL)}
= 0 - \frac{v}{\Lambda^2} \left( \frac{m_\psi}{2} C_{\varphi D} + \frac{v}{\sqrt{2}} C_{\psi \varphi}^{(\ast)} \right)\,.
\end{equation}

%%%%%%%%%%%%%%%%%%%%%%%%%%%%%%%%%%%%%
\section{Massless amplitudes}
\label{sec:massless}
%%%%%%%%%%%%%%%%%%%%%%%%%%%%%%%%%%%%%

\newcommand{\La}{\tilde{\Lambda}}
\newcommand{\cffs}[2]{\;c_{FF'\!S}^{#1#2}}
\newcommand{\mffs}[2]{\mathcal{M}(1_F^{#1},2_{F'}^{#2},3_S)}
\newcommand{\cffv}[3]{\;c_{FF'V}^{#1#2#3}}
\newcommand{\mffv}[3]{\mathcal{M}(1_F^{#1},2_{F'}^{#2},3_V^{#3})}
\newcommand{\cssv}{\;c_{SS'V}}
\newcommand{\mssv}[1]{\mathcal{M}(1_S,2_{S'},3_V^{#1})}
\newcommand{\mvvs}[2]{\mathcal{M}(1_V^{#1},2_{V'}^{#2},3_S)}
\newcommand{\cvvs}[2]{\;c_{VV'\!S}^{#1#2}}
\newcommand{\msss}{\mathcal{M}(1_S,2_{S'},3_{S''})}
\newcommand{\csss}{\;c_{SS'S''}}
\newcommand{\mvvv}[3]{\mathcal{M}(1_{V^a}^{#1},2_{V^b}^{#2},3_{V^c}^{#3})}
\newcommand{\cvvv}[3]{\;c_{V^aV^bV^c}^{#1#2#3}}
\newcommand{\cffvs}[3]{\;c_{FF'VS}^{#1#2#3}}
\newcommand{\mffvs}[4][]{\mathcal{M}(1_F^{#2},2_{F'}^{#3},3_V^{#4},4_S^{#1})}
For completeness, we collect in this section the massless three-point amplitudes.
These are fully determined by little-group covariance.
Recall that three-point amplitudes can only be written in terms of complex momenta, and involve either square or angle brackets (see \eg, ref.~\cite{Elvang:2015rqa}).
Furthermore, they should  vanish when all momenta tend to zero together (and become real).

\subsection{Two fermions and a boson}

\paragraph{\texorpdfstring{$FF'S$}{FF'S}}
\begin{equation}
\begin{aligned}
\mffs++	& = \sqb{12} \cffs++\,,
\\
\mffs--	& = \anb{12} \cffs--\,,
\\
\mffs+-	& = 0\,,
\\
\mffs-+	& = 0\,.
\end{aligned}
\end{equation}

\paragraph{\texorpdfstring{$FF'V$}{FF'V}}
\label{Appffgamma}
\begin{equation}
\begin{aligned}
\mffv-+- &= -\anb{13}^2/\anb{12}	\cffv-+{}	\,,\\
\mffv-++ &= +\sqb{23}^2/\sqb{12}	\cffv-+{}	\,,\\
\mffv+-- &= +\anb{23}^2/\anb{12}	\cffv+-{}	\,,\\
\mffv+-+ &= -\sqb{13}^2/\sqb{12}	\cffv+-{}	\,,\\
\mffv+++ &= \sqb{13}\sqb{23}	\cffv+++ /\La	\,,\\
\mffv--- &= \anb{13}\anb{23}	\cffv--- /\La	\,,\\
\mffv++- &= 0\,,\\
\mffv-++ &= 0\,.
\end{aligned}
\end{equation}
The coefficients of the \amp+-+ and \amp+-- (or \amp-++ and \amp-+-) amplitudes are identical since the two helicities of the massless vector are part of the same irreducible representation of the Lorentz group.
The relative sign between them is fixed by re-expressing the amplitudes in terms of polarization vectors.
The $\langle 1\varepsilon^\pm_3 2]$ structures should for instance have identical coefficients.
The $\sqb{12}^2/\sqb{13}\sqb{23}$ spinorial structure that one could have written down for the \amp++- amplitude is forbidden as it does not vanish as all momenta tend to zero.
The same reasoning applies also to the $\anb{12}^2/\anb{12}\anb{23}$ structure and the \amp--+ amplitude.

%%%%%%%%%%%%%%%%%%%%%%%%%%%%%%%%%%%%%%
\subsection{Three bosons}
%%%%%%%%%%%%%%%%%%%%%%%%%%%%%%%%%%%%%%

\paragraph{\texorpdfstring{$SS'S''$}{SS'S''}}
\begin{equation}
\msss	= \La \csss\,,
\end{equation}
This amplitude only arises when the theory possesses a scale $\La$.

\paragraph{\texorpdfstring{$SS'V$}{SS'V}}
\begin{equation}
\begin{aligned}
\mssv+	&= \sqb{13}\sqb{23}/\sqb{12}	\cssv	\,,\\
\mssv-	&= \anb{13}\anb{23}/\anb{12}	\cssv	\,,
\end{aligned}
\label{eqn:ssv}
\end{equation}
These amplitudes are antisymmetric under $1\leftrightarrow2$ exchange and therefore vanish if $S$ and $S'$ are identical scalars.
Again, we have imposed that the two amplitudes, expressed in terms of polarization vectors (as $(p_1-p_2)\cdot\varepsilon_3^\pm$) arise with the same coefficient since the two vector helicities are part of the same irreducible representation of the Lorentz group.
This implies that parity is conserved in the massless $SS'V$ amplitude.

\paragraph{\texorpdfstring{$VV'S$}{VV'S}}
\begin{equation}
\begin{aligned}
\mvvs++ &= \sqb{12}^2	\cvvs++/\La	\,,\\
\mvvs--	&= \anb{12}^2	\cvvs--/\La	\,.
\end{aligned}
\end{equation}
Here again, the two amplitudes that one can re-express in terms of polarization vectors as $(p_1\cdot\varepsilon_2^\pm)(p_2\cdot\varepsilon_1^\pm)$ are forced to have identical coefficients.

\paragraph{\texorpdfstring{$V^aV^bV^c$}{Va Vb Vc}}
\begin{equation}
\begin{aligned}
\mvvv+++	&= \sqb{12}\sqb{13}\sqb{23} \cvvv+++ / \La^2	\,,\\
\mvvv---	&= \anb{12}\anb{13}\anb{23} \cvvv--- / \La^2	\,,\\
\mvvv--+ &= {\anb{12}^3}/{\anb{13}\anb{23}} \; \cvvv{}{}{} \,,\\
\mvvv-+- &= {\anb{13}^3}/{\anb{12}\anb{23}} \; \cvvv{}{}{} \,,\\
\mvvv+-- &= {\anb{23}^3}/{\anb{12}\anb{13}} \; \cvvv{}{}{} \,,\\
\mvvv-++ &= {\sqb{23}^3}/{\sqb{12}\sqb{13}} \; \cvvv{}{}{} \,,\\
\mvvv+-+ &= {\sqb{13}^3}/{\sqb{12}\sqb{23}} \; \cvvv{}{}{} \,,\\
\mvvv++- &= {\sqb{12}^3}/{\sqb{13}\sqb{23}} \; \cvvv{}{}{} \,.
\end{aligned}
\label{eqn:massless_VVV}
\end{equation}
The \amp+++, \amp--- amplitudes, and that involving vectors of both helicities arise from three different contractions of $\varepsilon^\pm_1, \varepsilon^\pm_2, \varepsilon^\pm_3$ polarization vectors and therefore have independent couplings.
The relevant structures are displayed in \subappref{sec:matching_wwz} for massive vectors.

%%%%%%%%%%%%%%%%%%%%%%%%%%%%%%%%%%%%%
\subsection{Two fermions, one vector and one scalar}
\label{sec:massless_ffzh}
%%%%%%%%%%%%%%%%%%%%%%%%%%%%%%%%%%%%%

In this section, we derive the non-factorizable parts of the massless $FF'VS$ four-point amplitude.
It can be spanned by only one type of bracket, square or angle.
We choose to span the amplitude by 
\begin{align}
    \mffvs{h_1}{h_2}{h_3} &=
    \sqb{12}^{h_{1}+h_{2}-h_{3}}\sqb{13}^{h_{1}+h_{3}-h_{2}}\sqb{23}^{h_{3}+h_{2}-h_{1}}
    \cffvs{h_1}{h_2}{h_3}/\La^{h_1+h_2+h_3}\eqncomma
\end{align}
where the $\cffvs{}{}{}$ are dimensionless functions of $s_{ij}/\La$.
We however exclude spinor bilinear in denominators and therefore implicitly use equalities of the type $1/\sqb{ij}=\anb{ji}/2(p_i\cdot p_j)$ to remove them.
We then obtain:
\begin{equation}
\begin{aligned}
\mffvs+++ &= \sqb{13}\sqb{23}	\cffvs+++/\La^2	\,,\\
\mffvs-++ &= \anb{12}\sqb{23}^2	\cffvs-++/\La^3	\,,\\
\mffvs+-+ &= \anb{12}\sqb{13}^2	\cffvs+-+/\La^3	\,,\\
\mffvs--+ &= \sqb{13}\sqb{23}\anb{12}^2	\cffvs--+/\La^4	\,,\\
\mffvs++- &= \anb{13}\anb{23}\sqb{12}^2	\cffvs++-/\La^4	\,,\\
\mffvs-+- &= \sqb{12}\anb{13}^2	\cffvs+-+/\La^3	\,,\\
\mffvs+-- &= \sqb{12}\anb{23}^2	\cffvs-++/\La^3	\,,\\
\mffvs--- &= \anb{13}\anb{23}	\cffvs---/\La^2	\,.
\end{aligned}
\label{eqn:massless_FFVS}
\end{equation}

%%%%%%%%%%%%%%%%%%%%%%%%%%%%%%%%%%%%%%
\section{Non-factorizable \texorpdfstring{$\psic\psi Zh$}{ffZh} for massless fermions}
\label{sec:ffZh_EFT_derivation}
%%%%%%%%%%%%%%%%%%%%%%%%%%%%%%%%%%%%%%
In this section, we derive the non-factorizable part of the $\mathcal{M}\parn{\psic \psi Z h}$ four-point amplitude with massless fermions. 
We begin by pulling out the polarization of the massive $Z$
\begin{align}
\mathcal{M}\parn{1_{\psi^{c}}^{h_1},2_{\psi}^{h_2},\bs 3_{Z},\bs 4_{h}}&=  
    \bra{\bs 3}^{\alpha}\bra{\bs 3}^{\beta}M_{\alpha\beta} = 
    \Bra{\bs 3}_{\dot{\alpha}}\Bra{\bs 3}_{\dot{\beta}}\bar{M}^{\dot{\alpha}\dot{\beta}}\eqncomma
\end{align}
where $M_{\alpha\beta}\, (\bar{M}^{\dot{\alpha}\dot{\beta}})$ is the reduced amplitude in the undotted (dotted) basis. 
We can span the ${SL}\parn{2,\mathbb{C}}$ space of the reduced amplitudes\footnote{Any other structure with an open ${SL}\parn{2,\mathbb{C}}$ index can be spanned by 
 $\ket 1_{\alpha},\ket 2_{\beta}$ or $\Ket 1^{\dot{\alpha}},\Ket 2^{\dot{\beta}}$.}
with the independent spinors  $\ket 1_{\alpha},\ket 2_{\beta}$ or $\Ket 1^{\dot{\alpha}},\Ket 2^{\dot{\beta}}$. 
There are four more helicity carrying structures which we can write
\begin{align}\label{eqn:ffZh_EFT_possible_helicity_structures}
    \anb{12}, \quad \sqb{12}, \quad \langle 1 \bs3 2], \quad \langle 2 \bs3 1]
    \eqncomma
\end{align}
and only two of these are independent.
Therefore, the reduced amplitudes can be spanned by two of the structures in~\autoref{eqn:ffZh_EFT_possible_helicity_structures}, and two helicity spinors for the ${SL}\parn{2,\mathbb{C}}$ part. 
The little-group transformation properties and mass dimension, then fix the form of the reduced amplitudes. Thus the reduced amplitudes in the undotted and dotted bases are given by 
\begin{equation}
\begin{aligned}
    M_{\alpha\beta}&=
   ( \ket 1^{2-B}\ket 2^{B})_{\alpha \beta}\, \anb{12}^{-h_1 - h_2 - 1}\langle  1 \bs3 2]^{B-h_1+h_2-1}
    f_{2B +3h_1-h_2+1}\eqncomma \\
    \bar{M}^{\dot{\alpha}\dot{\beta}} &= (\Ket 1^{2-B}\Ket 2^{B})^{\dot \alpha \dot \beta}\, \sqb{12}^{h_1+h_2-1}
    \langle 2 \bs3 1]^{B+h_1-h_2-1}f_{2B-3h_1-h_2+1}\,,
    \end{aligned}
\end{equation}
where $B=0,1,2$, and $f_{-\ell}$ is a function of Mandelstam variables of mass dimension $-\ell$. 
We can choose a convenient basis for each helicity amplitude, dotted or undotted, and obtain
\begin{equation}
\label{eqn:ffZh_EFT_initial_form}\begin{aligned}
\mathcal{M}\parn{1_{\psi^{c}}^{+},2_{\psi}^{+},\bs 3_{Z},\bs 4_{h}} &=
    \frac{\sqb{1\bs 3}^{2}}{ \langle 2 \bs3 1]}\bar{F}_{0}^{\parn{++,1}} + 
    \sqb{1\bs 3}\sqb{2\bs 3}\bar{F}_{-2}^{\parn{++,2}} + 
    \sqb{2\bs 3}^{2}\langle 2 \bs3 1]\bar{F}_{-4}^{\parn{++,3}}\eqncomma\\
\mathcal{M}\parn{1_{\psi^{c}}^{+},2_{\psi}^{-},\bs 3_{Z},\bs 4_{h}} &=
    \frac{\sqb{1\bs 3}^{2}}{\sqb{12}}\bar{F}_{-1}^{\parn{+-,1}} +
    \frac{\sqb{1\bs 3}\sqb{2\bs 3}\langle 2 \bs3 1]}{\sqb{12}}\bar{F}_{-3}^{\parn{+-,2}} +
    \frac{\sqb{2\bs 3}^{2}\langle 2 \bs3 1]^{2}}{\sqb{12}}\bar{F}_{-5}^{\parn{+-,3}}\eqncomma\\
\mathcal{M}\parn{1_{\psi^{c}}^{-},2_{\psi}^{+},\bs 3_{Z},\bs 4_{h}} &=
    \frac{\anb{1\bs 3}^{2}}{\anb{12}}F_{-1}^{\parn{-+,1}} +
    \frac{\anb{1\bs 3}\anb{2\bs 3}\langle 1 \bs3 2]}{\anb{12}}F_{-3}^{\parn{-+,2}} +
    \frac{\anb{2\bs 3}^{2}\langle 1 \bs3 2]^{2}}{\anb{12}}F_{-5}^{\parn{-+,3}}\eqncomma\\
\mathcal{M}\parn{1_{\psi^{c}}^{-},2_{\psi}^{-},\bs 3_{Z},\bs 4_{h}} &=
    \frac{\anb{1\bs 3}^{2}}{ \langle 1 \bs3 2]}F_{0}^{\parn{--,1}} +
    \anb{1\bs 3}\anb{2\bs 3}F_{-2}^{\parn{--,2}} +
    \anb{2\bs 3}^{2}\langle 1 \bs3 2] F_{-4}^{\parn{--,3}}\eqndot
\end{aligned}
\end{equation}
The forms in \autoref{eqn:ffZh_EFT_initial_form} make it difficult to determine the EFT contribution. 
With some simplifications, we can recast them into the more useful form 
\begin{equation}
\label{eqn:ffZh_EFT_helicity_amp_appendix}\begin{aligned}
    \mathcal{M}\parn{1_{\psi^{c}}^{+},2_{\psi}^{+},\bs 3_{Z},\bs 4_{h}} & =\frac{c_{1}^{++}}{\bar{\Lambda}^{2}}\sqb{1\bs 3}\sqb{2\bs 3}+\frac{\sqb{12}}{\bar{\Lambda}^{3}}\parn{c_{2}^{++}\langle {\bs 3}(1 + 2){\bs 3}] +c_{3}^{++}\langle {\bs 3}(1 - 2){\bs 3}]}\eqncomma\\
\mathcal{M}\parn{1_{\psi^{c}}^{+},2_{\psi}^{-},\bs 3_{Z},\bs 4_{h}} & =\frac{c_{1}^{+-}}{\bar{\Lambda}^{2}}\sqb{1\bs 3}\anb{2\bs 3}+\frac{c_{2}^{+-}}{\bar{\Lambda}^{3}}\anb{12}\sqb{1\bs 3}^{2}+\frac{c_{3}^{+-}}{\bar{\Lambda}^{3}}\sqb{12}\anb{2\bs 3}^{2}\eqncomma\\
\mathcal{M}\parn{1_{\psi^{c}}^{-},2_{\psi}^{+},\bs 3_{Z},\bs 4_{h}} & =\frac{c_{1}^{-+}}{\bar{\Lambda}^{2}}\anb{1\bs 3}\sqb{2\bs 3}+\frac{c_{2}^{-+}}{\bar{\Lambda}^{3}}\sqb{12}\anb{1\bs 3}^{2}+\frac{c_{3}^{-+}}{\bar{\Lambda}^{3}}\anb{12}\sqb{2\bs 3}^{2}\eqncomma\\
\mathcal{M}\parn{1_{\psi^{c}}^{-},2_{\psi}^{-},\bs 3_{Z},\bs 4_{h}} & =\frac{c_{1}^{--}}{\bar{\Lambda}^{2}}\anb{1\bs 3}\anb{2\bs 3}+\frac{\anb{12}}{\bar{\Lambda}^{3}}\parn{c_{2}^{--}\langle {\bs 3}(1 + 2){\bs 3}] +c_{3}^{--}\langle {\bs 3}(1 - 2){\bs 3}]}\,.
\end{aligned}
\end{equation}

%%%%%%%%%%%%%%%%%%%%%%%%%%%%%%%%%%%%%
%%%%%%%%%%%%%%%%%%%%%%%%%%%%%%%%%%%%%

\end{fmffile}

\bibliographystyle{apsrev4-1_title}
\bibliography{on-shell-ewsb}

\end{document}